\documentclass[english,aps,twocolumn,showpacs,showkeys,preprintnumbers,prd,floatfix,nofootinbib,superscriptaddress,notitlepage,hidelinks]{revtex4-1}

\usepackage{graphicx}
\usepackage{amsmath}
\usepackage{amsfonts}
\usepackage{hyperref}
\usepackage{color}
\usepackage{placeins}
\usepackage{tabu}
\usepackage{nicefrac}
\usepackage{ulem}
\usepackage{float}
\usepackage{subfig}
\newcommand{\orcid}[1]{\,\href{https://orcid.org/#1}{\includegraphics[width=9pt]{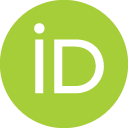}}\,}

\newcommand{\chidof}{\chi^2/N_{dof}}

\newcommand{\orcidTJ}{0000-0002-1334-7607} %
\newcommand{\orcidPD}{0000-0001-7960-7953} %
\newcommand{\orcidMK}{0000-0002-4665-3088} %
\newcommand{\orcidKK}{0000-0003-1412-447X} %
\newcommand{\orcidAK}{0000-0002-4090-0084} %
\newcommand{\orcidFM}{0000-0002-3888-1697} %
\newcommand{\orcidFO}{0000-0001-6799-2436} %
\newcommand{\orcidIS}{0000-0003-0373-474X} %
\newcommand{\orcidJY}{0000-0001-8366-0968} %
\newcommand{\orcidRR}{0000-0002-3316-2175} %
\begin{document}
\preprint{
\vbox{
\null \vspace{0.3in}
\hbox{MS-TP-22-10, IFJPAN-IV-2022-5}
}}

\title{Impact of heavy quark and quarkonium data on nuclear gluon PDFs}

\author{P.~Duwent\"aster\orcid{\orcidPD}}
\email{pit.duw@uni-muenster.de}
\affiliation{Institut f{ü}r Theoretische Physik, Westf{ä}lische Wilhelms-Universit{ä}t
M{ü}nster, Wilhelm-Klemm-Stra{ß}e 9, D-48149 M{ü}nster, Germany }

\author{T.~Je\v{z}o\orcid{\orcidTJ}}
\email{tomas.jezo@uni-muenster.de }
\affiliation{Institut f{ü}r Theoretische Physik, Westf{ä}lische Wilhelms-Universit{ä}t
M{ü}nster, Wilhelm-Klemm-Stra{ß}e 9, D-48149 M{ü}nster, Germany }

\author{M.~Klasen\orcid{\orcidMK}}
\affiliation{Institut f{ü}r Theoretische Physik, Westf{ä}lische Wilhelms-Universit{ä}t
M{ü}nster, Wilhelm-Klemm-Stra{ß}e 9, D-48149 M{ü}nster, Germany }

\author{K.~Kova\v{r}\'{\i}k\orcid{\orcidKK}}
\affiliation{Institut f{ü}r Theoretische Physik, Westf{ä}lische Wilhelms-Universit{ä}t
M{ü}nster, Wilhelm-Klemm-Stra{ß}e 9, D-48149 M{ü}nster, Germany }

\author{A.~Kusina\orcid{\orcidAK}}
\affiliation{Institute of Nuclear Physics Polish Academy of Sciences, PL-31342 Krakow, Poland}

\author{K.~F.~Muzakka\orcid{\orcidFM}}
\affiliation{Institut f{ü}r Theoretische Physik, Westf{ä}lische Wilhelms-Universit{ä}t
M{ü}nster, Wilhelm-Klemm-Stra{ß}e 9, D-48149 M{ü}nster, Germany }

\author{F.~I.~Olness\orcid{\orcidFO}}
\affiliation{Southern Methodist University, Dallas, TX 75275, USA }

\author{R.~Ruiz\orcid{\orcidRR}}
\affiliation{Institute of Nuclear Physics Polish Academy of Sciences, PL-31342 Krakow, Poland}

\author{I.~Schienbein\orcid{\orcidIS}}
\affiliation{Laboratoire de Physique Subatomique et de Cosmologie, Université
Grenoble-Alpes, CNRS/IN2P3, 53 avenue des Martyrs, 38026 Grenoble,
France }

\author{J.~Y.~Yu\orcid{\orcidJY}}
\affiliation{Laboratoire de Physique Subatomique et de Cosmologie, Université
Grenoble-Alpes, CNRS/IN2P3, 53 avenue des Martyrs, 38026 Grenoble,
France }

\date{\today}

\begin{abstract}
\vspace*{0.5cm}

A clear understanding of nuclear parton distribution functions (nPDFs) plays a crucial role in the interpretation of collider data taken at the Relativistic Heavy Ion Collider (RHIC), the Large Hadron Collider (LHC) and in the near future at the Electron-Ion Collider (EIC). 
Even with the recent inclusions of vector boson and light meson production data, the uncertainty of the gluon PDF remains substantial and limits the interpretation of heavy ion collision data.
To obtain new constraints on the nuclear gluon PDF, we extend our recent nCTEQ15WZ+SIH analysis to inclusive quarkonium and open heavy-flavor meson production data from the LHC.
This vast new data set covers a wide kinematic range and puts strong constraints on the nuclear gluon PDF down to $x\lesssim 10^{-5}.$
The theoretical predictions for these data sets are obtained from a data-driven approach, where proton-proton data are used to determine effective scattering matrix elements. 
This approach is validated with detailed comparisons to existing next-to-leading order (NLO) calculations in non-relativistic QCD (NRQCD) for quarkonia and in the general-mass variable-flavor-number scheme (GMVFNS) for the open heavy-flavored mesons. In addition, the uncertainties from the data-driven approach are determined using the Hessian method and accounted for in the PDF fits.
This extension of our previous analyses represents an important step toward the next generation of PDFs not only by including new data sets, but also by exploring new methods for future analyses.

\vspace*{2.5cm}
\end{abstract}

\maketitle
\tableofcontents{}

\section{Introduction}\label{sec:intro}

Parton distribution functions (PDFs) are fundamental quantities required to calculate predictions for processes involving hadronic initial states. The underlying theoretical framework is based on factorization theorems, which have been proven from first principles of QCD for a number of collider processes in $ep$ and $pp$ collisions~\cite{Collins:1989gx}. This formalism provides both a field theoretical definition of the PDFs,
and the definition of the short distance hard scattering cross sections at the partonic level.
Additionally, it includes a statement about the error of the collinear factorization formula, which is inversely proportional to some power of the hard scale of the process.
The predictive power of this formalism lies in the fact that the PDFs are universal, \textit{i.e.} process independent, whereas the process dependent short distance cross sections can be systematically calculated in perturbation theory. 
This approach has been widely used in analyses of proton PDFs, which have been constrained with great precision~\cite{Hou:2019efy, Ball:2017nwa, Khalek:2018mdn, Gao:2017yyd, Kovarik:2019xvh, Alekhin:2017olj, Nadolsky:2008zw, Sato:2019yez, Harland-Lang:2014zoa, Thorne:2019mpt, Ball:2009mk, Lin:2017snn, Lin:2020rut}.
Assuming that the twist-2 collinear factorization remains valid also in the case of $eA$ and $pA$ collisions, nuclear parton distribution functions (nPDFs) have been determined~\cite{Kovarik:2015cma, Eskola:2016oht, AbdulKhalek:2019mzd, AbdulKhalek:2020yuc, Ethier:2020way, Guzey:2019kik, Klasen:2017kwb, Klasen:2018gtb, Kovarik:2019xvh, Armesto:2015lrg, Eskola:2021nhw} as well, albeit with significantly larger uncertainties compared to the case of proton PDFs.
In particular, the poorly constrained nuclear gluon PDF has been the focus of recent nCTEQ studies~\cite{Kusina:2020lyz, Duwentaster:2021ioo}, but unfortunately the uncertainties have remained substantial. Through DGLAP evolution, the gluons also produce significant uncertainties on other flavors. Even the vector boson production and single inclusive hadron production data sets, that were included in our recent nCTEQ15WZ(+SIH) studies~\cite{Kusina:2020lyz,Duwentaster:2021ioo}, do not constrain the gluon below $x\approx10^{-3}$.

\begin{figure*}[htb!]
	\centering
	\includegraphics[width=0.48\textwidth]{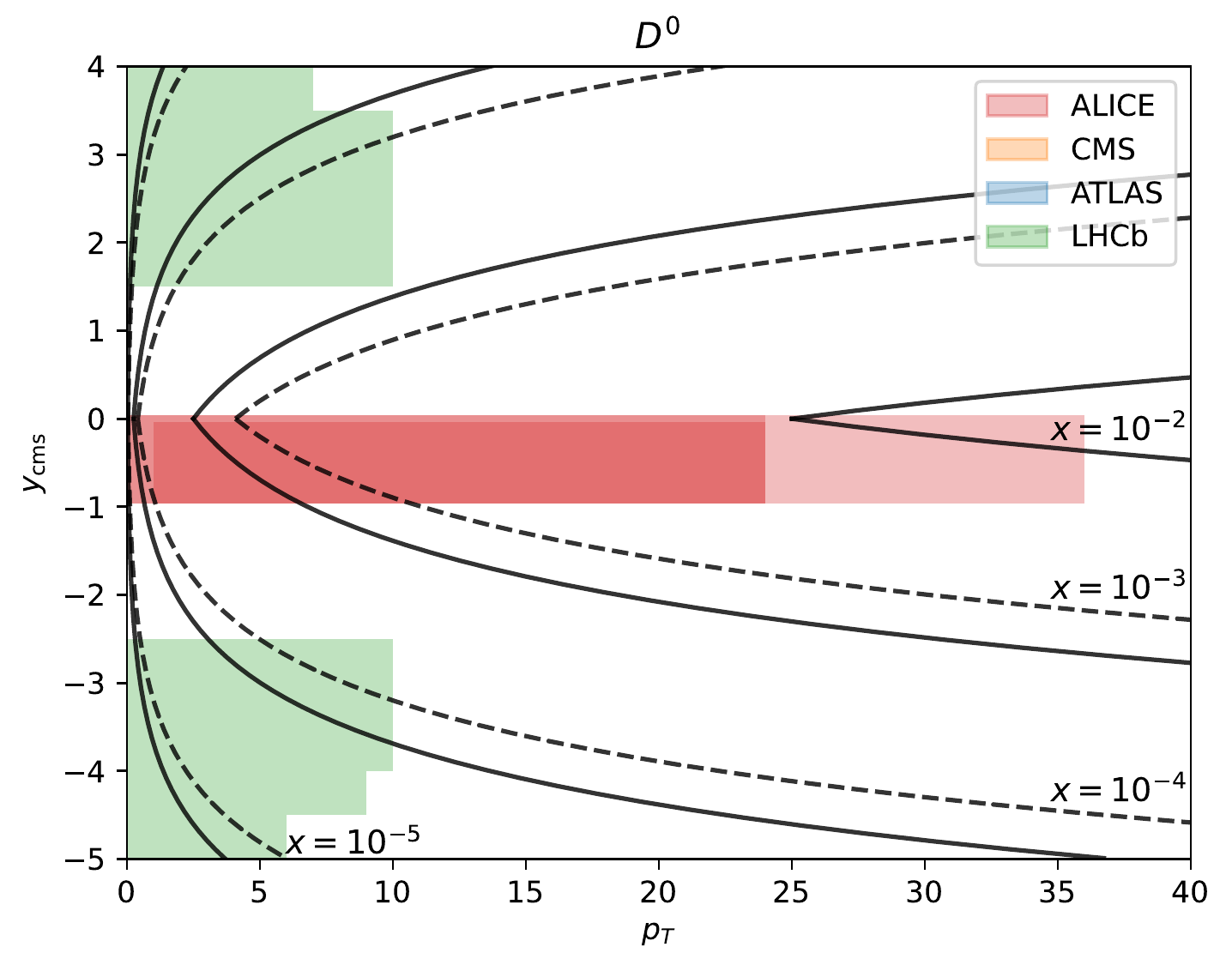}
	\includegraphics[width=0.48\textwidth]{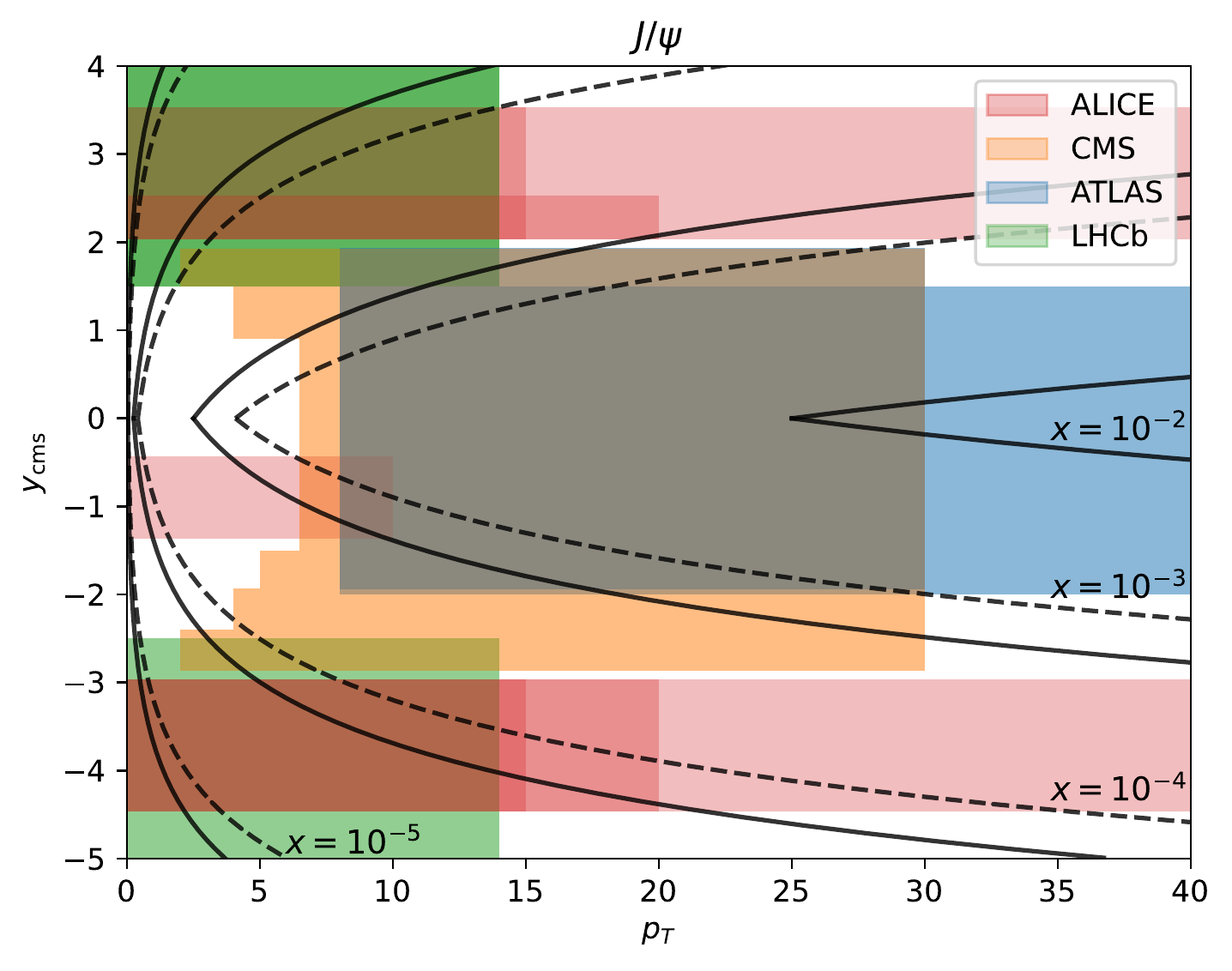}
	\includegraphics[width=0.48\textwidth]{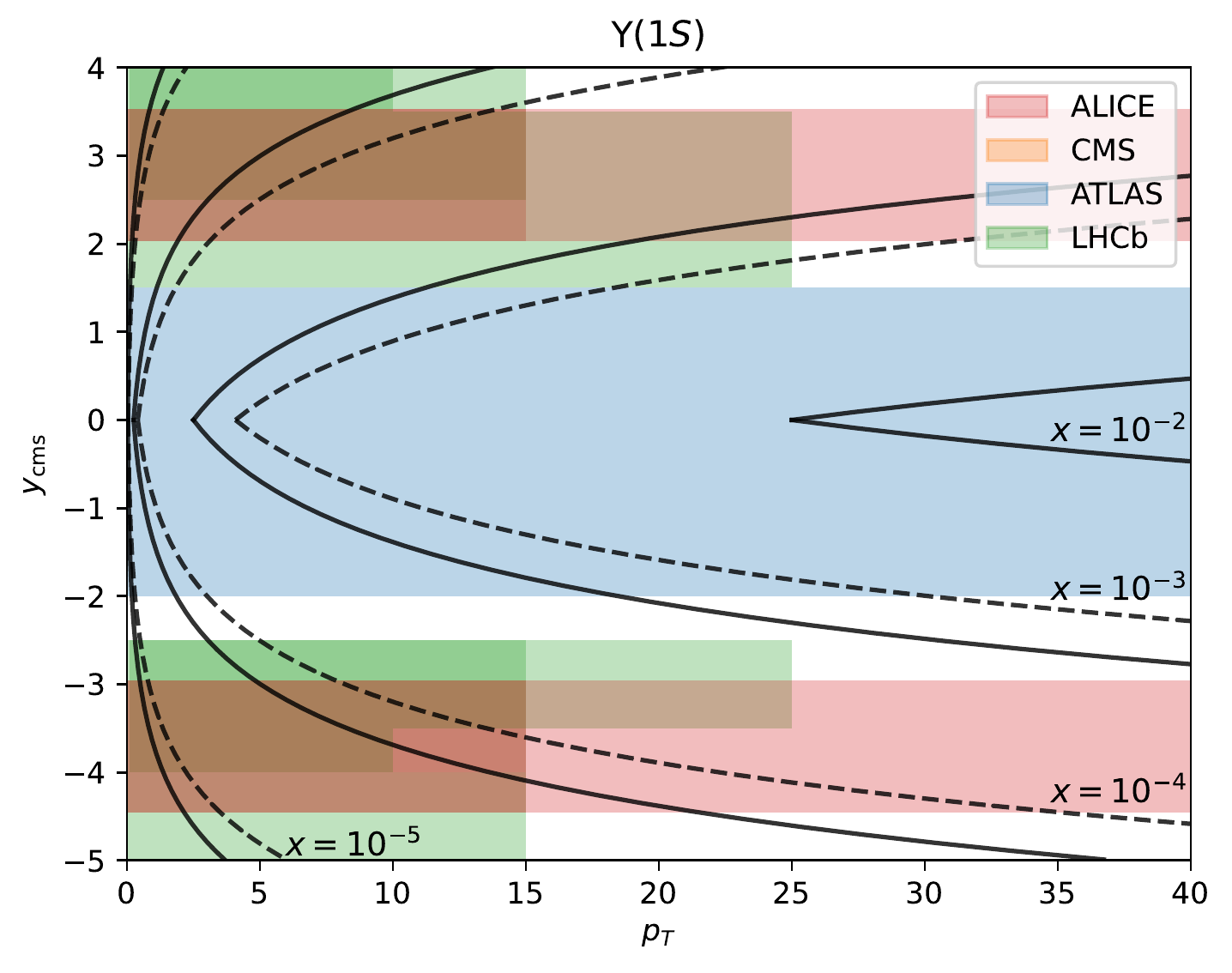}
	\includegraphics[width=0.48\textwidth]{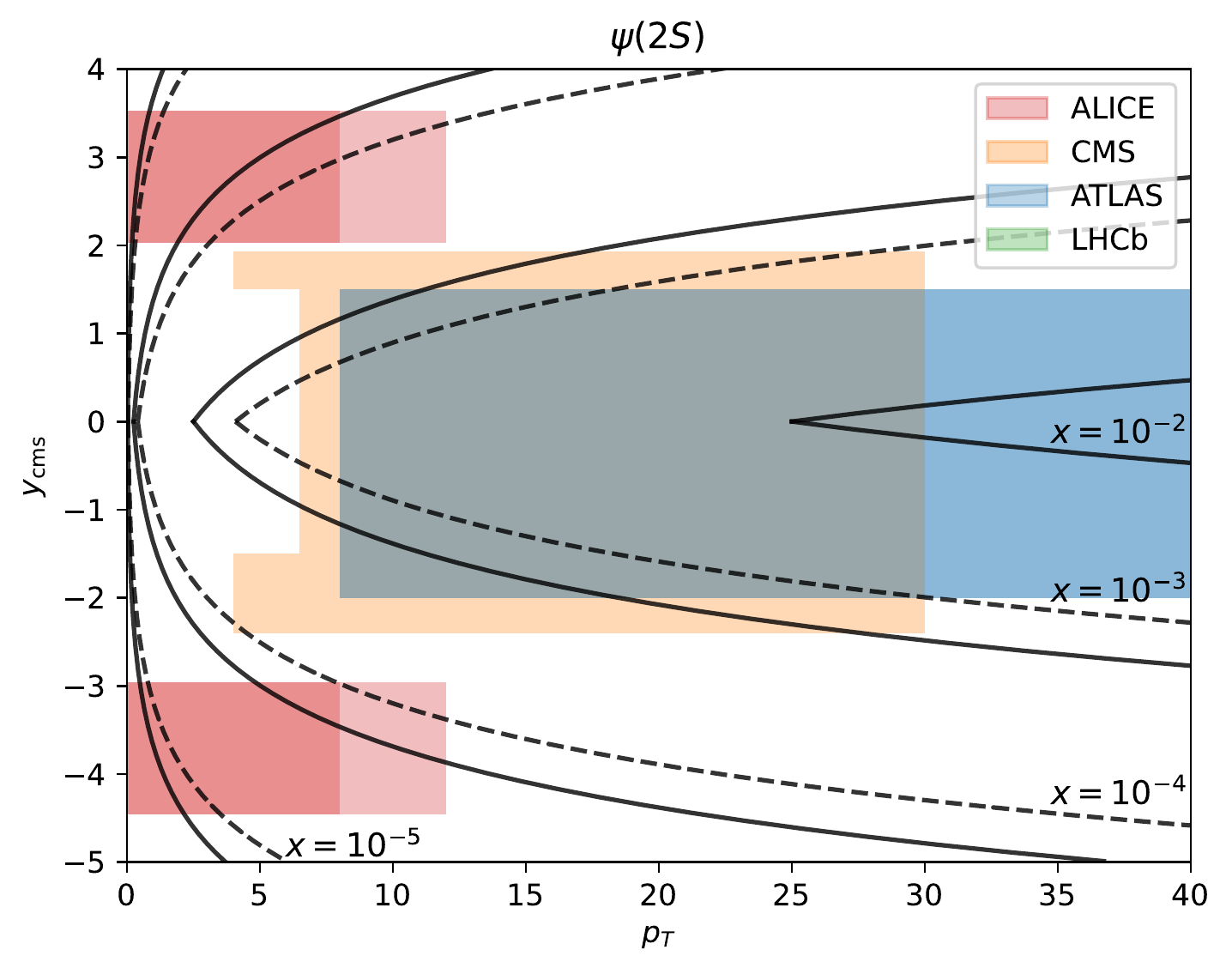}
	\caption{Coverage of the kinematic $(p_T,y_{\mathrm{cms}})$-plane of the quarkonium and open heavy quark production data sets from proton-lead collisions. ALICE data is shown in red, ATLAS in blue, CMS in orange and LHCb in green. The dashed and solid contours show the estimated $x$-dependence for $\sqrt{s}=5$ and 8\,TeV, respectively.}
	\label{fig:pPbKinematics}
\end{figure*}

In this paper, we perform a global analysis of nuclear PDFs in the nCTEQ framework, including open heavy flavor meson and heavy quarkonium data. Heavy quarkonium and open heavy-flavor meson production data (in the following collectively called heavy quark production, HQ) has the potential to yield new constraints on the gluon PDF, because the gluon-gluon channel contributes the dominant part to the overall cross section of these processes. This was first shown in the case of proton PDFs in Ref.~\cite{PROSA:2015yid}, before the reweighting study presented in Ref.~\cite{Kusina:2017gkz} was the first to demonstrate that the LHC heavy-quark data is also useful to constrain the nuclear gluon PDF at small $x$.
The $x$-dependence of the data can be estimated at leading order as 
\begin{align}
    x\approx \frac{2p_T}{\sqrt{s}} \exp(-|y|) ,
\end{align}
where $y$ is the rapidity in the centre-of-mass frame\footnote{All rapidities mentioned in the following are in the centre-of-mass frame and are therefore denoted simply by $y$.}. 
Fig.~\ref{fig:pPbKinematics} shows the kinematic coverage of the available data with contours for the corresponding $x$ dependence according to the estimate above.
This shows the data to be sensitive to $x$ values below $10^{-5}$ for the most forward or backward rapidity at low $p_T$.

$D$-meson production data in particular has been used in recent nuclear PDF analyses to reduce the uncertainty on the gluon PDF~\cite{Eskola:2021nhw, Khalek:2022zqe}. In Ref.~\cite{Eskola:2021nhw}, the ratio of double differential cross sections between proton-proton and proton-lead is compared to pQCD predictions in the GMVFNS~\cite{Helenius:2018uul, Kniehl:2004fy, Kniehl:2005mk} and the data is included directly in the global analysis. In contrast, Ref.~\cite{Khalek:2022zqe} compares the same data to predictions in a fixed-flavor number scheme with POWHEG and Pythia8 and uses this to perform a Bayesian reweighting of the PDFs.

In this investigation we will study single inclusive\footnote{The meaning of the word ``inclusive" can be somewhat ambiguous in this context. Here, we use it to denote processes, where the hadron is produced together with an arbitrary number of other particles, as opposed to exclusive production. This is the case for all processes studied in this paper. Later we will therefore use the word ``inclusive" to denote the sum of prompt and non-prompt production of a particle.} production of open heavy-flavor mesons and heavy quarkonia in proton-proton and proton-lead collisions. Including these processes in a global PDF fit is not as straight-forward as others, since there is no universally accepted theoretical model for quarkonia. Therefore we introduce a data-driven approach, that relies only on the following assumptions:
\begin{itemize}
    \item the gluon-gluon channel contributes the dominant fraction of the total cross section,
    \item it is sufficient to focus on subprocesses with 2 $\rightarrow$ 2 kinematics and subprocesses with more than 2 hard final state particles can be neglected.
\end{itemize}
If these assumptions are valid, the data-driven approach has considerable advantages over the other available calculations in perturbative QCD. Firstly, the studied processes can be accurately described across a large kinematic region with well controlled uncertainties. The second advantage is the speed of the calculation, which is significantly faster than the available pQCD calculations~\cite{Kniehl:2009ar}. Perturbative calculations at NLO would require very efficient gridding to be included in a fit, where the predictions for hundreds of data points need to be evaluated thousands of times.

The focus of this study will be on incorporating this process into a new global analysis, while accounting for the uncertainty of the data-driven theory, to determine the nuclear gluon PDF with greater accuracy than previously possible. 
A precise knowledge of the nuclear PDFs is important for the following reasons:
Firstly, they provide a description of the hadronic or nuclear structure in terms of quark and gluon degrees of freedom. In the context of the standard pQCD formalism the PDFs are universal and therefore required to make predictions for a wide range of collider observables. Finally, they provide a starting point for comparisons with microscopic models predicting the nuclear modifications (at the twist-2 level) in different $x$ regions.
This includes microscopic models for nuclear effects on PDFs in the shadowing region~\cite{Armesto:2006ph,Frankfurt:2011cs, Kopeliovich:2012kw, Kulagin:2004ie}, the antishadowing region~\cite{Brodsky:1989qz, Brodsky:2004qa, Kulagin:2004ie}, or the EMC effect~\cite{Geesaman:1995yd, Norton:2003cb, Hen:2013oha, Malace:2014uea, Hen:2016kwk, Kulagin:2004ie}.
At small-$x$ and moderately hard scales, the density of gluons becomes very large such that the assumptions underlying collinear factorization are expected to break down. This kinematic region is described by the theory of Color Glass Condensates~\cite{Iancu:2000hn, Gelis:2010nm} and there are also promising unified approaches which interpolate between the CGC at small $x$ and collinear factorization at large $x$, see, \textit{e.g.} Ref.~\cite{Jalilian-Marian:2020xvg} and references therein.
Nevertheless, it is fair to say that for now there is no unambiguous microscopic picture of the inner workings of heavier nuclei.

It should be stressed again, that throughout this paper, our main underlying assumption is that the twist-2 collinear factorization remains valid also in the case of $eA$ and $pA$ collisions for the same observables. As it has been discussed in Refs.~\cite{Qiu:2003cg, Accardi:2004be} this is reasonable, even if higher twist terms may be enhanced in the nuclear case up to higher hard scales $(\propto A^{1/3})$. We impose kinematic cuts on the data to effectively reduce the impact of these higher twist effects and confirm phenomenologically that all remaining data is well described.
In the future, such higher twist effects could be modelled to extend the reach towards data with lower hard scales. One example is the effects due to fully coherent energy loss~\cite{Arleo:2012hn,Arleo:2012rs, Arleo:2021bpv}. These contributions are formally higher twist (twist 3), but have been shown to be relevant for hard process data up to moderately large transverse momenta $p_T\approx10\,\mathrm{GeV}$. It could therefore be interesting to include such effects in future global analyses, however more work would be needed both on the conceptual and the phenomenological side. 

The next section provides an overview of the nCTEQ framework and the integration of the new data-driven approach. 
Following that, we perform and evaluate the fit of the proton-proton baseline for the theory in Sec.~\ref{sec:ppbaseline}.
In Sec.~\ref{sec:pdffit} we present the fits obtained using the HQ data and evaluate the compatibility between the new and old data.
Finally, in Sec.~\ref{sec:conclusion} we summarize our findings and give an outlook for future work.

\section{Theoretical approach}
\subsection{The nCTEQ framework}\label{sec:framework}
The nCTEQ project expands upon the foundation of the proton PDF global fitting analysis by including the nuclear dimension. In early proton PDF analyses~(\textit{e.g.}~Ref.~\cite{Olness:2003wz}), the nuclear data was used to calculate correction factors which were then applied to the proton PDF fit without any uncertainties. In contrast, the nCTEQ framework enables full communication between nuclear and proton data, which means that observed tensions between data sets can be investigated through the lens of nuclear corrections. 

The details of the nCTEQ15 nPDFs are presented in Ref.~\cite{Kovarik:2015cma}. The current analysis, along with the other recent nCTEQ analyses, such as  nCTEQ15WZ~\cite{Kusina:2020lyz}, nCTEQ15HIX~\cite{Segarra:2020gtj} and nCTEQ15WZ+SIH~\cite{Duwentaster:2021ioo}, is performed with a new \texttt{C++}-based code \texttt{nCTEQ++}. This allows us to easily interface external programs such as HOPPET~\cite{Salam:2008qg}, \mbox{APPLgrid}~\cite{Carli:2010rw}, and INCNLO~\cite{INCNLO}. In particular, we work at leading twist and next-to-leading order (NLO) of QCD for both the PDF and FF evolution equations as well as the hard scattering coefficients. The calculation code for the quarkonia and open heavy quarks is a partial \texttt{C++} adaption of HELAC-Onia 2.0~\cite{Shao:2015vga} and uses the data-driven approach explained in Sec.~\ref{sec:approach} instead of a pQCD calculation.

For the fits in this investigation, we use the same 19 parameters as for the nCTEQ15WZ(+SIH) sets. These 19 parameters include the 16 free parameters of the nCTEQ15 analysis, with an additional 3 open parameters for the strange distribution. For the nCTEQ15 set, the strange PDF was constrained by the relation $s = \bar{s} = {(\kappa/2)(\bar{u} {+} \bar{d})}$ at the initial scale $Q_0 = 1.3$\,GeV, which forces it into the same form as the other sea quarks. 

Our PDFs are parameterized at the initial scale
$Q_0 = 1.3$ GeV as
\begin{align}
    xf_i^{p/A}(x,Q_0)=c_0x^{c_1}(1-x)^{c_2}e^{c_3x}(1+e^{c_4}x)^{c_5} \ ,
\end{align}
and the nuclear $A$ dependence is encoded in the coefficients as
\begin{align}
    c_k \longrightarrow c_k(A) \equiv p_{k}+ a_{k}(1-A^{-b_{k}}) \ ,
\end{align}
where $k = \{1, ..., 5\}$.
The 16 free parameters used for the nCTEQ15 set describe the $x$-dependence of the 
\{$g, u_v, d_v, \bar{d} {+} \bar{u}$\} PDF combinations, and we do not vary the $\bar{d}/\bar{u}$ parameters; see Ref.~\cite{Kovarik:2015cma} for details. 
As in the nCTEQ15WZ(+SIH) analysis, we have added three strange PDF parameters: $\{a_{0}^{s+\bar{s}}, a_{1}^{s+\bar{s}}, a_{2}^{s+\bar{s}}\}$; these parameters correspond to the nuclear modification of the overall normalization, the low-$x$ exponent and the large-$x$ exponent of the strange quark distribution,
respectively. 

In total,  the 19 open parameters are:
\begin{align*}
    \{a^{u_v}_{1},\ a^{u_v}_{2},\ a^{u_v}_{4},\ a^{u_v}_{5},\ a^{d_v}_{1},\ a^{d_v}_{2},\ a^{d_v}_{5},\ a^{\bar{u}+\bar{d}}_{1},\  a^{\bar{u}+\bar{d}}_{5},\\ a^{g}_{1},\ a^{g}_{4},\ a^{g}_{5},\
    b^{g}_{0},\ b^{g}_{1},\ b^{g}_{4},\  b^{g}_{5},\ \bm{a^{s+\bar{s}}_{0},\ a^{s+\bar{s}}_{1},\ a^{s+\bar{s}}_{2}}\} .
\end{align*}

All the fixed parameters are kept as they were in nCTEQ15.

\subsection{The data-driven approach}\label{sec:approach}
Instead of performing the cross section calculations of the heavy mesons in perturbative QCD, we take the data-driven approach outlined initially in Ref.~\cite{Kom:2011bd} and used for a reweighting study in Refs.~\cite{Kusina:2017gkz, Kusina:2020dki}. In this approach, the cross section for two nuclei $A$ and $B$ scattering and producing a quarkonium or open heavy-flavor meson $\mathcal{Q}$ is calculated as the convolution integral 
of the two initial state gluon PDFs $f_{1,g}(x_1, \mu), f_{2,g}(x_2, \mu)$ and a fitted effective scattering matrix element $\overline{\left|\mathcal{A}_{gg\rightarrow \mathcal{Q} + X}\right|^2}$ over the $A B \rightarrow \mathcal{Q}$ phase space
\begin{align*}
&\sigma(A B \rightarrow \mathcal{Q}+X)= \\
&\int \mathrm{d} x_{1} \mathrm{~d} x_{2} f_{1,g}\left(x_{1}, \mu\right) f_{2,g}\left(x_{2}, \mu\right) \frac{1}{2 \hat{s}} \overline{\left|\mathcal{A}_{g g \rightarrow \mathcal{Q}+X}\right|^{2}} \mathrm{dPS}.
\end{align*}
The effective scattering matrix element is parameterized with the Crystal Ball function 
\begin{align}
\begin{split}
&\overline{\left|\mathcal{A}_{g g \rightarrow \mathcal{Q}+X}\right|^{2}}= \frac{\lambda^2\kappa\hat{s}}{M_\mathcal{Q}^2} e^{a|y|}\\ 
&\times\begin{cases} e ^{ -\kappa \frac{p_{T}^{2}}{M_{\mathcal{Q}}^{2}}} & \text { if } p_{T} \leq\left\langle p_{T}\right\rangle \\ e ^ {-\kappa \frac{\left\langle p_{T}\right\rangle^{2}}{M_{\mathcal{Q}}^{2}}}\left(1+\frac{\kappa}{n} \frac{p_{T}^{2}-\left\langle p_{T}\right\rangle^{2}}{M_{\mathcal{Q}}^{2}}\right)^{-n} & \text { if } p_{T}>\left\langle p_{T}\right\rangle\end{cases} , \label{eqn:CrystalBall}
\end{split}
\end{align}
where the five parameters\footnote{The parameter name $``\left<p_T \right>"$ is somewhat misleading. The parameterization was initially invented for a different purpose, where this parameter did have the physical meaning of the particle's average transverse momentum, but this interpretation is lost in the current context. However, we decided to keep the name to keep consistency with previous works.} $\lambda$, $\kappa$, $\left<p_T \right>$, $n$ and $a$ are then fitted for each final state $\mathcal{Q}$. We have introduced the fifth parameter $a$, which was not present in the original parameterization~\cite{Gaiser:1982}, to allow for a more accurate reproduction of the rapidity dependence~\cite{ellis2003qcd}. 
The parameters are then fitted to $pp \rightarrow \mathcal{Q}+X$ data. 
Once the optimal parameters are found, we can also determine the uncertainty of our Crystal Ball fit via the same Hessian method used to calculate our PDF uncertainties. We can then account for these uncertainties by adding them in quadrature to the systematic uncertainties of the $pPb \rightarrow \mathcal{Q}+X$ data. The included final states in this analysis are $D^0$, $J/\psi$, $\Upsilon(1S)$ and $\psi(2S)$ mesons. Note, however, that prompt and non-prompt production of the same particle need to be considered as two different final states. Inclusive production is generally not fitted separately, but calculated as the sum of the other two. The exception to this is $\Upsilon(1S)$, where all available data is for inclusive production. Other final states, like $D^\pm$ or higher excitations of $\Upsilon$, are excluded due to insufficient data, but in principle the approach can be used for any measurement of single inclusive hadrons, as long as sufficient proton-proton data is available for the baseline and the cross section is dominated by the gluon-gluon channel.

The default scales $\mu=\mu_0$ that enter the PDFs are chosen as they were in the previous reweighting study, as shown in Tab.~\ref{tab:scales}. The entire procedure is also repeated once with the scale doubled and once with the scale halved to obtain an estimate of the impact that this choice has on the final result.
\begin{table*}[!htbp]
    \renewcommand{\arraystretch}{1.8}
    \setlength\tabcolsep{5pt}
	\centering
	\caption{Scale choices for the different particles.}
	\begin{tabular}{|c||c|c|c|c|c|c|}
		\hline 
		     & $D^0$ & $J/\psi$ & $B \rightarrow J/\psi$ & $\Upsilon(1S)$  & $\psi(2S)$ & $B \rightarrow \psi(2S)$ \\
		\hline
		\hline
		 $\mu_0^2$ & 4$M_D^2+p_{T,D}^2$ & $M_{J/\psi}^2+p_{T,J/\psi}^2$ & 4$M_{B}^2+\frac{M_B^2}{M_{J/\psi}^2}p_{T,J/\psi}^2$ & $M_{\Upsilon(1S)}^2+p_{T,\Upsilon(1S)}^2$ & $M_{\psi(2S)}^2+p_{T,\psi(2S)}^2$ & 4$M_{B}^2+\frac{M_B^2}{M_{\psi(2S)}^2}p_{T,\psi(2S)}^2$ \\
		\hline
	\end{tabular}     	
	\label{tab:scales}
\end{table*}

\section{Proton-Proton baseline}\label{sec:ppbaseline}
In the first step of the analysis, we use the proton-proton data sets listed in Tabs.~\ref{tab:databaseline_D0}~-~\ref{tab:databaseline_Y1S} to determine the parameters of the Crystal Ball function. The largest portion of the proton-proton data comes from $J/\psi$ production, followed by $\Upsilon(1S)$, then $\psi(2S)$ and finally $D^0$.

\subsection{Cuts and excluded data}
Both in the baseline fit and the PDF fit, we cut all heavy-quark data with $p_T<3\,$GeV and outside of the rapidity range $-4<y_{cms}<4$. These cuts allow predictions with reasonable $\chi^2$ for all remaining data points. Relaxing the cuts introduces data points with $\chidof=\mathcal{O}(10)$. 
We also exclude the lowest $p_T$ bin of the 2011 CMS $J/\psi$ data sets, because the corresponding paper mentions that there may be large acceptance effects in this bin that are not included in the uncertainties. The lowest rapidity bin of the 7~TeV $\Upsilon(1S)$ production data set from LHCb is included in the fit, but the normalization is determined separately from the rest of the data set. This $y$-bin contains 20 $p_T$ bins (17 after cuts), which qualitatively agree with the remaining fit, but the normalization is off by 25\%. Therefore, keeping the normalization of this bin the same as the remainder of the data set causes the $\chidof$ for the entire $\Upsilon(1S)$ fit to increase from 0.92 to 1.6. This particular bin is also not described by other models based on the color-octet mechanism \cite{LHCb:2015log}. 

We also limit the study to LHC data only, as the available RHIC data taken at very different energies is not guaranteed to work with the same fit parameters and would also likely not provide additional strong constraints.

\subsection{Baseline fit}
A comparison of the fits with the data is shown in Figs.~\ref{fig:ppBaselineJPSI} - \ref{fig:ppBaselinePSI2S}. The uncertainties of the fit are determined via the same Hessian method that is used in our PDF fits. All predictions show very close agreement with the data across the region included in the fit. The large number of data points produces very small uncertainties for $J/\psi$ and $\Upsilon(1S)$ (not visible on the logarithmic scale). While the uncertainties of $D^0$ and $\psi(2S)$ are somewhat larger, due to the lower number of data points, they are still small compared to the experimental uncertainties.

The obtained parameters and values of $\chidof{}$ for each process are given in Tab.~\ref{tab:chi2baseline}. The parameters for $J/\psi$ and $\psi(2S)$ are each obtained in a combined fit with prompt, non-prompt and inclusive data, which results in only one value of $\chidof$ for each. There is overall very good agreement between the data and the fitted theory with $\chidof$ values slightly below one for $J/\psi$, $\psi(2S)$ and $\Upsilon(1S)$ and a particularly low $\chidof$ value of 0.25 obtained for $D^0$ production. 

Alternative baseline fits have been performed with a variety of parameterizations including more degrees of freedom, like the extended Crystal Ball function (CB2) \cite{SignalExtractionALICE}, which has a polynomial tail on both sides of the Gaussian for a total of seven parameters. In those fits, there is no significant improvement in the description of the currently included data and relaxing the cuts still leads to unreasonable $\chi^2$ values for the data points that would be introduced. Therefore, we keep the parameterization as written in Eq.~\eqref{eqn:CrystalBall} as a good compromise between a reasonable number of parameters and accuracy of the data description.

\begin{table*}[htbp!]
    \renewcommand{\arraystretch}{1.5}  %
    \setlength\tabcolsep{5pt}
	\caption{Crystal Ball parameters and $\chi^2/d.o.f.$ values for the Crystal Ball function for the different processes.}
	\centering
	\vspace{0.2cm}
	\begin{tabular}{|c||c|c|c|c|c|c|}
		\hline 
		     & $D^0$ & $J/\psi$ & $B \rightarrow J/\psi$ & $\Upsilon(1S)$  & $\psi(2S)$ & $B \rightarrow \psi(2S)$ \\
		\hline
		\hline
		 $\kappa$               & 0.33457 & 0.47892 & 0.15488 & 0.94524 & 0.21589 & 0.45273 \\
		\hline
		 $\lambda$              & 1.82596 & 0.30379 & 0.12137 & 0.06562 & 0.07528 & 0.13852 \\
		\hline
		 $\left<p_T \right>$    & 2.40097 & 5.29310 &-7.65026 & 8.63780 & 8.98819 & 7.80526 \\
		\hline
		 $n$                    & 2.00076 & 2.17366 & 1.55538 & 1.93239 & 1.07203 & 1.64797 \\
		\hline 
		 $a$                    &-0.03295 & 0.02816 &-0.08083 & 0.22389 &-0.10614 & 0.06179 \\
		\hline 
		\hline 
		$N_{\mathrm{points}}$ & 34 & \multicolumn{2}{c|}{501}  & 375  & \multicolumn{2}{c|}{55} \\
		\hline 
		$\chidof$ & 0.25 & \multicolumn{2}{c|}{0.88}  & 0.92  & \multicolumn{2}{c|}{0.77} \\
		\hline 
	\end{tabular}     	
	\label{tab:chi2baseline}
\end{table*}

\begin{table*}[htbp!]
    \renewcommand{\arraystretch}{1.4}
    \setlength\tabcolsep{4pt}
	\caption{Overview of the available  $pp \  \longrightarrow \  D^0+X$  production data sets and their number of data points.}
	\centering	
	\begin{tabular}{|c|c|c|c|c|c|}
		\hline 
		Group & Year & Ref. & ID & Type & Points after / before cuts \\ 
		\hline
		\hline
		ALICE  & 2012 & \cite{ALICE:2012mhc} & 3008 & Prompt & 7 / 9  \\ 
		\hline 
		LHCb  & 2013 & \cite{LHCb:2013xam} & 3014 & Prompt & 22 / 38  \\ 
		\hline 
		ALICE  & 2016 & \cite{ALICE:2016yta} & 3021 & Prompt & 7 / 10  \\ 
		\hline 
	\end{tabular}     	
	\label{tab:databaseline_D0}
	
	\caption{Overview of the available  $pp \  \longrightarrow \  J/\psi+X$  production data sets and their number of data points. Note that the ALICE 2017 data set is split into two for technical reasons (one part is taken at 5~TeV and one at 13~TeV).}
	\centering	
	\begin{tabular}{|c|c|c|c|c|c|}
		\hline 
		Group & Year & Ref. & ID & Type & Points after / before cuts \\ 
		\hline
		\hline
		ATLAS  & 2011 & \cite{ATLAS:2011aqv} & 3003 & Prompt & 63 / 64  \\ 
		\hline 
		CMS  & 2011 & \cite{CMS:2011rxs} & 3001 & Prompt & 39 / 44  \\ 
		\hline 
		LHCb  & 2011 & \cite{LHCb:2011zfl} & 3002 & Prompt & 50 / 66  \\ 
		\hline 
		CMS  & 2017 & \cite{CMS:2017exb} & 3020 & Prompt & 50 / 52  \\
		\hline 
		ATLAS  & 2018 & \cite{ATLAS:2017prf} & 3017 & Prompt & 33 / 33  \\ 
		\hline 
		\hline 
		CMS  & 2010 & \cite{CMS:2010nis} & 3007 & Non-prompt & 11 / 14  \\ 
		\hline
		ATLAS  & 2011 & \cite{ATLAS:2011aqv} & 3006 & Non-prompt & 63 / 64  \\ 
		\hline 
		CMS  & 2011 & \cite{CMS:2011rxs} & 3005 & Non-prompt & 39 / 44  \\ 
		\hline 
		LHCb  & 2011 & \cite{LHCb:2011zfl} & 3004 & Non-prompt & 50 / 66  \\ 
		\hline 
		CMS  & 2017 & \cite{CMS:2017exb} & 3019 & Non-prompt & 50 / 52  \\ 
		\hline 
		ATLAS  & 2018 & \cite{ATLAS:2017prf} & 3018 & Non-prompt & 33 / 33  \\ 
		\hline 
		\hline
		ALICE  & 2015 & \cite{ALICE:2015pgg} & 3022 & Inclusive & 10 / 13  \\ 
		\hline 
		ALICE  & 2017 & \cite{ALICE:2017leg} & 3023 & Inclusive & 8 / 11  \\ 
		\hline 
		ALICE  & 2017 & \cite{ALICE:2017leg} & 3024 & Inclusive & 14 / 18  \\ 
		\hline 
		ALICE  & 2019 & \cite{ALICE:2019pid} & 3016 & Inclusive & 4 / 7  \\ 
		\hline 
	\end{tabular}     	
	\label{tab:databaseline_JPSI}
	
	\caption{Overview of the available  $pp \ \longrightarrow \ \Upsilon(1S)+X$  production data sets and their number of data points. Note that the LHCb 2015 data set is split into two for technical reasons (one half is taken at 7~TeV and one at 8~TeV).}
	\centering	
	\begin{tabular}{|c|c|c|c|c|c|}
		\hline 
		Group & Year & Ref. & ID & Type & Points after / before cuts \\ 
		\hline
		\hline 
		ATLAS  & 2012 & \cite{ATLAS:2012lmu} & 3012 & Inclusive & 88 / 100  \\ 
		\hline 
		LHCb  & 2012 & \cite{LHCb:2012aa} & 3025 & Inclusive & 55 / 75  \\ 
		\hline 
		CMS  & 2013 & \cite{CMS:2013qur} & 3013 & Inclusive & 30 / 42  \\ 
		\hline
		ALICE  & 2014 & \cite{ALICE:2014uja} & 3009 & Inclusive & 3 / 5  \\ 
		\hline 
		LHCb  & 2015 & \cite{LHCb:2015log} & 3011 & Inclusive & 89 / 109  \\ 
		\hline 
		LHCb  & 2015 & \cite{LHCb:2015log} & 3015 & Inclusive & 89 / 109  \\ 
		\hline 
		ALICE  & 2015 & \cite{ALICE:2015pgg} & 3031 & Inclusive & 3 / 5  \\ 
		\hline 
		ATLAS  & 2017 & \cite{ATLAS:2017prf} & 3031 & Inclusive & 18 / 24  \\ 
		\hline 
	\end{tabular}     	
	\label{tab:databaseline_Y1S}
	
	\caption{Overview of the available  $pp \ \longrightarrow \ \psi(2S)+X$  production data sets and their number of data points.}
	\centering	
	\begin{tabular}{|c|c|c|c|c|c|}
		\hline 
		Group & Year & Ref. & ID & Type & Points after / before cuts \\ 
		\hline
		\hline 
		ATLAS  & 2017 & \cite{ATLAS:2017prf} & 3026 & Prompt & 15 / 15  \\ 
		\hline 
		\hline 
		ATLAS  & 2017 & \cite{ATLAS:2017prf} & 3027 & Non-prompt & 15 / 15  \\ 
		\hline 
		\hline 
		ALICE  & 2015 & \cite{ALICE:2015pgg} & 3028 & Inclusive & 6 / 9  \\ 
		\hline 
		CMS  & 2018 & \cite{CMS:2018gbb} & 3029 & Inclusive & 9 / 10  \\ 
		\hline 
		ALICE  & 2017 & \cite{ALICE:2017leg} & 3030 & Inclusive & 9 / 12  \\ 
		\hline 
	\end{tabular}     	
	\label{tab:databaseline_PSI2S}
\end{table*}

\subsection{Comparison with $J/\psi$ production in NRQCD}
\begin{figure*}[htbp!]
	\centering
	\includegraphics[width=0.48\textwidth]{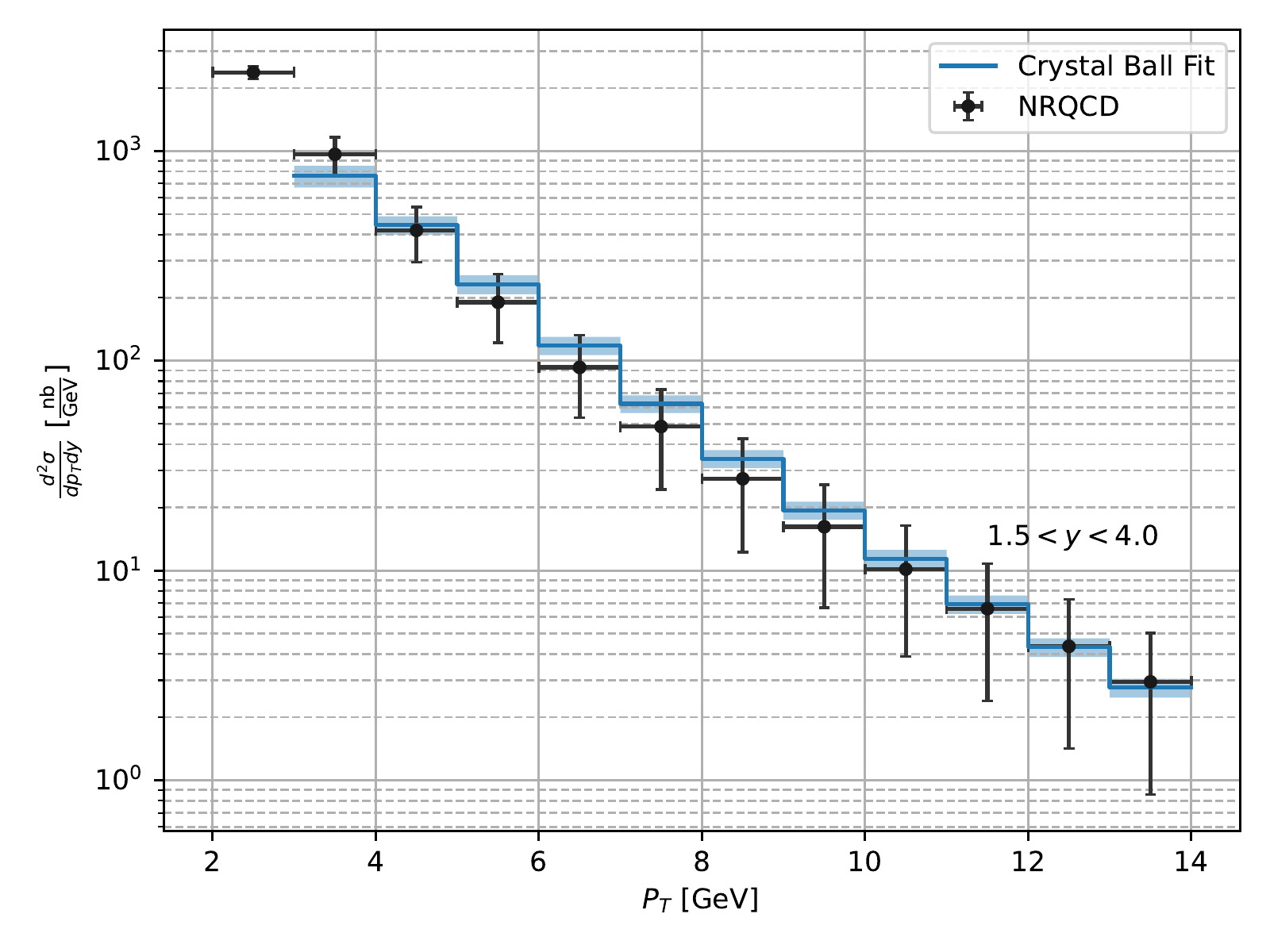}
	\includegraphics[width=0.48\textwidth]{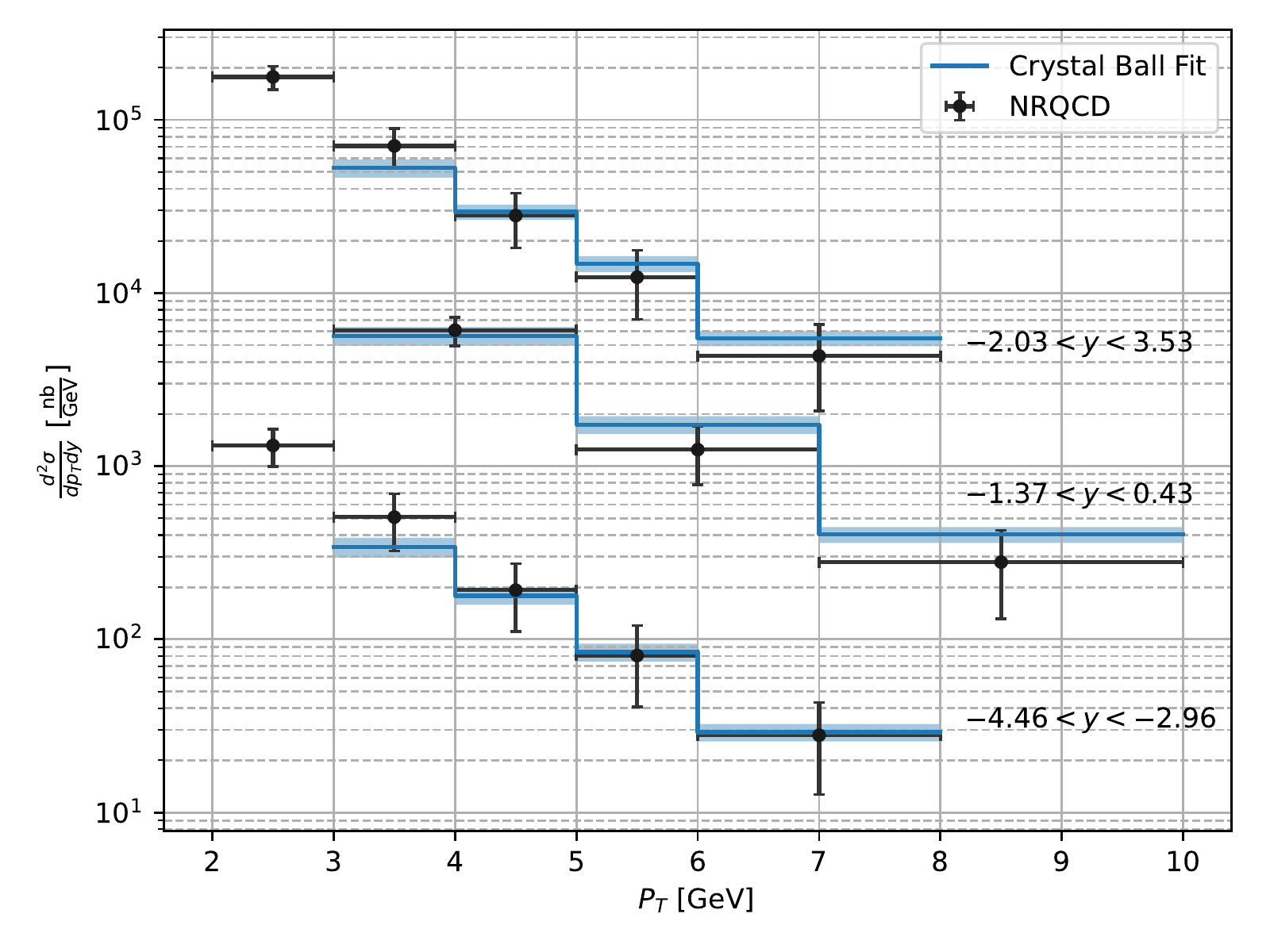}
	\includegraphics[width=0.48\textwidth]{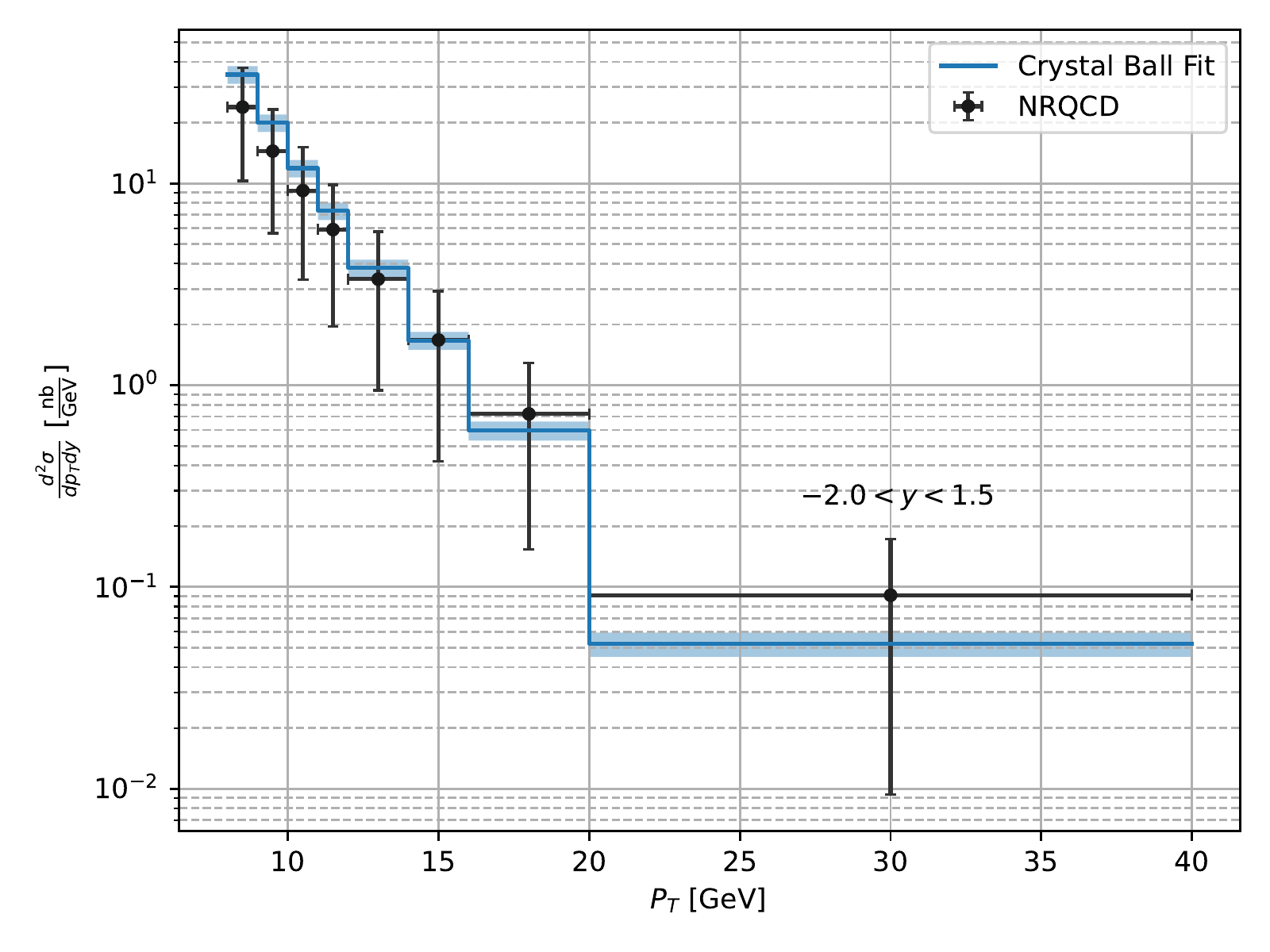}
	\caption{Comparison between prompt $J/\psi$ production in $pp$ collisions for LHCb\cite{LHCb:2017ygo}, ALICE\cite{ALICE:2018mml} and ATLAS\cite{ATLAS:2015mpz} kinematics as predicted by NRQCD and with the data-driven approach. The uncertainties of the NRQCD predictions come from scale variation $1/2 < \mu_r/\mu_{r,0} = \mu_f/\mu_{f,0} = \mu_{\mathrm{NRQCD}}/\mu_{\mathrm{NRQCD},0} < 2$ around the base scale $\mu_{r,0} = \mu_{f,0}$ = $\sqrt{p_T^2+4m_c^2}$ and $m_{\mathrm{NRQCD},0} = m_c$. Different rapidity bins are separated by multiplying the cross sections by powers of ten for visual clarity.}
	\label{fig:NRQCD}
\end{figure*}

The predictions for $J/\psi$ production from the data-driven method can also be compared to predictions made using non-relativistic QCD (NRQCD). The NRQCD framework assumes the speed of the heavy quarks to be slow compared to the speed of light to derive a factorization theorem where the physics is separated into short- and long-distance factors~\cite{Bodwin:1994jh}. The short-distance physics, \textit{i.e.} the production of the heavy quark-antiquark pair, can then be calculated perturbatively, while the non-perturbative long distance matrix elements parameterizing the formation of the bound state need to be determined empirically. This has been done by various independent groups including Ma et al.~\cite{Ma:2010yw, Ma:2010jj} and Butenschoen et al.~\cite{Butenschoen:2010rq, Butenschoen:2011yh}. The latter group has provided us with predictions for prompt $J/\psi$ production using their NLO calculation and our nCTEQ15 proton PDF, which is shown in Fig.~\ref{fig:NRQCD}. The uncertainties shown for NRQCD are calculated by varying all scales together by a factor of two around their default values $\mu_{r,0} = \mu_{f,0}$ = $\sqrt{p_T^2+4m_c^2}$ and $m_{\mathrm{NRQCD},0} = m_c = 1.5$\,GeV. There is a very good agreement between the two methods across the entire kinematic range, but the Crystal Ball method produces significantly smaller uncertainties. Varying the scales in the Crystal Ball fit produces only negligible differences, as the scale dependence is mostly absorbed into the parameters of the effective matrix element.

\subsection{Comparison with $D^0$ production in the GMVFNS}
\begin{figure*}[htbp!]
	\centering
	\includegraphics[width=0.48\textwidth]{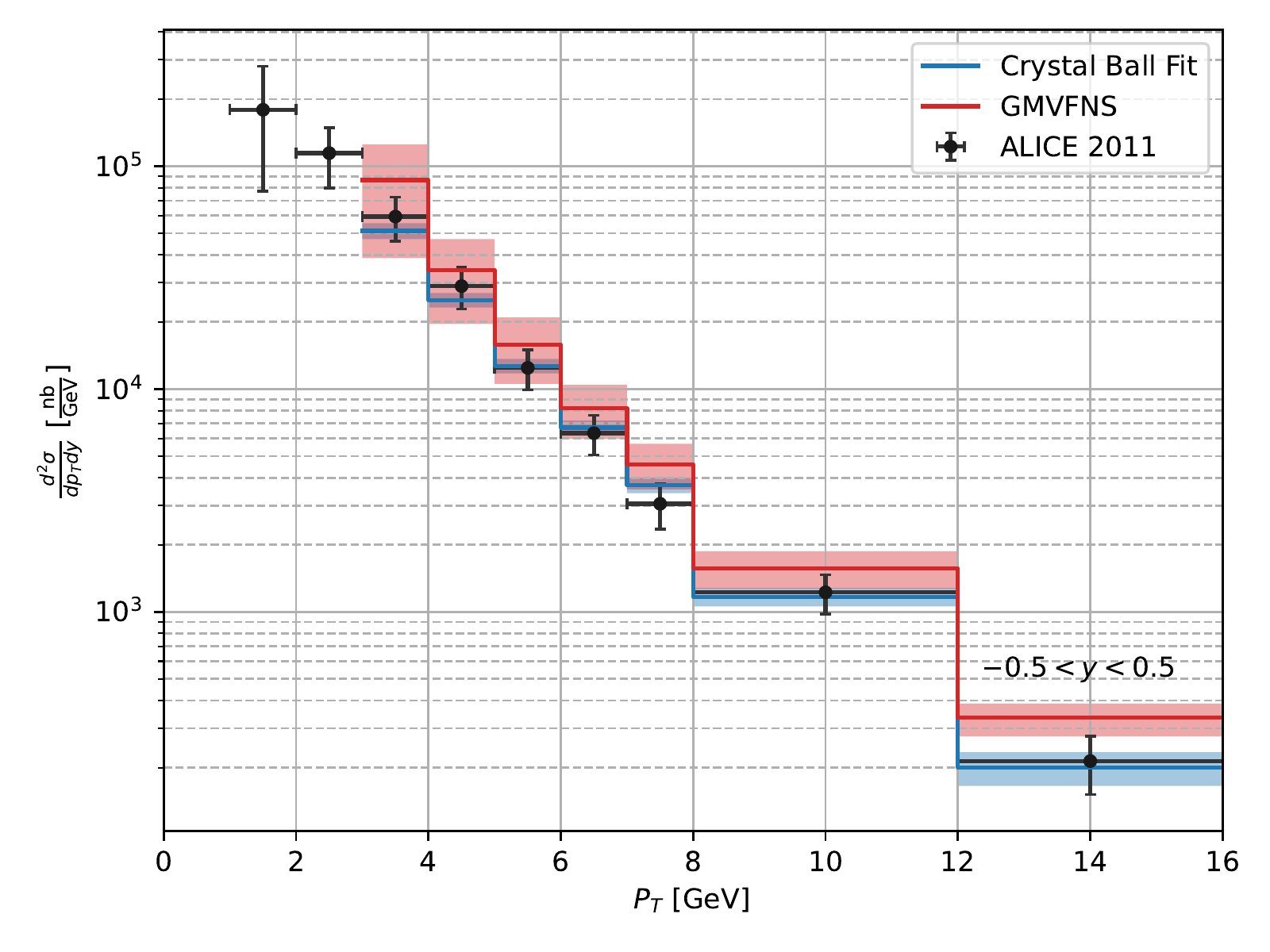}
	\includegraphics[width=0.48\textwidth]{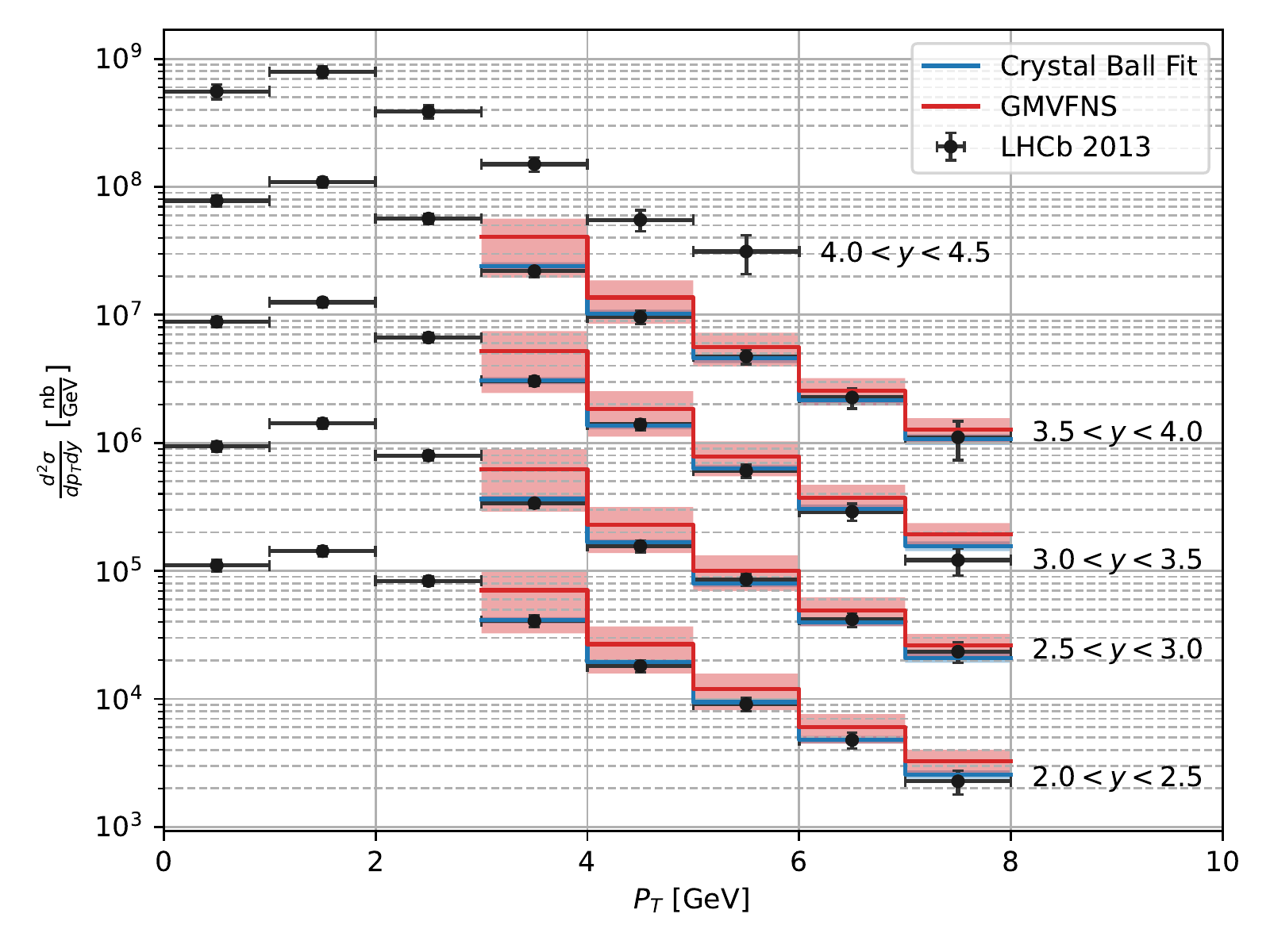}
	\includegraphics[width=0.48\textwidth]{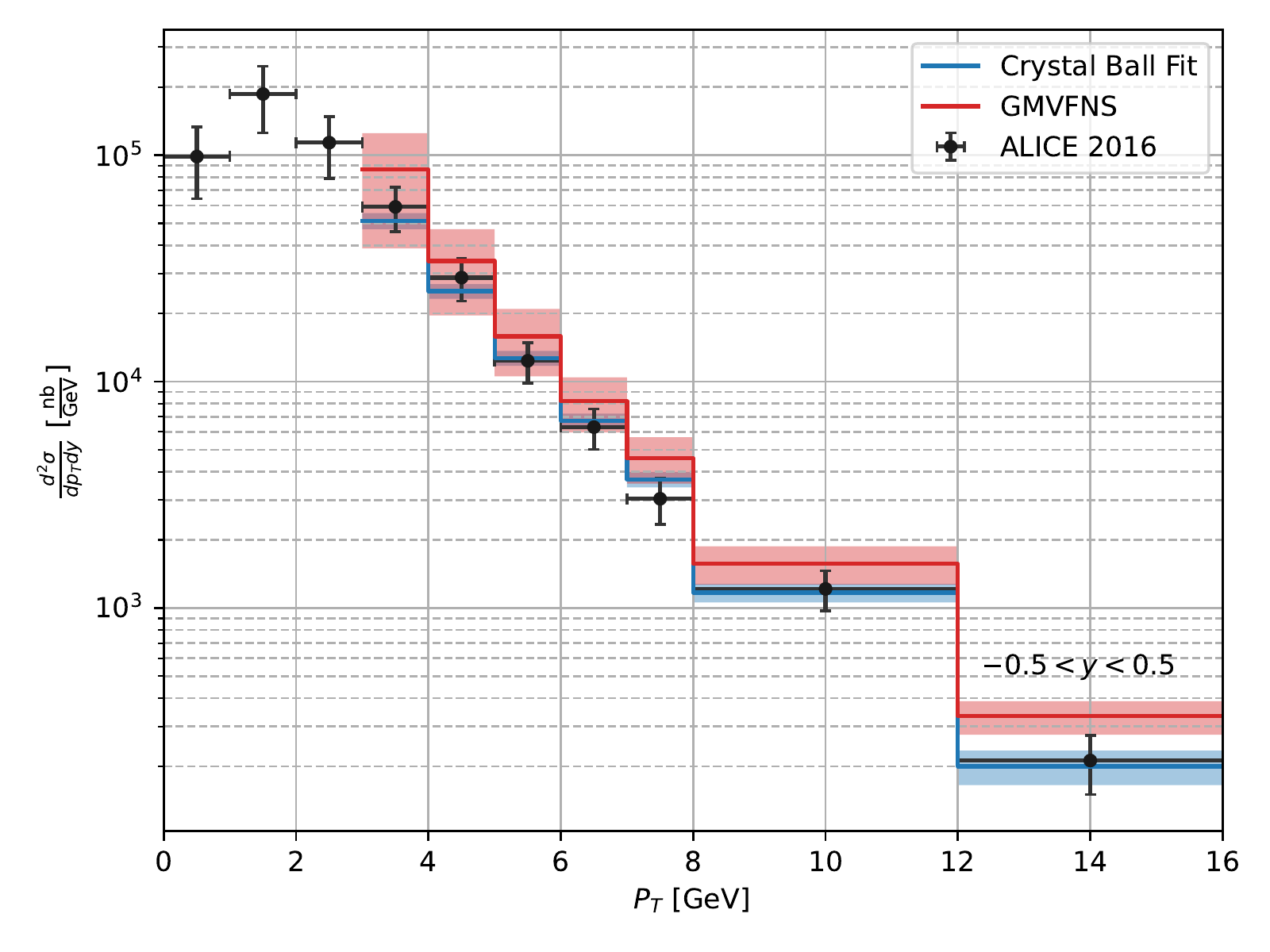}
	\caption{Comparison between prompt $D^0$ production as predicted in the GMVFNS (red) and with the data-driven approach (blue). The uncertainties of the GMVFNS predictions come from varying the scales individually by a factor of 2, such that there is never a factor 4 between two scales. Different rapidity bins are separated by multiplying the cross sections by powers of ten for visual clarity.}
	\label{fig:GMVFNS}
\end{figure*}

The predictions for $D^0$ production can also be compared with perturbative calculations. These calculations can be carried out using the General-Mass Variable-Flavor-Number-Scheme (GMVFNS) implementation of heavy quark production at NLO QCD by Kniehl et al.~\cite{Kniehl:2004fy, Kniehl:2005mk}. Fig.~\ref{fig:GMVFNS} shows a comparison of the predictions obtained from the GMVFNS code, with those from our Crystal Ball fit for the data sets used in the fit for all the $pp \to D^0 + X$ data used in the baseline fit. In the input of GMVFNS we use the nCTEQ15 proton PDF set, we set the $c$ quark mass to $m_c = 1.3$ GeV, and the renormalization and the initial/final factorization scales to $\mu_r = \mu_i = \mu_f = \sqrt{p_T^2 + 4 m_c^2}$. As a fragmentation function we use the one with identifier 712 from the KKKS08 set of fragmentation functions~\cite{Kneesch:2007ey} which was obtained in a global fit to Belle, CLEO, ALEPH and OPAL data. The uncertainties of the GMVFNS predictions are obtained by varying the three scales individually by a factor of two, such that there is never a factor four between two scales. These uncertainties are similar in size as the data uncertainty, except for the low-$p_T$ region, where they are somewhat larger. Overall the central prediction of the GMVFNS calculation slightly overshoots the data. This can perhaps be attributed to the contribution from largely unconstrained gluon component of the fragmentation function, which contributes at almost 50\%. However, there is always overlap between the data and GMVFNS theory uncertainty. The uncertainty of the Crystal Ball fit is similar in size as the GMVFNS one for large $p_T$, but contrary to the latter it decreases for lower $p_T$ values. The central values are very close to the data points, as indicated by the low $\chidof$ value seen in Tab.~\ref{tab:chi2baseline}. Overall the two methods are in very good agreement with only minor discrepancies seen in the highest $p_T$ bins.

We also compared our Crystall Ball fit and GMVFNS predictions against more recent data of $D^0$ production in $pp$ collisions from ALICE~\cite{ALICE:2017olh, ALICE:2019nxm} and LHCb~\cite{LHCb:2016ikn}, which we have not been used in the present analysis. We do not show the comparisons here, but we report that both the Crystall Ball fit as well as GMVFNS reproduce the ALICE and LHCb data well. This data could provide further constraints on the $D^0$ Crystall Ball parameters. 
\FloatBarrier

\section{Impact of HQ data on nPDF fits}\label{sec:pdffit}
Using the Crystal Ball parameters determined in the previous section we can now perform a new global nPDF fit using the available heavy-quark data. 
The new fits are using the same framework as nCTEQ15WZ+SIH, including all settings like open parameters, scales and cuts for the previously included data. We do not include the changes made for nCTEQ15HIX~\cite{Segarra:2020gtj} and nCTEQ15$\nu$~\cite{nCTEQ15nu} as these developments are mostly orthogonal to those made in this study and do not affect the low-$x$ gluon PDF. One minor change from the previous analyses is the treatment of normalizations. Previously, $\chi^2$-penalties were assigned individually for each affected data set, whereas now they are applied only once per normalization parameter. 

\subsection{Data selection}

We add the heavy-quark data sets shown in Tabs.~\ref{tab:datapPb_D0} - \ref{tab:datapPb_Y1S} to the new PDF fit for a total of 1484 (548 new, 936 old) data points. Similar to the fragmentation function uncertainties of the SIH data in Ref.~\cite{Duwentaster:2021ioo}, we can compensate for the theoretical uncertainty of the data-driven approach by adding the uncertainty from the Crystal Ball fit as a systematic uncertainty to all new data sets.

\begin{table*}[p!]
    \renewcommand{\arraystretch}{1.4}
    \setlength\tabcolsep{4pt}
	\caption{Overview of the available  $pPb \  \longrightarrow \  D^0+X$  production data sets and their number of data points.}
	\centering	
	\begin{tabular}{|c|c|c|c|c|c|}
		\hline 
		Group & Year & Ref. & ID & Type & Points after / before cuts \\ 
		\hline
		\hline
		ALICE  & 2014 & \cite{ALICE:2014xjz} & 3101 & Prompt & 8 / 10  \\ 
		\hline 
		ALICE  & 2016 & \cite{ALICE:2016yta} & 3123 & Prompt & 8 / 11  \\ 
		\hline 
		LHCb  & 2017 & \cite{LHCb:2017yua} & 3102 & Prompt & 53 / 92  \\ 
		\hline 
		ALICE  & 2019 & \cite{ALICE:2019fhe} & 3122 & Prompt & 13 / 21  \\ 
		\hline 
	\end{tabular}     	
	\label{tab:datapPb_D0}
	
	\caption{Overview of the available  $pPb \  \longrightarrow \  J/\psi+X$  production data sets and their number of data points.}
	\centering	
	\begin{tabular}{|c|c|c|c|c|c|}
		\hline 
		Group & Year & Ref. & ID & Type & Points after / before cuts \\ 
		\hline
		\hline
		LHCb  & 2013 & \cite{LHCb:2013gmv} & 3108 & Prompt & 25 / 40  \\ 
		\hline
		ATLAS  & 2015 & \cite{ATLAS:2015mpz} & 3118 & Prompt & 10 / 10  \\ 
		\hline 
		LHCb  & 2017 & \cite{LHCb:2017ygo} & 3105 & Prompt & 88 / 140  \\ 
		\hline
		CMS  & 2017 & \cite{CMS:2017exb} & 3120 & Prompt & 51 / 53  \\ 
		\hline 
		ATLAS  & 2018 & \cite{ATLAS:2017prf} & 3117 & Prompt & 8 / 8  \\ 
		\hline
		\hline
		LHCb  & 2013 & \cite{LHCb:2013gmv} & 3107 & Non-prompt & 25 / 40  \\ 
		\hline 
		ATLAS  & 2015 & \cite{ATLAS:2015mpz} & 3119 & Non-prompt & 10 / 10  \\ 
		\hline
		LHCb  & 2017 & \cite{LHCb:2017ygo} & 3106 & Non-prompt & 88 / 140  \\ 
		\hline 
		CMS  & 2017 & \cite{CMS:2017exb} & 3121 & Non-prompt & 51 / 53  \\ 
		\hline 
		ATLAS  & 2018 & \cite{ATLAS:2017prf} & 3116 & Non-prompt & 8 / 8  \\ 
		\hline
		\hline
		ALICE  & 2013 & \cite{ALICE:2013snh} & 3103 & Inclusive & 0 / 12  \\ 
		\hline 
		ALICE  & 2015 & \cite{ALICE:2015sru} & 3104 & Inclusive & 10 / 25  \\ 
		\hline
		ALICE  & 2018 & \cite{ALICE:2018mml} & 3112 & Inclusive & 9 / 24  \\ 
		\hline 
	\end{tabular}     	
	\label{tab:datapPb_JPSI}
	
	\caption{Overview of the available  $pPb \ \longrightarrow \ \Upsilon(1S)+X$  production data sets and their number of data points.}
	\centering	
	\begin{tabular}{|c|c|c|c|c|c|}
		\hline 
		Group & Year & Ref. & ID & Type & Points after / before cuts \\ 
		\hline
		\hline
		ALICE  & 2014 & \cite{ALICE:2014ict} & 3110 & Inclusive & 0 / 4  \\ 
		\hline 
		LHCb  & 2014 & \cite{LHCb:2014rku} & 3111 & Inclusive & 0 / 2  \\ 
		\hline 
		ATLAS  & 2018 & \cite{ATLAS:2017prf} & 3109 & Inclusive & 6 / 8  \\ 
		\hline 
		LHCb  & 2018 & \cite{LHCb:2018psc} & 3113 & Inclusive & 36 / 66  \\ 
		\hline 
		ALICE  & 2019 & \cite{ALICE:2019qie} & 3114 & Inclusive & 3 / 10  \\ 
		\hline 
	\end{tabular}     	
	\label{tab:datapPb_Y1S}
	
	\caption{Overview of the available  $pPb \ \longrightarrow \ \psi(2S)+X$  production data sets and their number of data points.}
	\centering	
	\begin{tabular}{|c|c|c|c|c|c|}
		\hline 
		Group & Year & Ref. & ID & Type & Points after / before cuts \\ 
		\hline
		\hline
		ALICE  & 2014 & \cite{ALICE:2014cgk} & 3127 & Inclusive & 2 / 8  \\ 
		\hline 
		ALICE  & 2020 & \cite{ALICE:2020vjy} & 3126 & Inclusive & 3 / 10  \\ 
		\hline 
		ATLAS  & 2017 & \cite{ATLAS:2017prf} & 3124 & Prompt & 8 / 8  \\ 
		\hline 
		CMS  & 2018 & \cite{CMS:2018gbb} & 3115 & Prompt & 17 / 17  \\ 
		\hline 
		ATLAS  & 2017 & \cite{ATLAS:2017prf} & 3125 & Non-prompt & 8 / 8  \\ 
		\hline 
	\end{tabular}     	
	\label{tab:datapPb_PSI2S}
\end{table*}

For the new HQ data, we use the same cuts as in the proton-proton baseline and additionally exclude $D^0$ data points with $p_T>15$\,GeV, because there is no baseline data. Furthermore, we remove two individual points from the 2018 LHCb $\Upsilon(1S)$ data set that are described very poorly with $\chi^2$ values of 66 and 26, respectively. Both points are at the high-$p_T$ edge of the experiment's kinematic range, which makes systematic errors a likely explanation, since the remaining 36 data points of the set can be well described.

To estimate the impact of the scale choice on the final PDFs, we repeat the entire procedure, including the proton-proton baseline fit, with the scale of the heavy quark production processes set to $\frac12$ or 2 times their regular value. %

\subsection{Resulting PDFs}
\begin{figure*}[p]
	\centering
	\includegraphics[width=\textwidth]{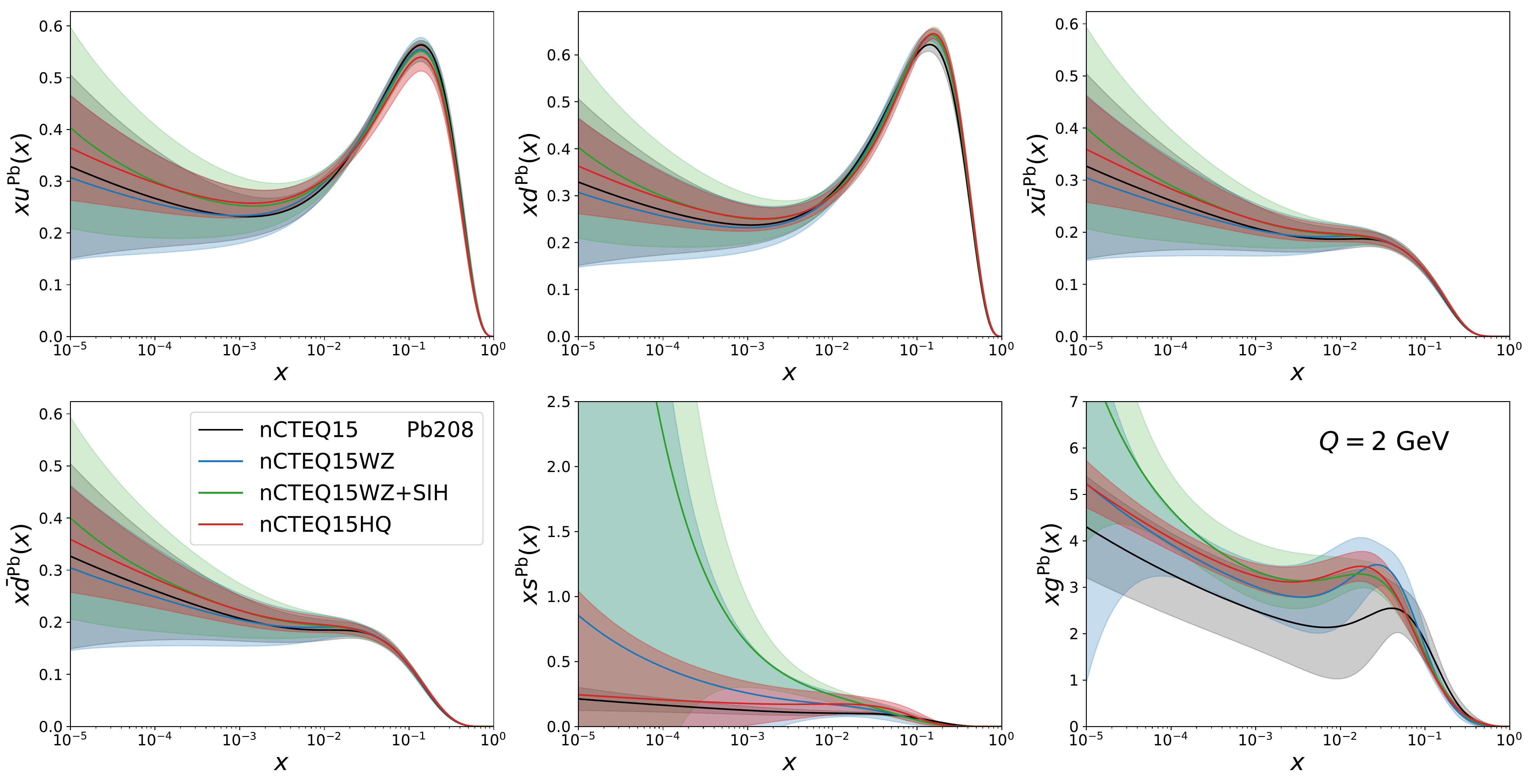}
	\caption{Lead PDFs from different nCTEQ15 versions. The baseline nCTEQ15 fit is shown in black, nCTEQ15WZ in blue, nCTEQ15WZSIH in green, and the new fit in red.}
	\label{fig:nCTEQ_PDFs}
\end{figure*}

\begin{figure*}[p]
	\centering
	\includegraphics[width=\textwidth]{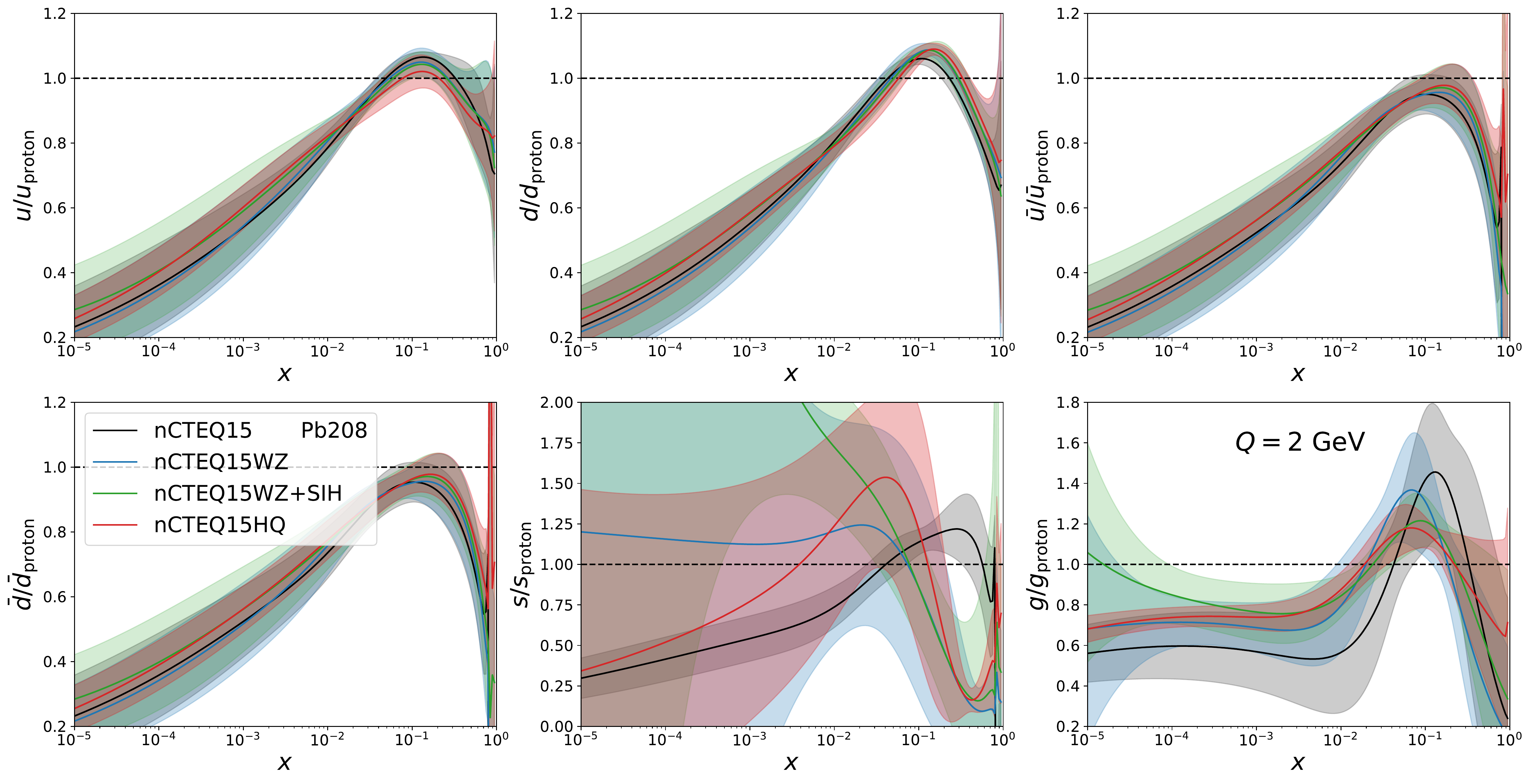}
	\caption{Ratio of lead and proton PDF from different nCTEQ15 versions. The baseline nCTEQ15 fit is shown in black, nCTEQ15WZ in blue, nCTEQ15WZSIH in green, and the new fit in red.}
	\label{fig:nCTEQ_PDFs_RpPb}
\end{figure*}

Fig.~\ref{fig:nCTEQ_PDFs} shows a comparison of the new fit, labelled nCTEQ15HQ\footnote{Since nCTEQ15WZ+SIH+HQ would be too long, we break this naming convention and shorten it to nCTEQ15HQ. This does not imply removal of the WZ and SIH data.}, with the PDFs obtained in previous nCTEQ analyses at a scale of $Q=2$\,GeV. The same  comparison is shown in Fig.~\ref{fig:nCTEQ_PDFs_RpPb} in terms of nuclear modification factors. The central value of the gluon PDF retains a similar shape as the nCTEQ15WZ+SIH fit in the $x>10^{-3}$ region. Below this point, the nCTEQ15WZ+SIH fit starts diverging towards higher values, while the new fit moves towards similar values as is nCTEQ15WZ. From $x\approx0.2$ downwards the uncertainties of the new gluon PDF are reduced significantly compared to the previous analyses, particularly below $x<10^{-4}$, where the previous uncertainties start to increase rapidly. 
The ratio plots underline that there are no significant changes to the up- and down-quark distributions, but there is a minor reduction in $u_{\mathrm{valence}}$ around $x\approx0.1$. The uncertainties of the new fit still include the central values of all previous fits. The gluon ratio on the other hand shows a reduction in the nuclear modification for $x>0.008$. In this region, the gluon ratio is mostly compatible with unity, but the central value goes from 0.8 at the lowest end of this region to 1.2 at its peak and then down to 0.6 at very high $x$. At lower $x$-values, however, the reduced uncertainties clearly show a suppression of the gluon in lead between 20\% and 40\% compared to the proton case. The modification of the strange quark is similar to the other quark flavours in the new fit, but the uncertainties are still larger than the modification for all $x<0.3$.
The central values of the new up and down quark PDFs are mostly very close to those of nCTEQ15WZ+SIH, with a minor downward shift at low-$x$ and reduced uncertainties in the same region due to the better constraints on the gluon.
The strange quark PDF stays similar to previous fits in the $x>0.01$ region, but reverts to a low value similar to the original nCTEQ15, while nCTEQ15WZ and particularly nCTEQ15WZ+SIH start increasing. It is important to note however, that the relative uncertainty of the strange quark is still larger than 100\% at low $x$, so that it is impossible to draw any strong conclusions from this analysis. We refer the reader to our recent neutrino analysis~\cite{nCTEQ15nu} for complementary information.

\begin{figure*}[t]
	\centering
	\includegraphics[width=\textwidth]{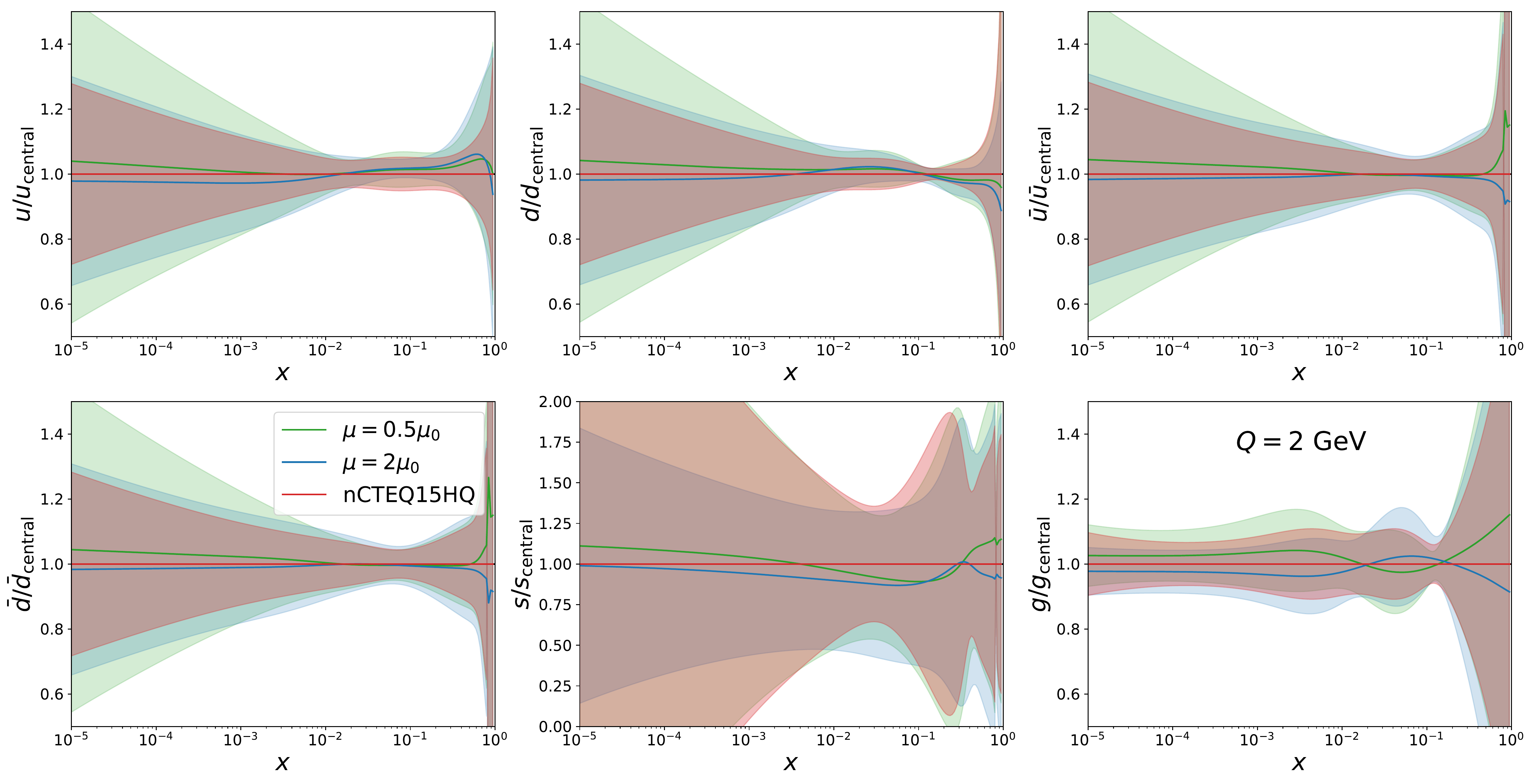}
	\caption{Comparisons of fits, where the scale for heavy quark production is varied by a factor two around the central value $\mu_0$.}
	\label{fig:scale_PDFs}
	\centering
	\includegraphics[width=\textwidth]{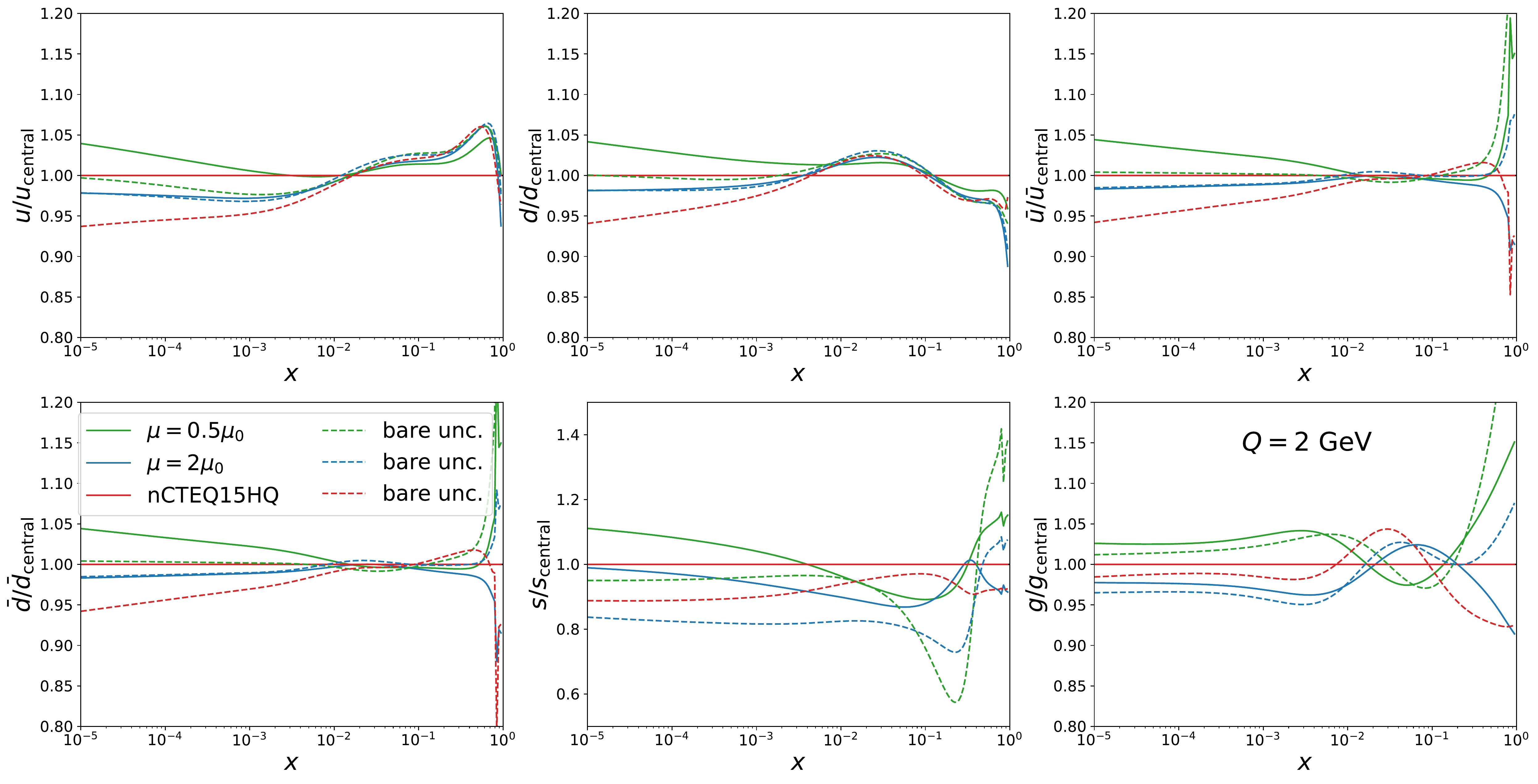}
	\caption{Comparisons of PDF central values with Crystal Ball uncertainties included in the fit (solid lines) and without theory uncertainties (dashed lines). }
	\label{fig:unc_PDFs}
\end{figure*}

The impact of different scale choices is shown in Fig.~\ref{fig:scale_PDFs}, where each flavor $i$ is shown as a ratio over the corresponding flavor $i_\mathrm{central}$ from the fit with the central scale choice.
The central values of the up, down and gluon PDFs with modified scales lie close together with differences no larger than a few percent. The only exception to this is the high-$x$ region, where the gluon PDF vanishes and therefore the relative uncertainties blow up. The gluon PDFs cross each other at $x=0.02$ and $x=0.15$, with the lower scale being above the central scale in the high- and low-$x$ regions and the upper scale showing the opposite behaviour. The PDFs extracted with the modified scale stay well within the uncertainties of the regular fit and show slightly larger uncertainties themselves, which are likely due to a slightly worse fit. The strange quark sees variations of up to 15\%, but given the very large uncertainties, this is still well within expectations. This conclusion is somewhat different from the one reached in the previous reweighting analysis~\cite{Kusina:2017gkz, Kusina:2020dki}, where the scale choice resulted in significant differences. Two updates in the methodology are responsible for this change: Firstly, the fact that normalizations of the data sets are now included as nuisance parameters in the fit, mitigates the scale dependence due to the large normalization uncertainties on many data sets. Secondly, the more restrictive kinematic cuts remove the data that is most sensitive to the scale choice.

A final set of alternative fits is shown in Fig.~\ref{fig:unc_PDFs}, where the previous three fits are compared to equivalent fits where the heavy-quark data sets keep their bare uncertainties, \textit{i.e.} the uncertainties from the Crystal Ball fit are not added to the systematic uncertainties. For the up- and down-quark flavours this leads to shifts on a similar order as the scale variation, while the changes in the gluon are even smaller. The changes in the strange quark PDF are somewhat larger due to it's weakly constrained nature. The difference between the dashed and solid lines gives an estimate of the potential enhancement of the impact from the heavy-quark data if further proton-proton data is added to the baseline to reduce the Crystal Ball uncertainty.

\subsection{Fit quality}
Figs.~\ref{fig:pPbFitJPSI} - \ref{fig:pPbFitPSI2S} show a comparison of the proton-lead data with predictions from the new nCTEQ15HQ PDFs including PDF uncertainties. For all four particle species, there is very close agreement across the entire kinematic range and the PDF uncertainties are barely visible. Note however, that the uncertainties from the Crystal Ball fit, which are not shown here, would still be of the same size as in the baseline predictions in Fig.~\ref{fig:ppBaselineJPSI} - \ref{fig:ppBaselinePSI2S}.

\begin{figure*}[htb!]
	\centering
	\includegraphics[width=\textwidth]{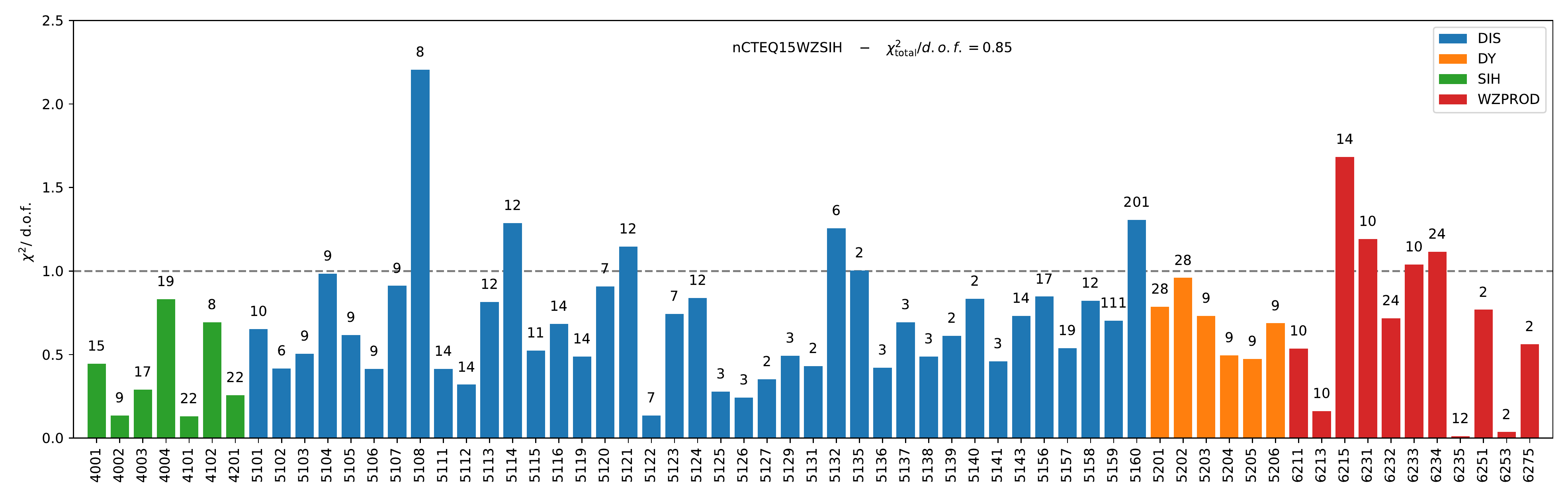}
	\includegraphics[width=\textwidth]{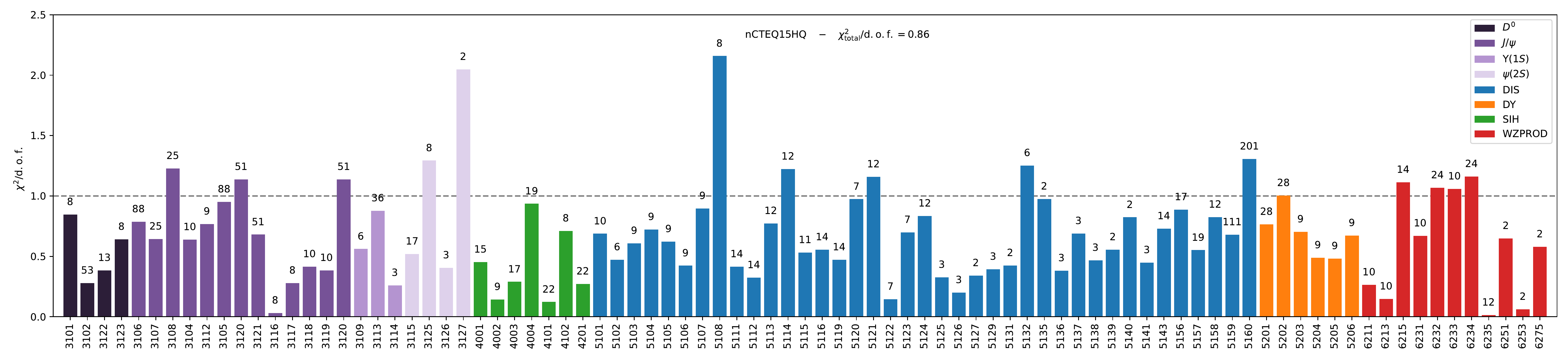}
	\caption{$\chidof{}$ values for each data set in the previous nCTEQ15WZ+SIH fit (upper panel) and the new nCTEQ15HQ fit (lower panel).}
	\label{fig:chi2perdata}
\end{figure*}

For a more quantitative evaluation of the fit quality, one can take a look at Fig.~\ref{fig:chi2perdata}, which shows the $\chidof$ for each data set of the previous nCTEQ15WZ+SIH and the new nCTEQ15HQ fits in the upper and lower panels, respectively. The comparison shows no significant rise in $\chi^2$ for any of the established data sets. The $\chi^2$ of the outlier 6215 (ATLAS Run I, $Z$ production) among the vector boson production data sets is even reduced due to the new normalization treatment. This means that there are no incompatibilities between the previous and the new data sets. The new heavy-quark data sets themselves mostly show $\chidof$ values below or around one with a distribution similar to that of the DIS data sets. The only major outlier among the new sets is the $\psi(2S)$ production from a 2014 measurement by ALICE with just two data points.

Tab.~\ref{tab:chi2table_mainfits} shows a comparison of the $\chidof$ values for each process of the new fit and the preceding three nCTEQ15 generations. The original nCTEQ15 is the only set of PDFs that does not give a good description of the heavy-quark data, especially for $J/\psi$ mesons. nCTEQ15WZ already displays significant improvement for $D^0$ and $J/\psi$, thus giving a reasonable $\chidof$ of 0.92 for the new data sets. nCTEQ15WZ+SIH retains a very similar overall $\chi^2$, but does worse than nCTEQ15WZ for $D^0$ and $\Upsilon(1S)$, while improving the description of the $J/\psi$ data. Finally, the new nCTEQ15HQ fit gives a good $\chi^2$ for all processes with only minor increases for DIS and WZ production compared to %
both nCTEQ15WZ and nCTEQ15WZ+SIH, 
which are outweighed by a significant improvement in the large sample of new heavy-quark data. Interestingly, the $\chidof$ value for $\psi(2S)$ production barely changes between the four fits, but since it is the smallest of the new data sets with large theory uncertainties from the baseline fit, the small impact is to be expected.
\begin{table*}[htb!]	
    \renewcommand{\arraystretch}{1.4}
    \setlength\tabcolsep{4pt}
	\caption{$\chidof{}$ values for the individual heavy-quark final states, the individual processes {DIS, DY, WZ, SIH, HQ}, and the total. The shown $\chi^2$ is the sum of regular $\chi^2$ and normalization penalty. Excluded processes are shown in parentheses. 
	Note that both nCTEQ15 AND nCTEQ15WZ included the neutral pions from STAR and PHENIX.\\
	}
	\centering	
	\begin{tabular}{|c||c|c|c|c||c|c|c|c|c|c|c|}
		\hline
		 & $D^0$ & $J/\psi$ & $\Upsilon(1S)$ & $\psi(2S)$ & DIS  & DY & WZ & SIH & HQ & \textbf{Total}\\ 
		\hline 
		\hline
		nCTEQ15           & (0.56) & (2.50) & (0.82) & (1.06) & 0.86 & 0.78 & (2.19) & (0.78) & (1.96) & \textbf{1.23}  \\ 
		\hline 
		nCTEQ15WZ         & (0.32) & (1.04) & (0.76) & (1.02) & 0.91 & 0.77 & 0.63 & (0.47) & (0.92) & \textbf{0.90}  \\
		\hline 
		nCTEQ15WZ+SIH     & (0.46) & (0.84) & (0.90) & (1.07) & 0.91 & 0.77 & 0.72 & 0.40 & (0.93) & \textbf{0.92}  \\
		\hline 
		nCTEQ15HQ         & 0.35 & 0.79 & 0.79 & 1.06 & 0.93 & 0.77 & 0.78 & 0.40 & 0.77 & \textbf{0.86}  \\
		\hline 
	\end{tabular}     
	\label{tab:chi2table_mainfits}
\end{table*}

\begin{figure*}[t!]
	\centering
	\includegraphics[width=0.31\textwidth]{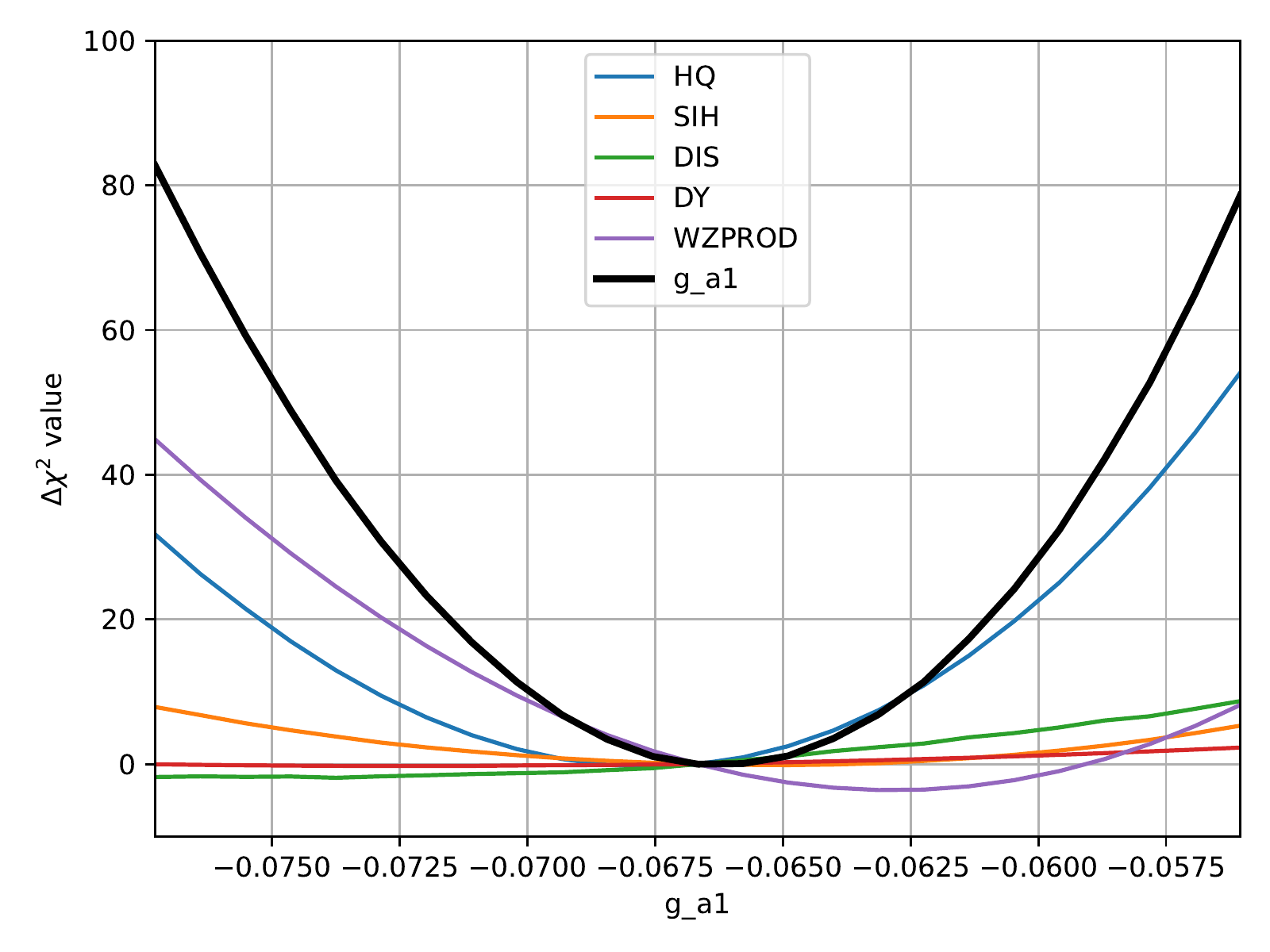}
	\includegraphics[width=0.31\textwidth]{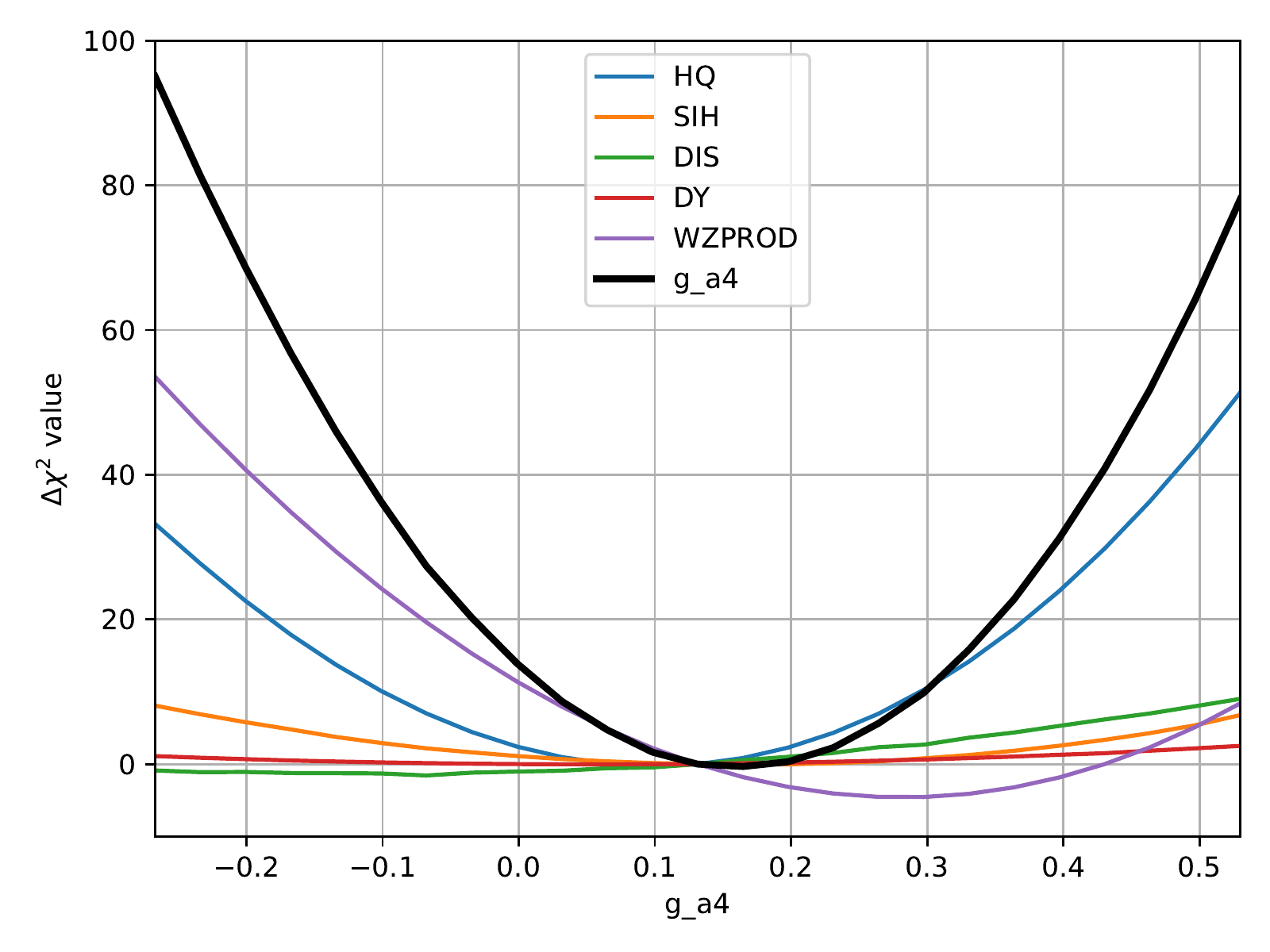}
	\includegraphics[width=0.31\textwidth]{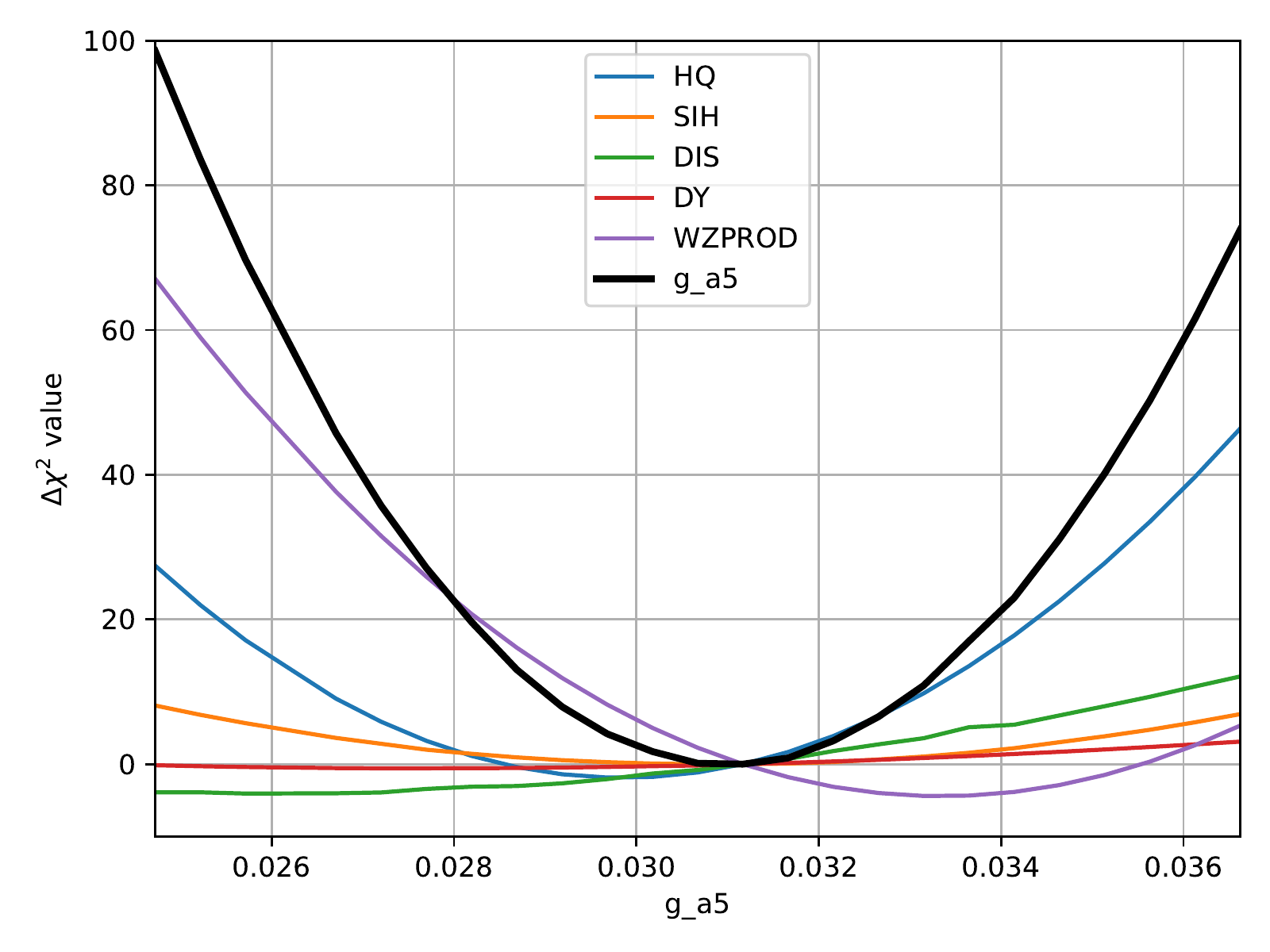}
	\includegraphics[width=0.31\textwidth]{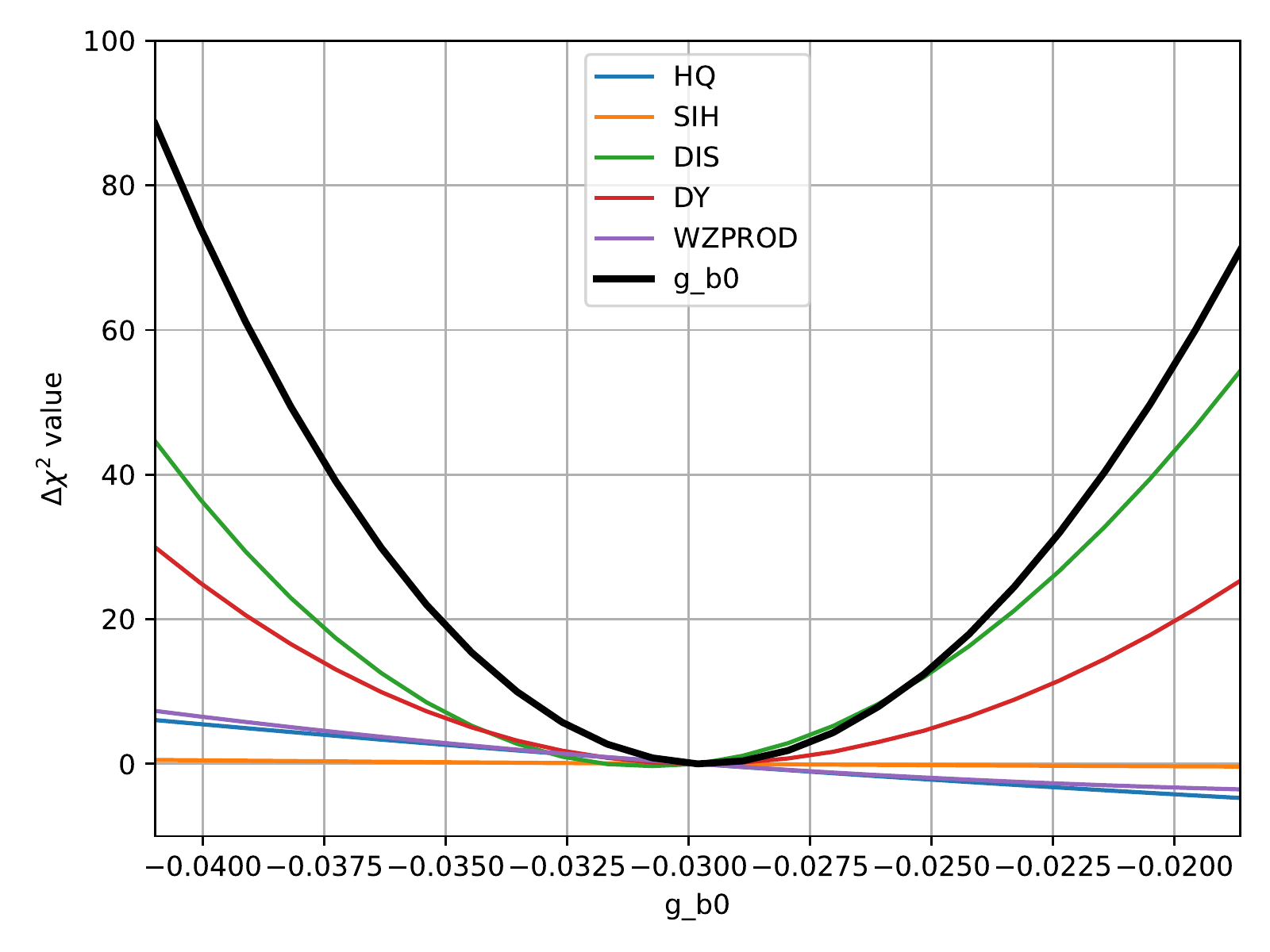}
	\includegraphics[width=0.31\textwidth]{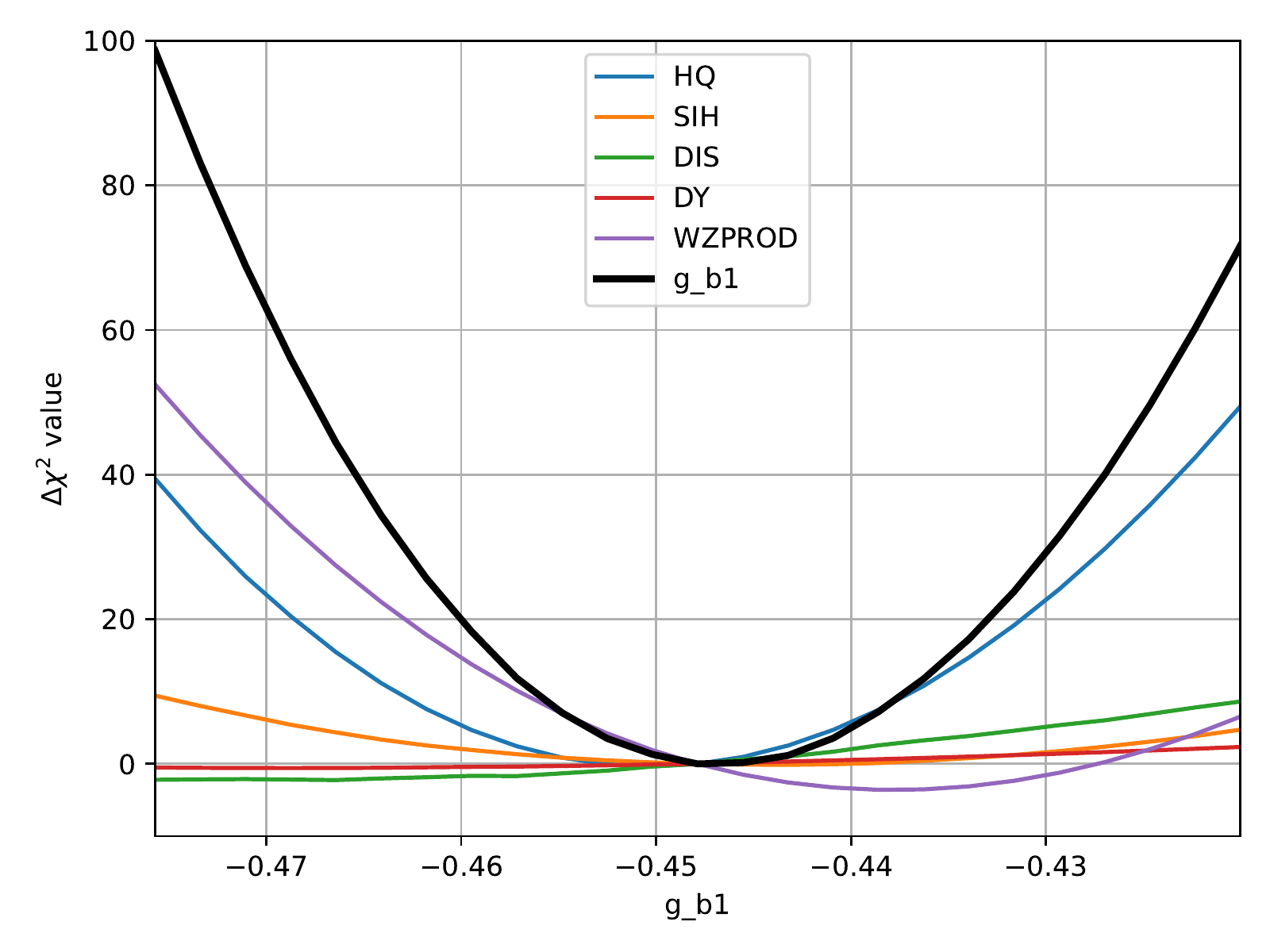}
	\includegraphics[width=0.31\textwidth]{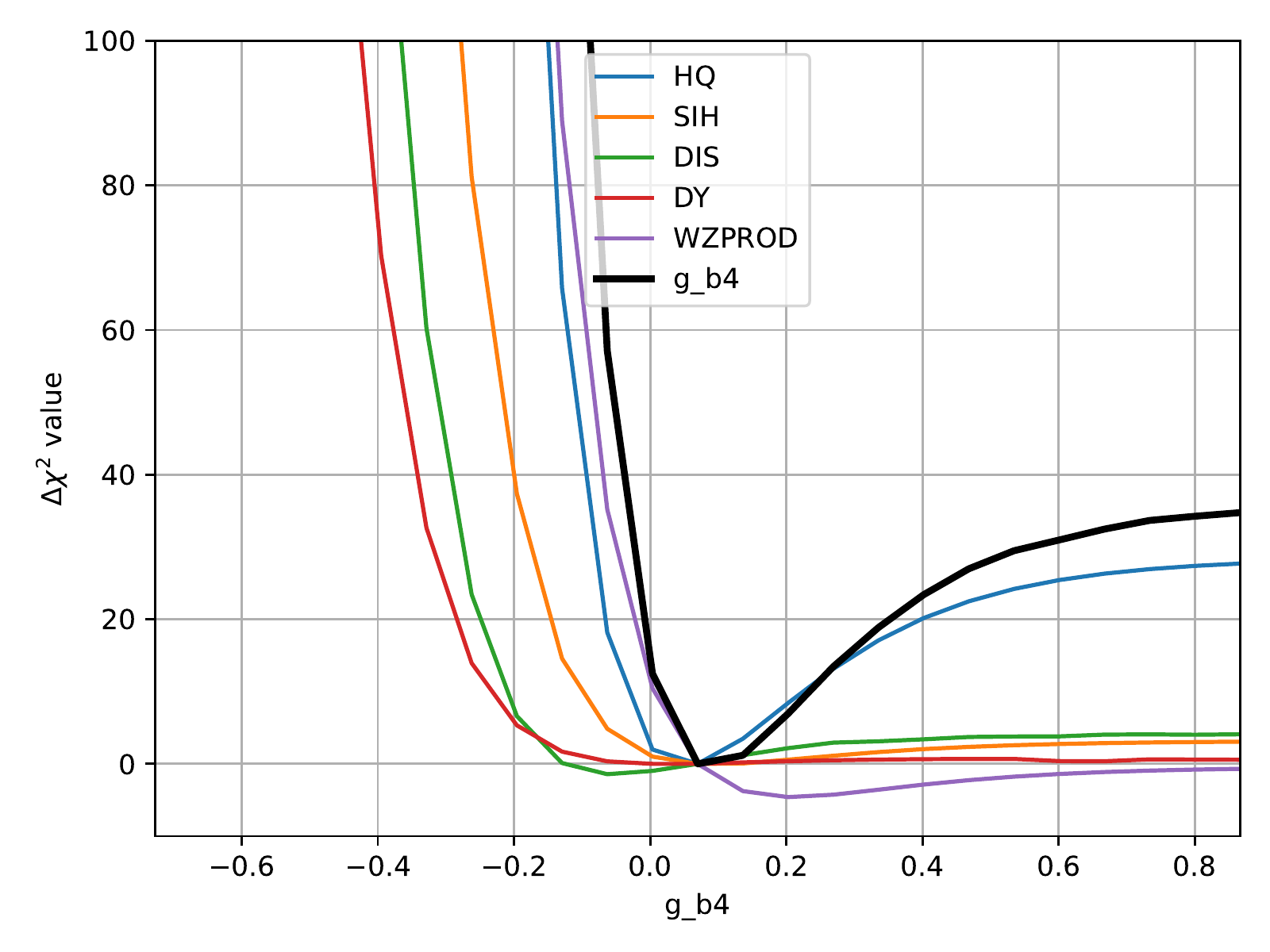}
	\includegraphics[width=0.31\textwidth]{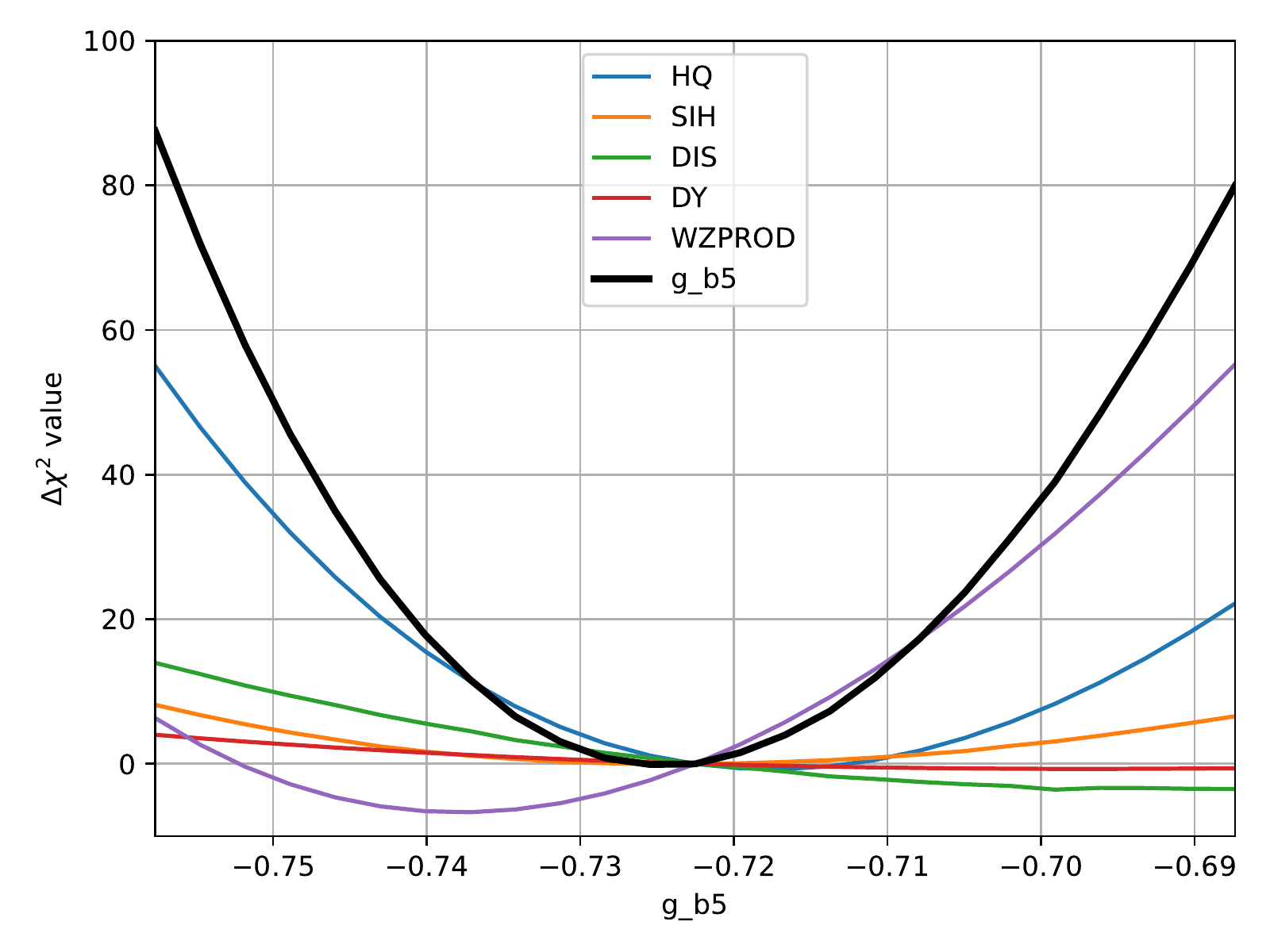}
	\caption{Scans along the seven open gluon parameters, divided by experiment types with the total shown in black.}
	\label{fig:scans}
\end{figure*}

To estimate the impact of the new data on the PDF parameters, it is instructive to look at the $\Delta\chi^2$ profiles along specific parameters. Fig.~\ref{fig:scans} shows parameter scans for all of the seven gluon parameters included in the fit with the $\Delta\chi^2$ split into processes. It is immediately evident that the new heavy-quark data is the dominant constraint on most of the parameters with the WZ production always constraining one side, because its minimum is somewhat off-center. This indicates some slight tensions with the WZ production data, but looking back at Tab.~\ref{tab:chi2table_mainfits} this tension largely comes from the SIH data and is still well within expectations. The only parameter that is not constrained mostly by vector boson and heavy quark production is $b_{0}^g$. This parameter determines the $A$ dependence of the gluon normalization and minor shifts can therefore be compensated through the normalization parameters of the data sets.
Additionally, since parameters with $b$ indices are related to the nuclear $A$ dependence, they may cancel out when looking only at a singular $A$ as is the case in the LHC data, which is only taken in proton-lead collisions. 

\section{Conclusions}\label{sec:conclusion}

In conclusion, we have incorporated a large new data set of heavy quarkonium and open heavy-flavor production in the nCTEQ++ framework and extended our PDF analysis of the gluon to significantly lower $x$ values than was previously possible.

We employed a data-driven approach to determine the theory and investigated the advantages and limitations of this approach in detail. In particular, we determined the kinematic range where the approach is applicable and verified the predictions with those from rigorous pQCD calculations for $J/\psi$ and $D^0$ production. 

We obtained good $\chidof$ values for the new data sets without compromising those of the established sets. The new data has a tremendous impact on the gluon PDF,  especially in the region $x<0.01$, and lowers the uncertainty considerably. Through the DGLAP evolution, this impact on the gluon also has consequences for the quark PDFs at low $x$, which remain at similar central values, but also experience a reduction of their uncertainties. 

The data-driven approach has proven to be both reliable with a reasonable estimation of its uncertainties, and also a powerful tool for the difficult task of constraining the nuclear gluon PDF. Therefore, the presented analysis and the PDFs obtained therein showcase a considerable advancement on the path towards a precise understanding of nuclear structure.

\section*{Acknowledgements}
The work of P.D., T.J., M.K.\ and K.K.\ was funded by the Deutsche Forschungsgemeinschaft (DFG, German Research Foundation) – project-id 273811115 – SFB 1225. 
We thank Peter Braun-Munzinger and Johanna Stachel for stimulating this research and all our SFB collaborators for very useful discussions. 
P.D., T.J., M.K., K.K.\ and K.F.M.\ also acknowledge support of the DFG through the Research Training Group GRK 2149. 
The work of I.S. was supported in part by the French National Centre for Scientific Research CNRS through IN2P3 Project GLUE@NLO.
F.O.\   acknowledges  support through US DOE grant DE-SC0010129. 
A.K. acknowledges the support of Narodowe Centrum Nauki under Sonata Bis Grant No. 2019/34/E/ST2/00186.

\appendix
\renewcommand{\textfraction}{0.001} 

\section{Data comparisons}
\begin{figure*}[htbp!]
	\centering
	\includegraphics[width=0.48\textwidth]{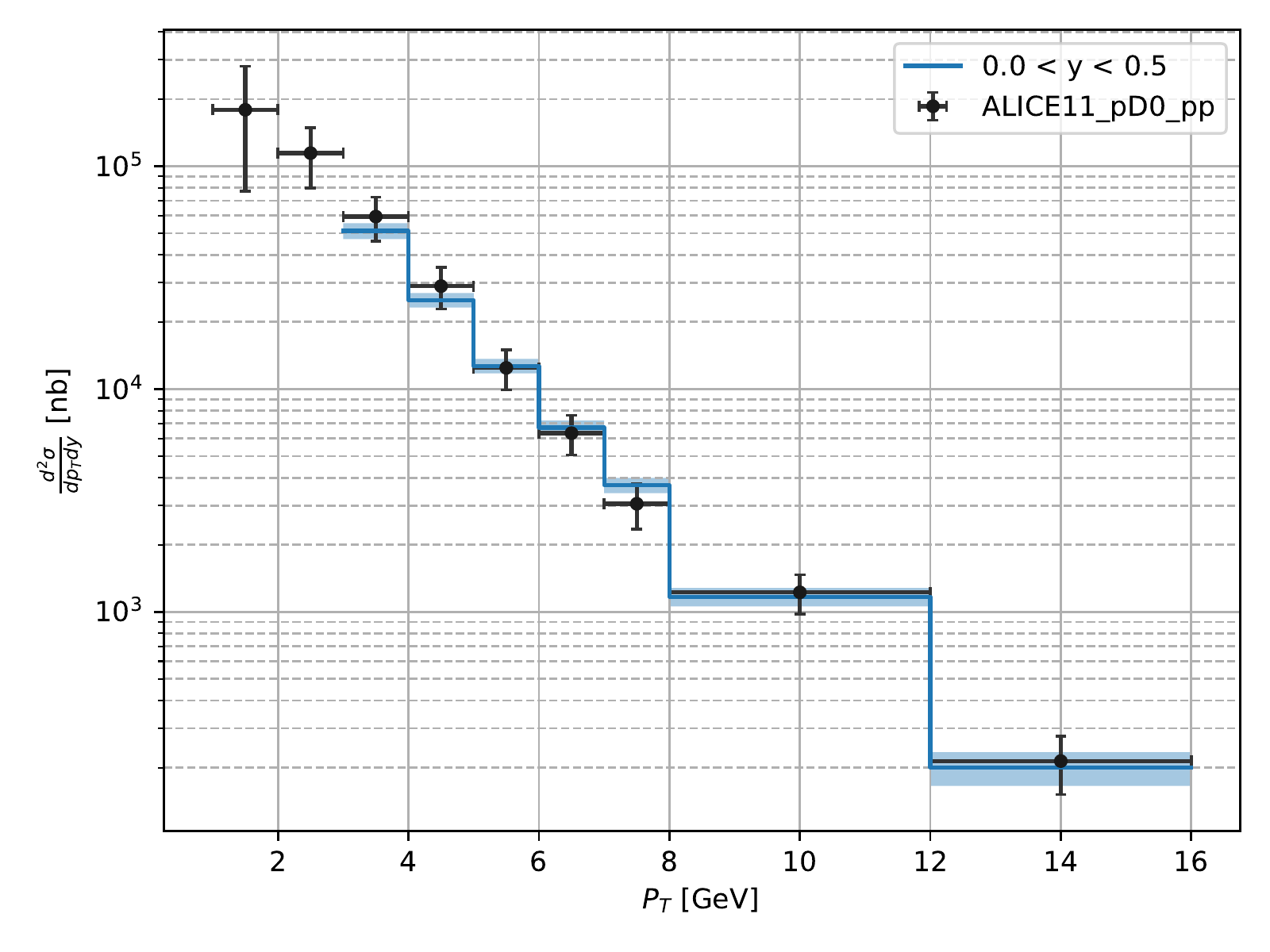}
	\includegraphics[width=0.48\textwidth]{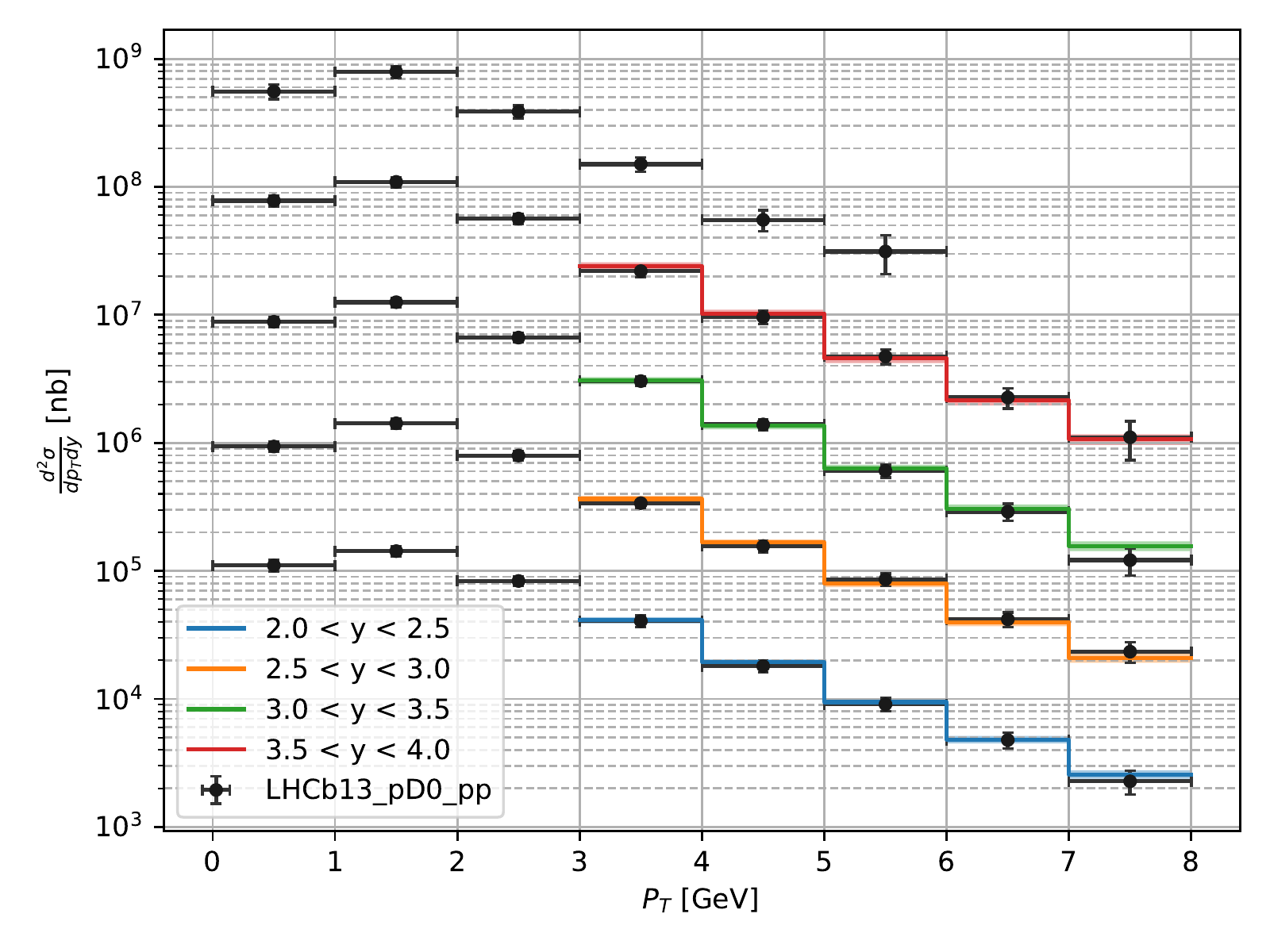}
	\includegraphics[width=0.48\textwidth]{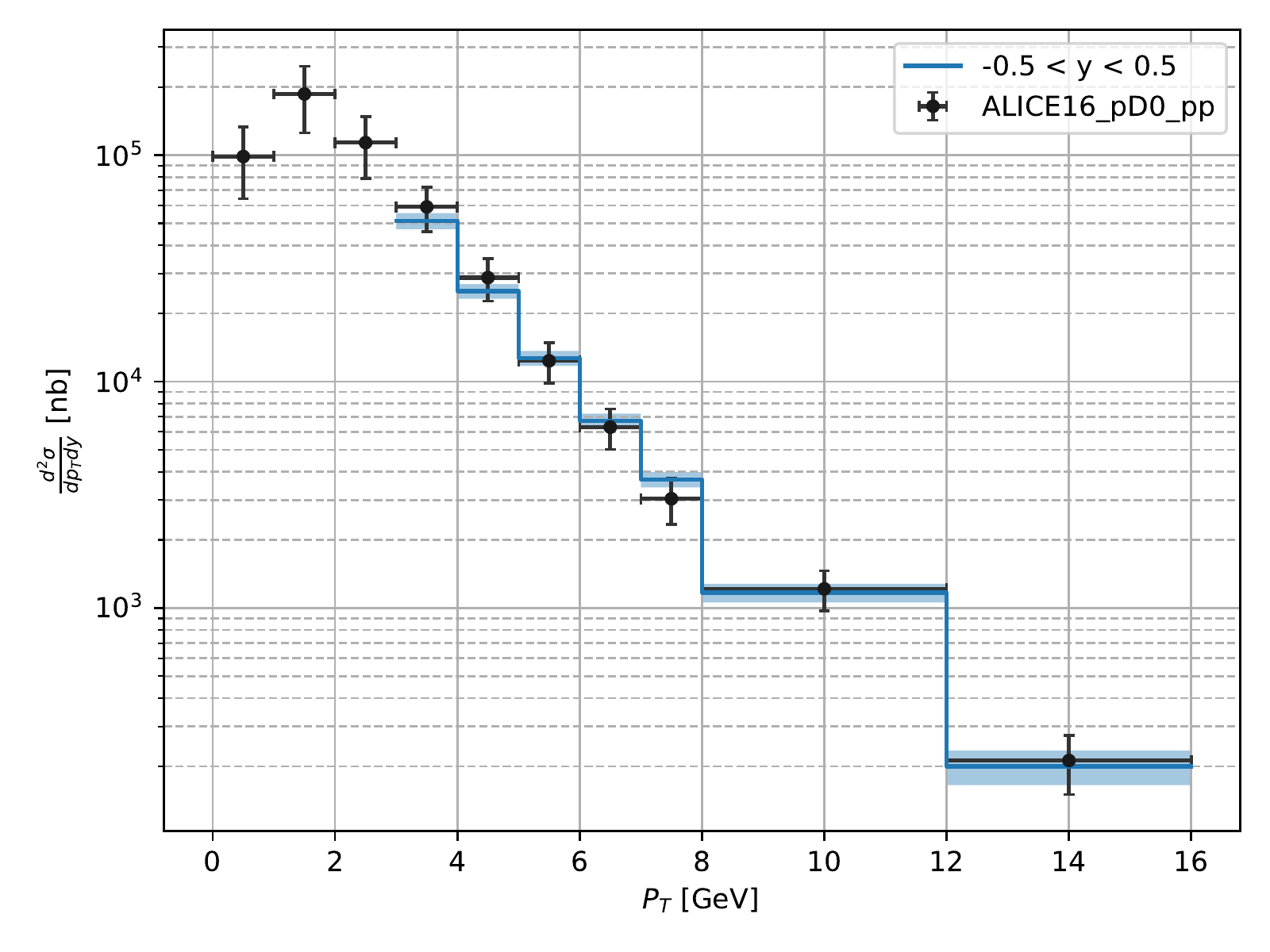}
	\caption{Predictions for $D^0$ production in proton-proton collisions with uncertainties from the Crystal Ball fit. Different rapidity bins are separated by multiplying the cross sections by powers of ten for visual clarity.}
	\label{fig:ppBaselineD0}
\end{figure*}

\begin{figure*}[htbp!]
	\centering
	\includegraphics[width=0.48\textwidth]{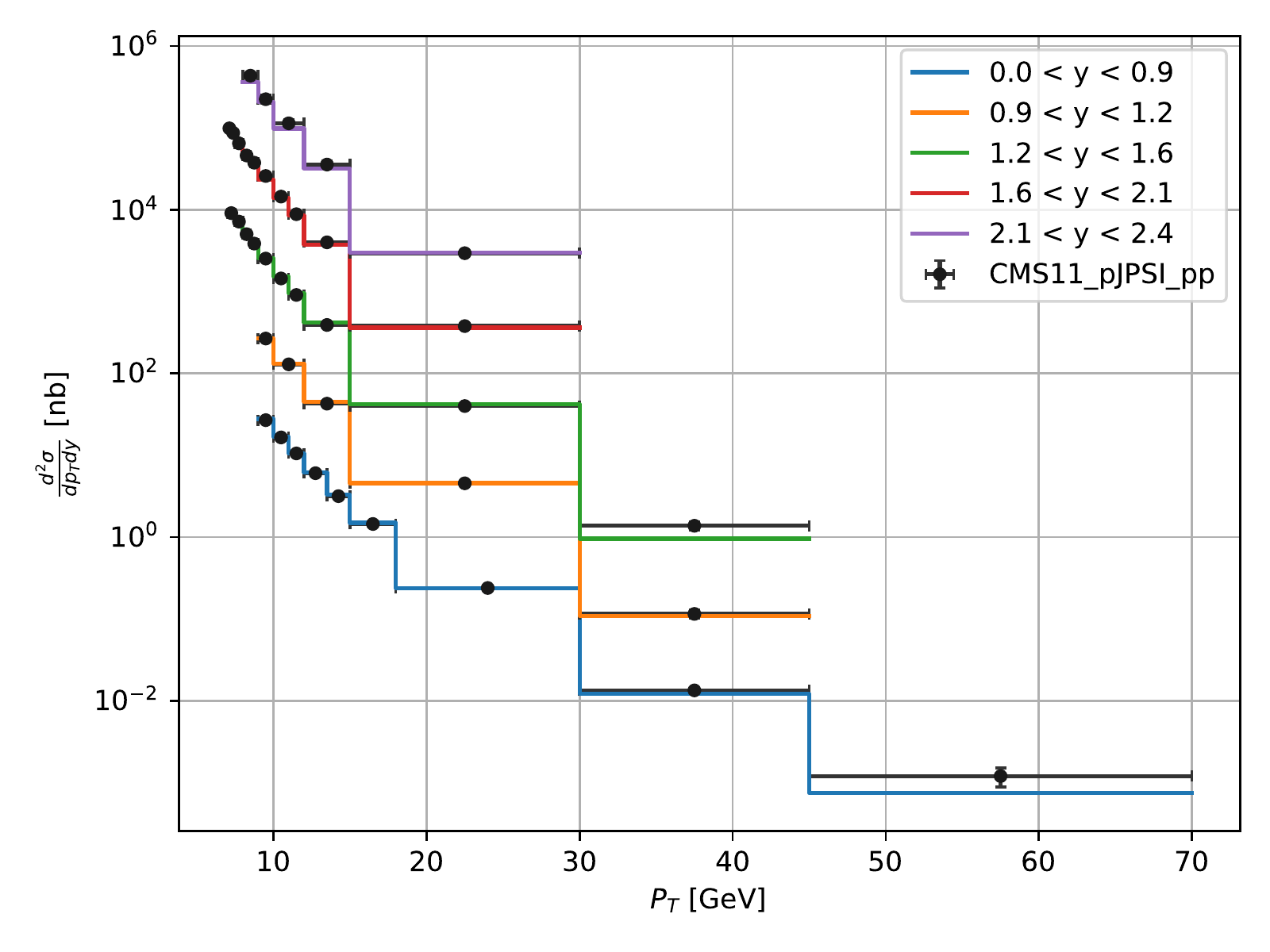}
	\includegraphics[width=0.48\textwidth]{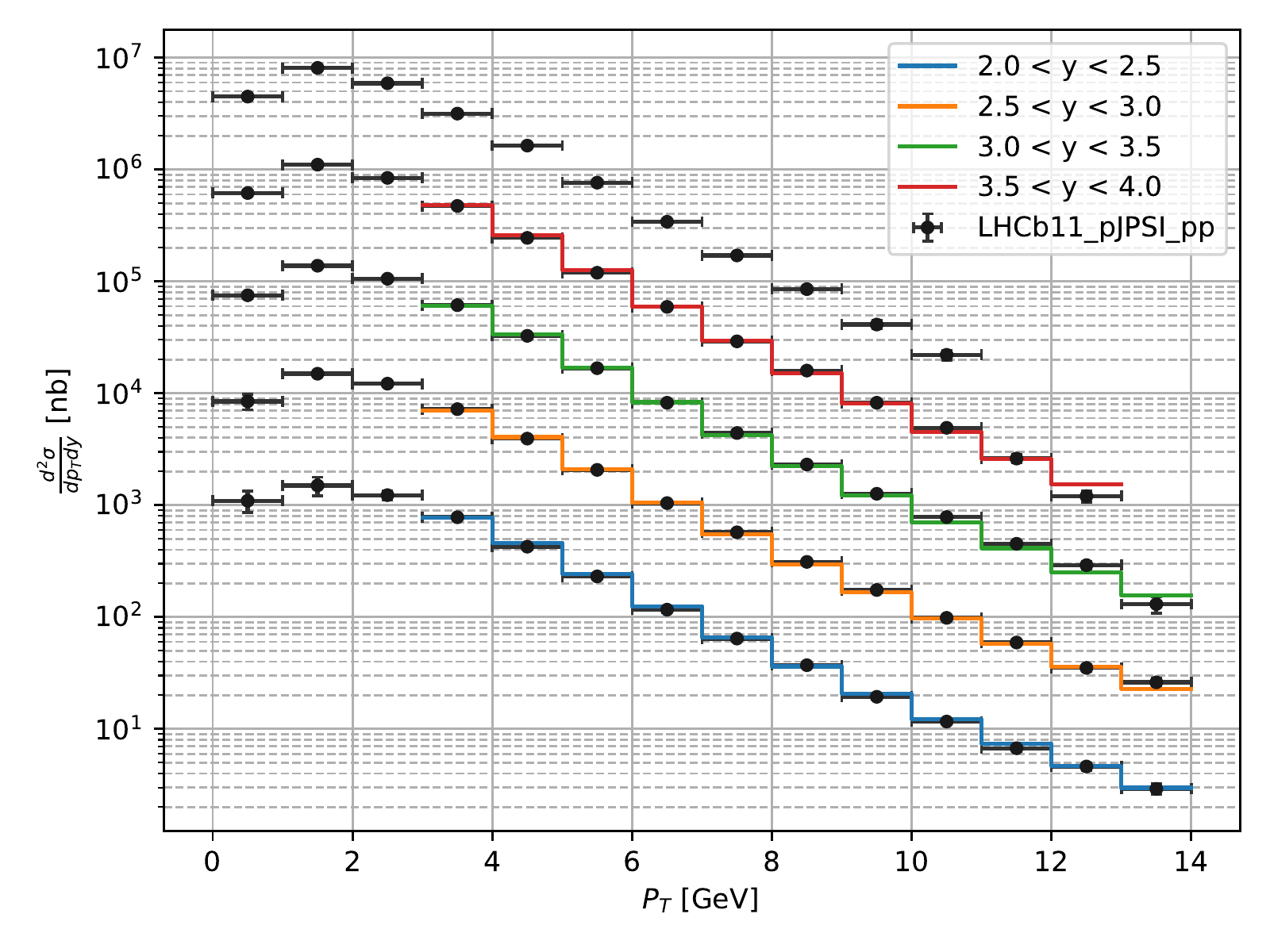}
	\includegraphics[width=0.48\textwidth]{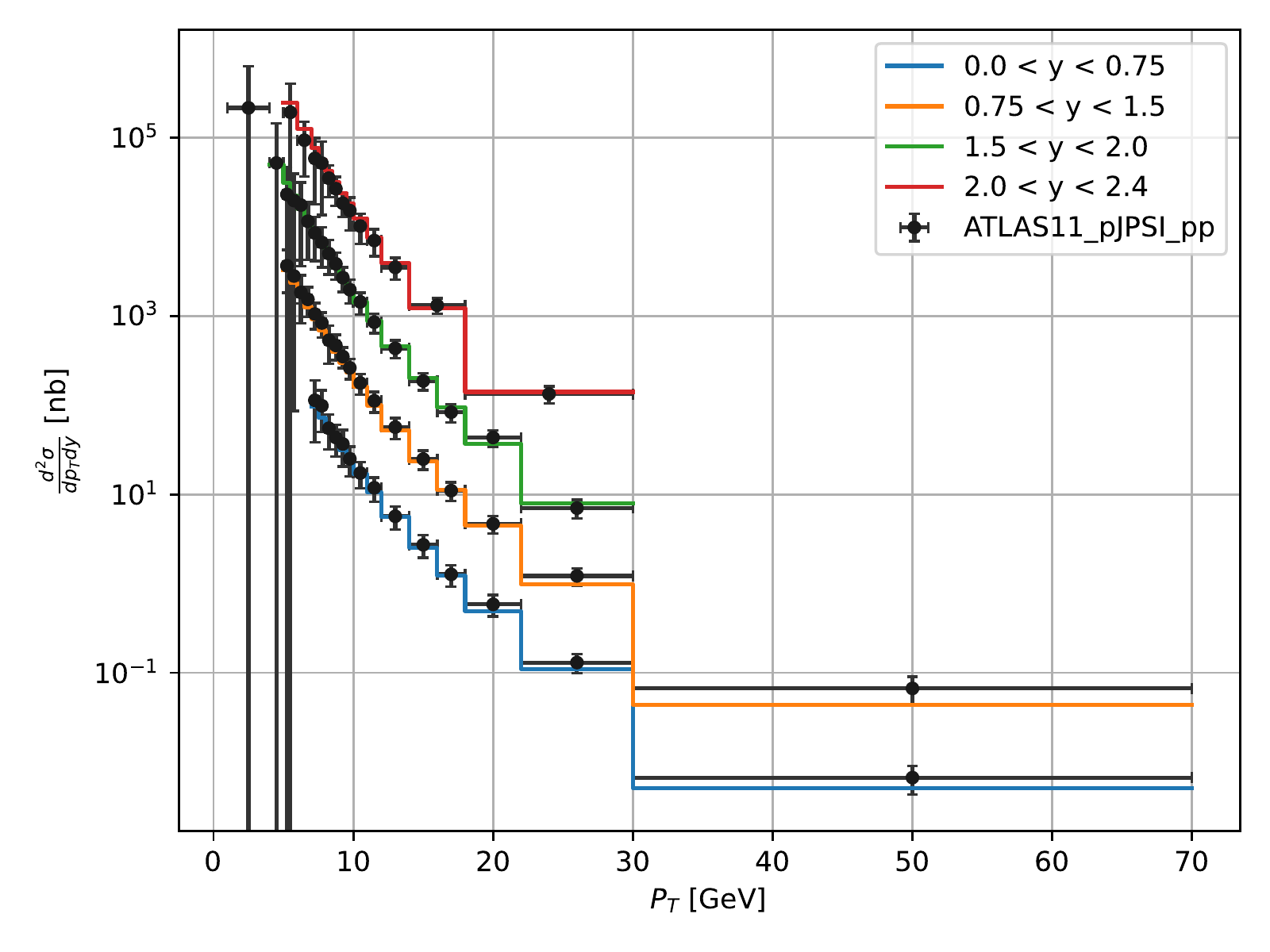}
	\includegraphics[width=0.48\textwidth]{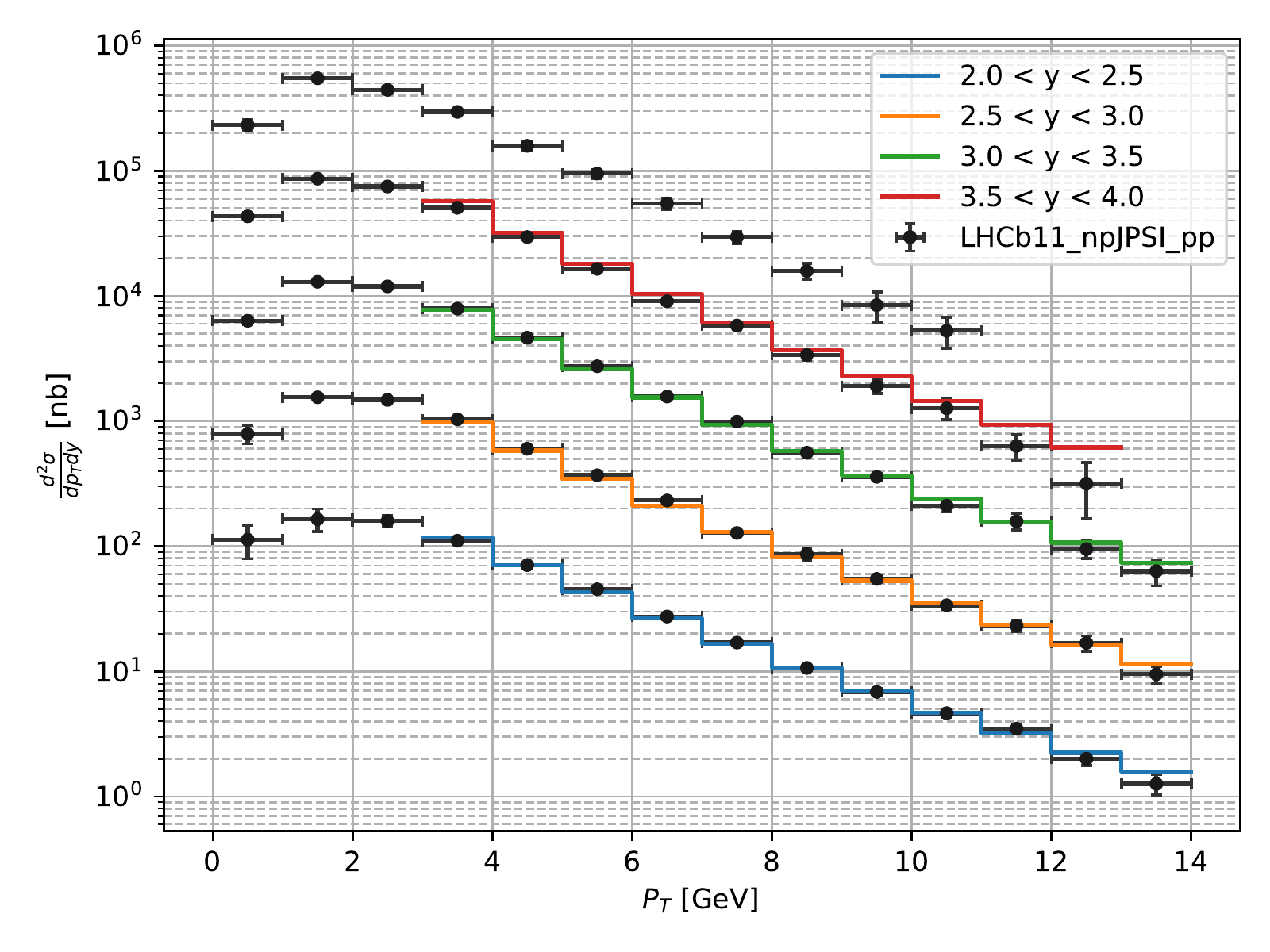}
	\includegraphics[width=0.48\textwidth]{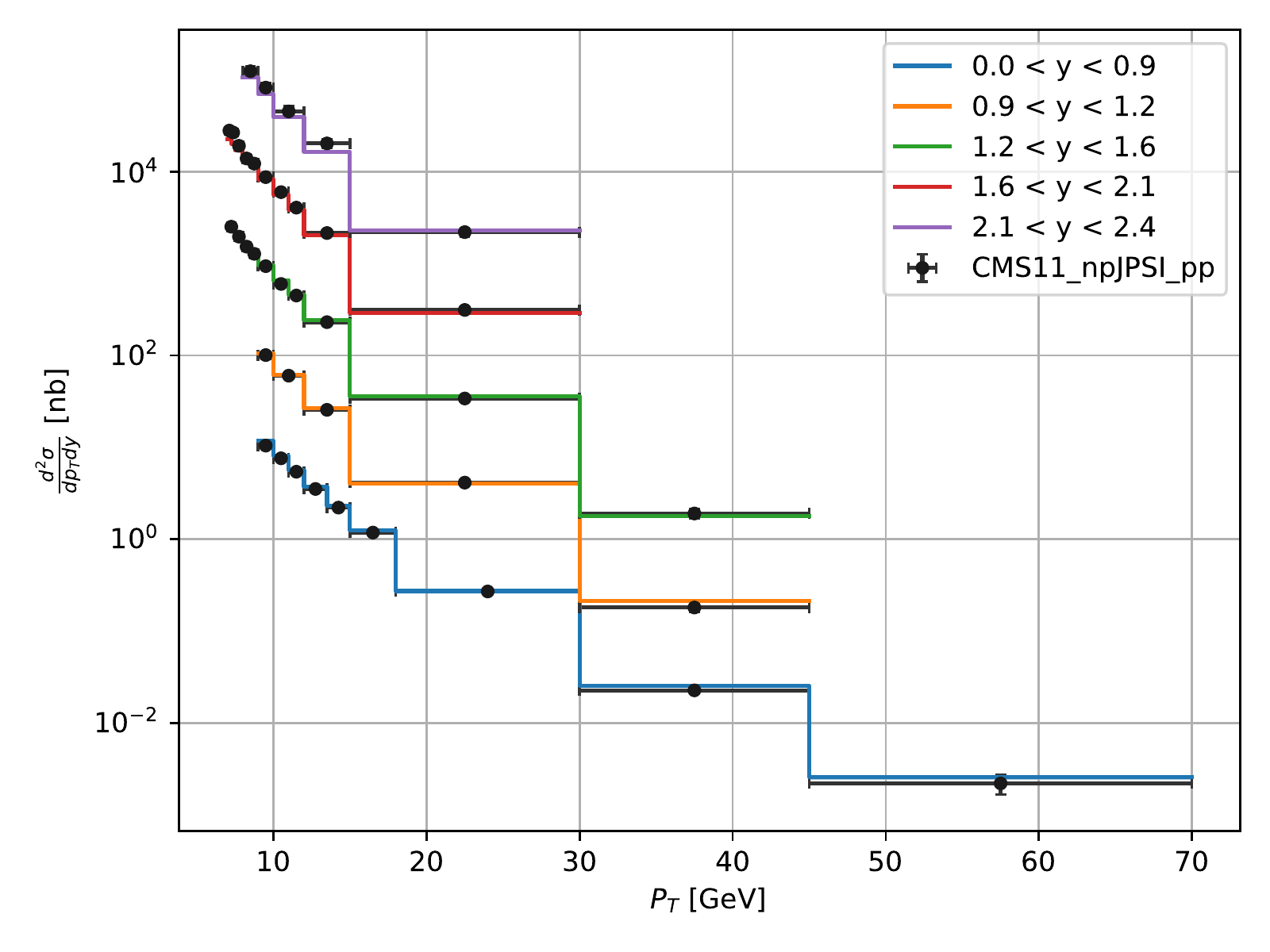}
	\includegraphics[width=0.48\textwidth]{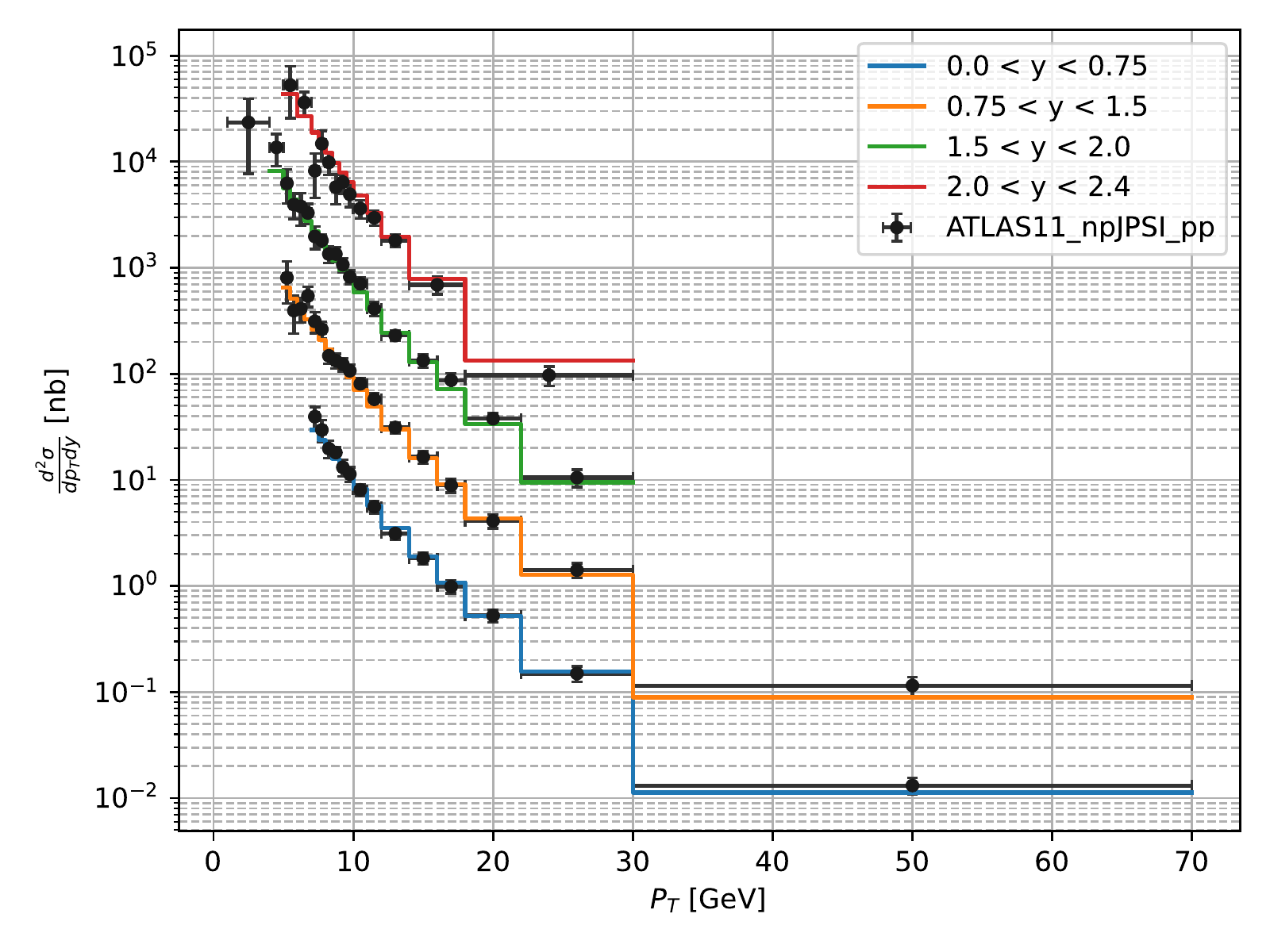}
	\caption{Predictions for $J/\psi$ production in proton-proton collisions with uncertainties from the Crystal Ball fit. Different rapidity bins are separated by multiplying the cross sections by powers of ten for visual clarity.}
	\label{fig:ppBaselineJPSI}
\end{figure*}

\begin{figure*}[htbp!]
	\centering
	\includegraphics[width=0.48\textwidth]{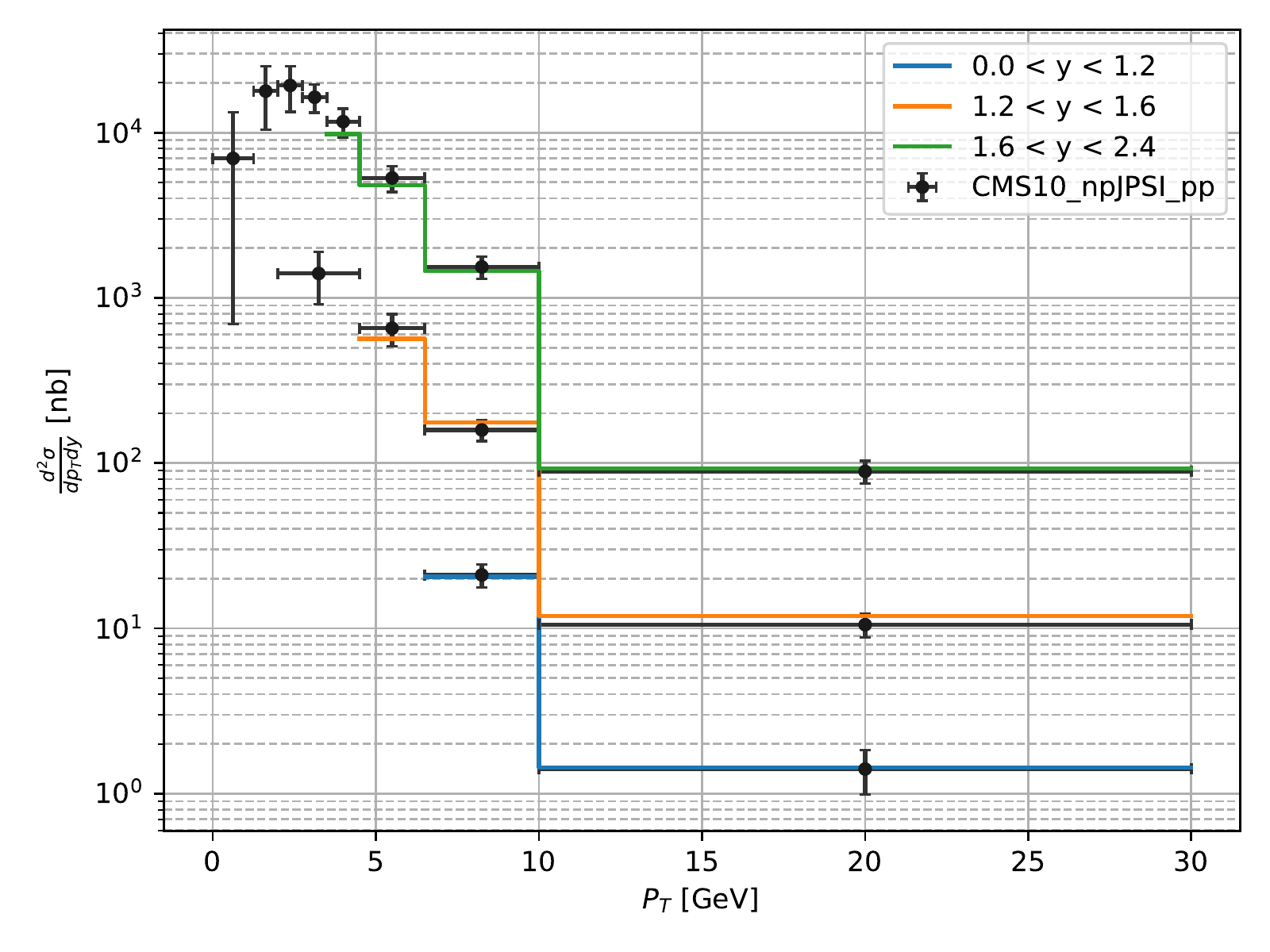}
	\includegraphics[width=0.48\textwidth]{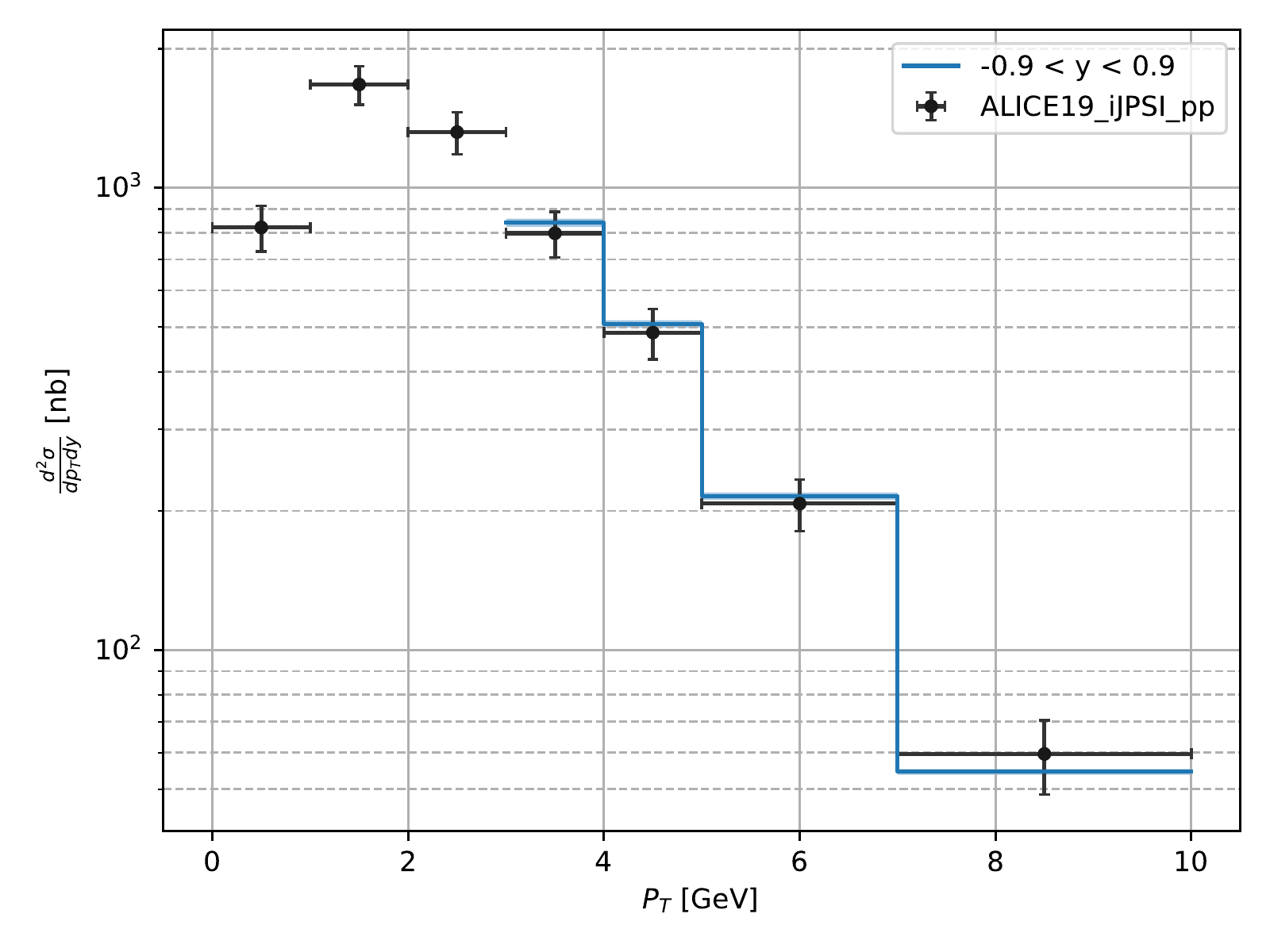}
	\includegraphics[width=0.48\textwidth]{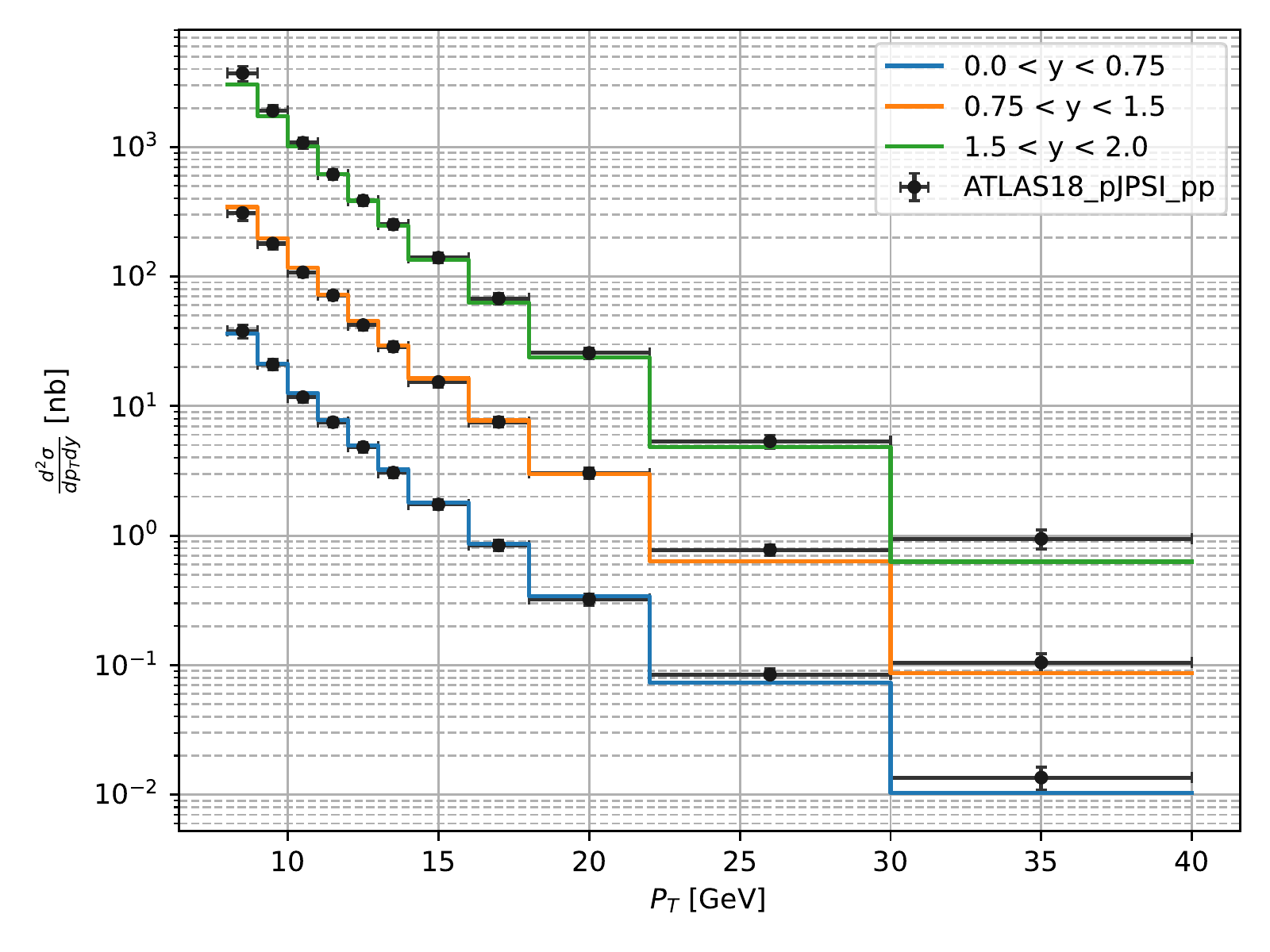}
	\includegraphics[width=0.48\textwidth]{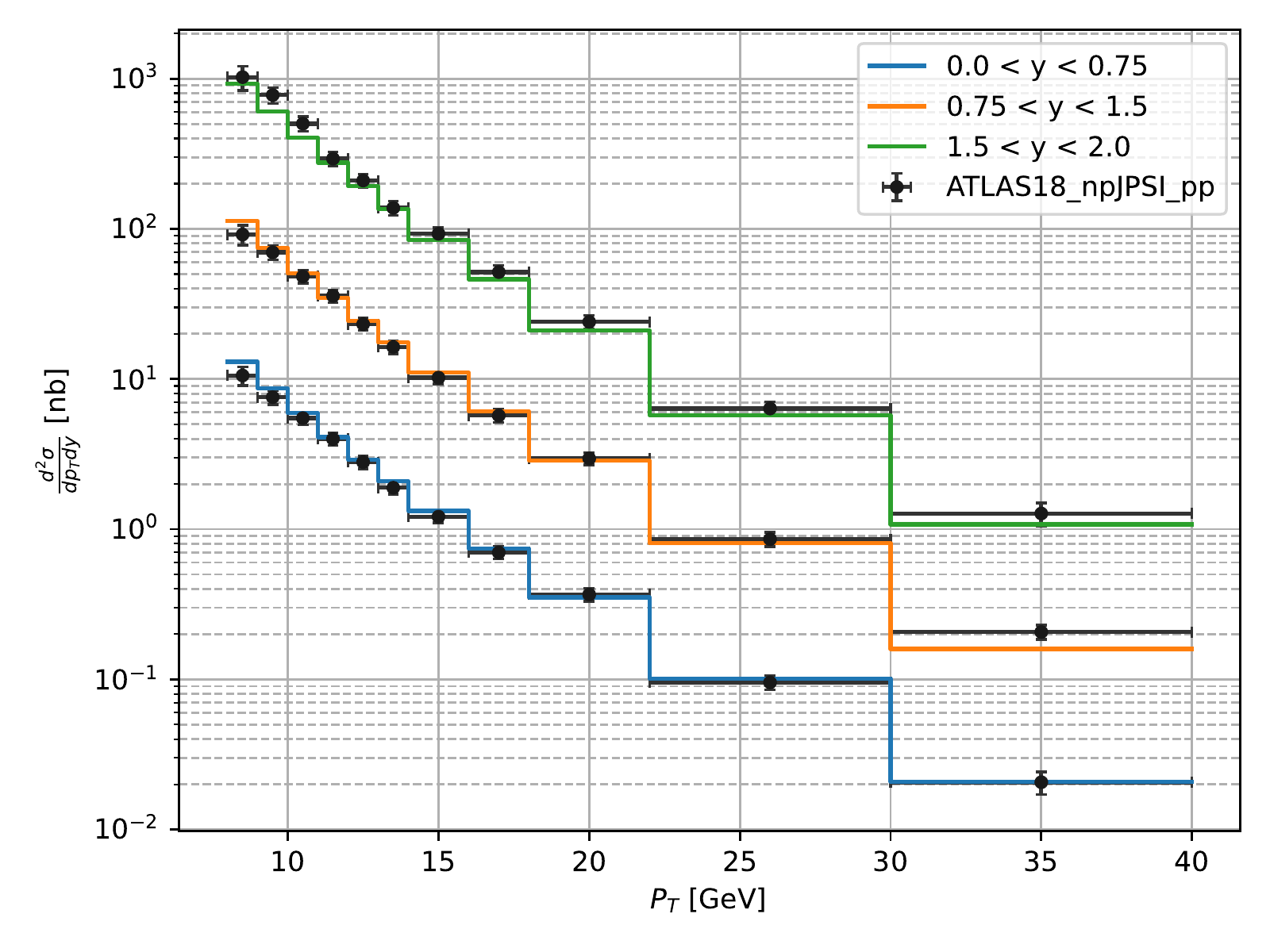}
	\includegraphics[width=0.48\textwidth]{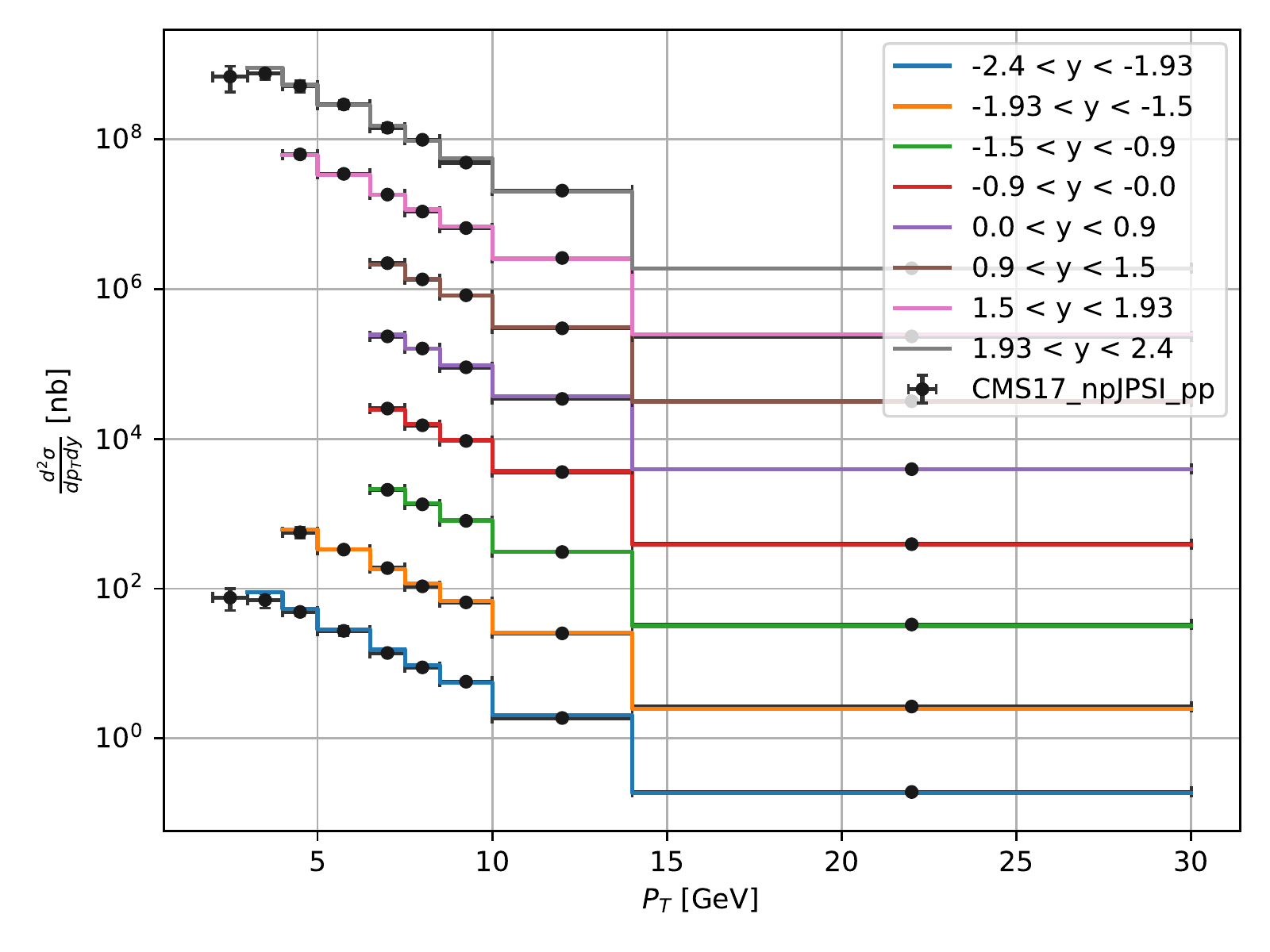}
	\includegraphics[width=0.48\textwidth]{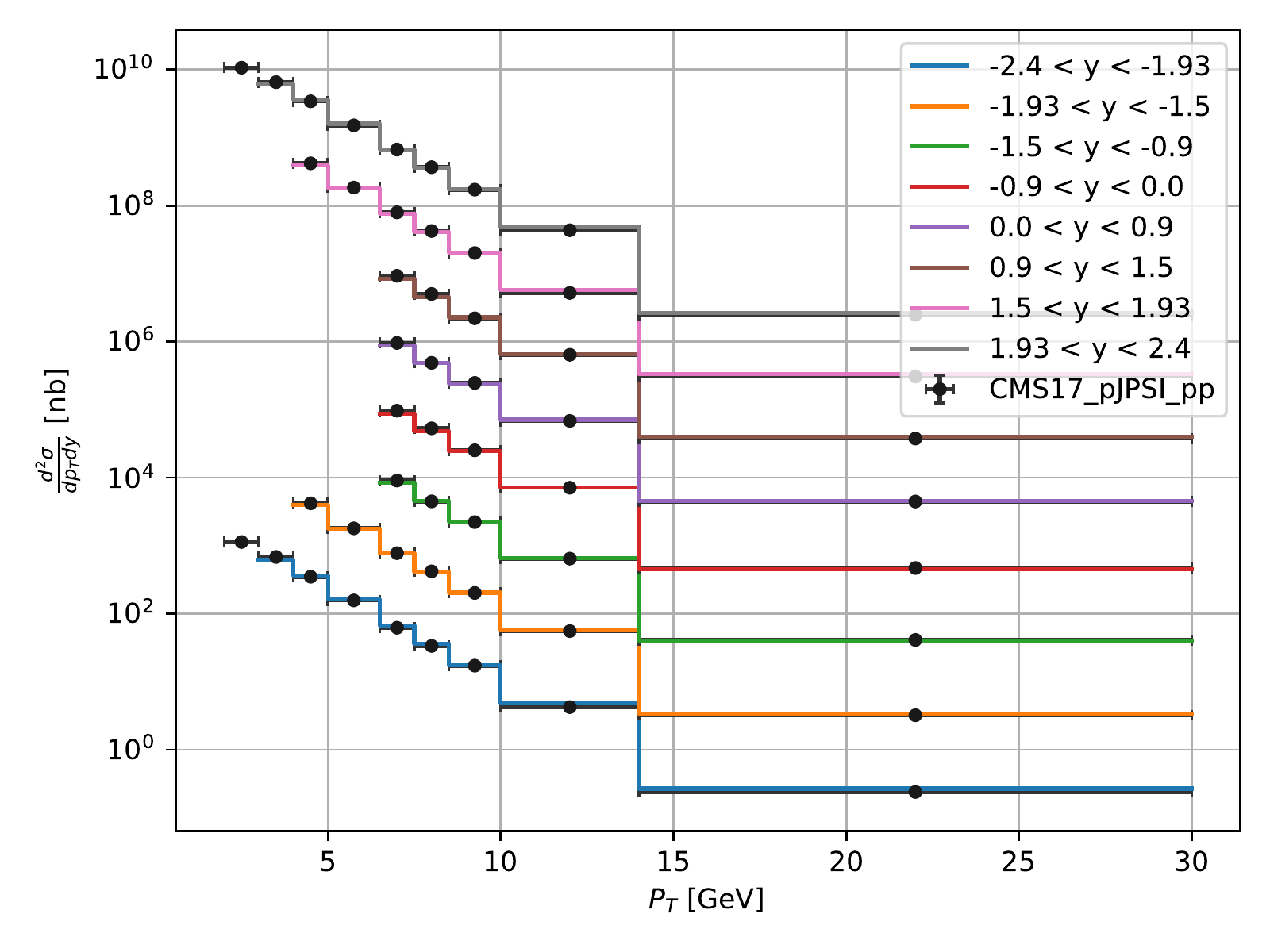}
	\caption{Predictions for $J/\psi$ production in proton-proton collisions with uncertainties from the Crystal Ball fit. Different rapidity bins are separated by multiplying the cross sections by powers of ten for visual clarity.}
	\label{fig:ppBaselineJPSI2}
\end{figure*}

\begin{figure*}[htbp!]
	\centering
	\includegraphics[width=0.48\textwidth]{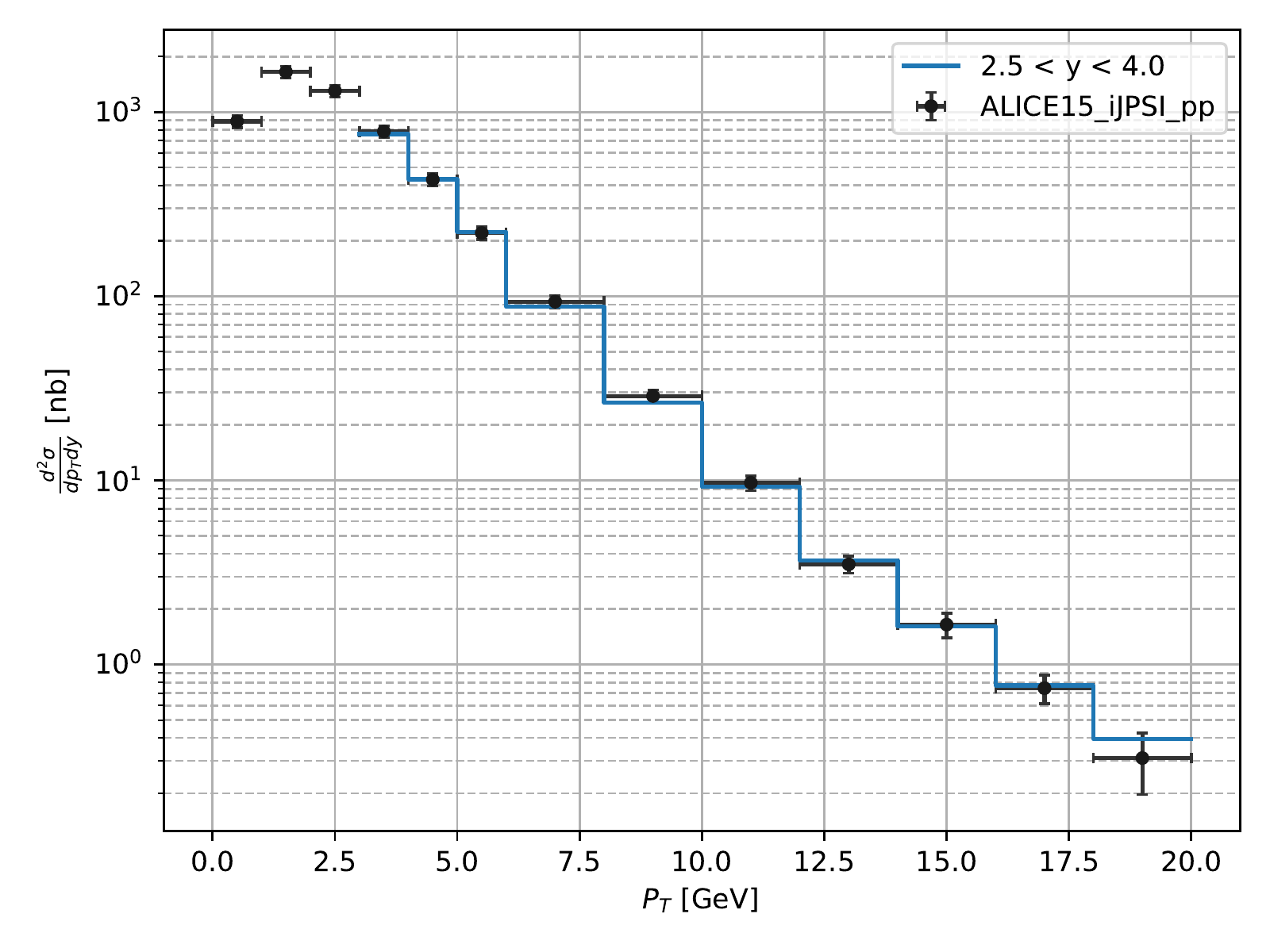}
	\includegraphics[width=0.48\textwidth]{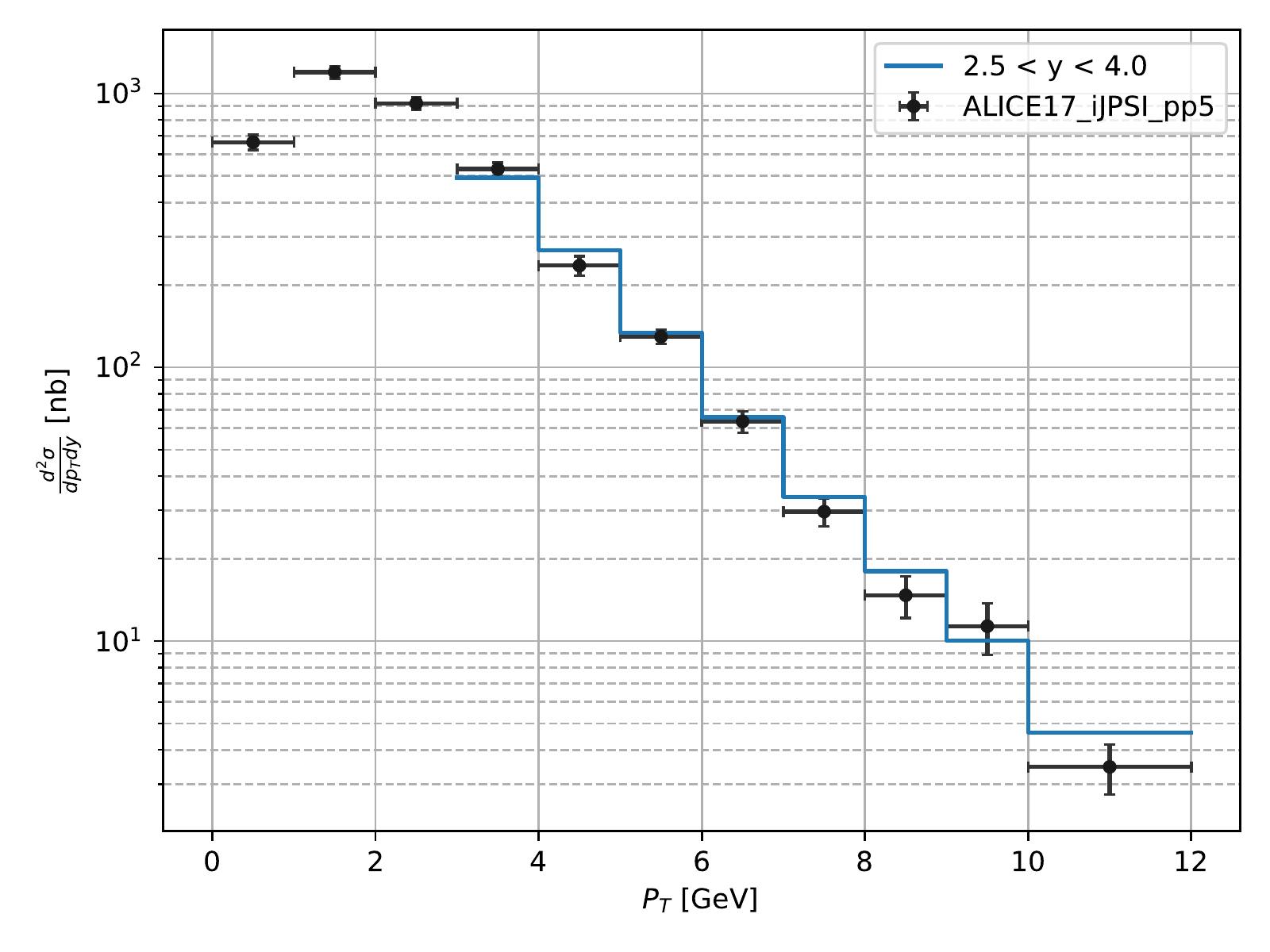}
	\includegraphics[width=0.48\textwidth]{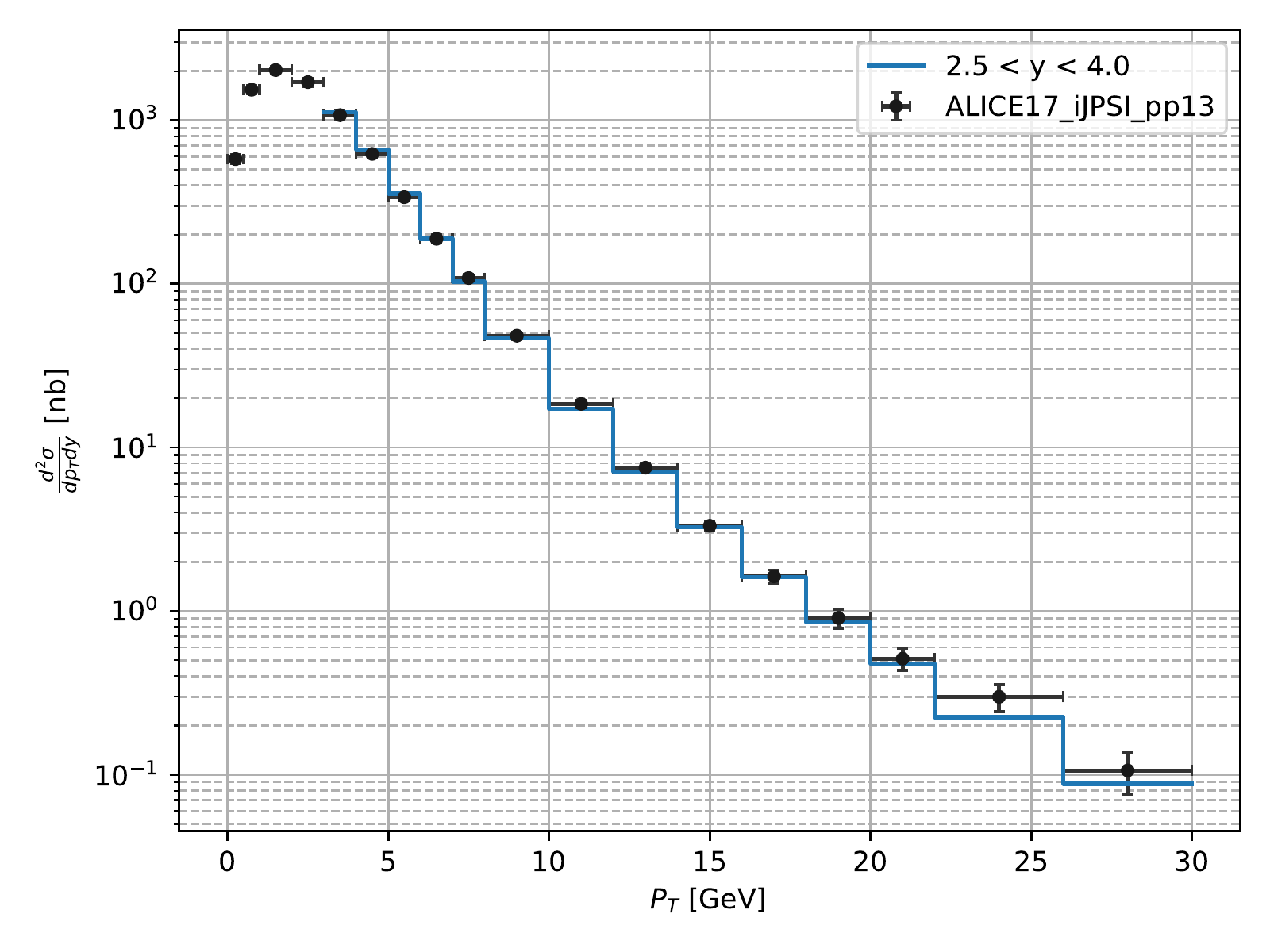}
	\caption{Predictions for $J/\psi$ production in proton-proton collisions with uncertainties from the Crystal Ball fit. Different rapidity bins are separated by multiplying the cross sections by powers of ten for visual clarity.}
	\label{fig:ppBaselineJPSI3}
\end{figure*}

\begin{figure*}[htbp!]
	\centering
	\includegraphics[width=0.48\textwidth]{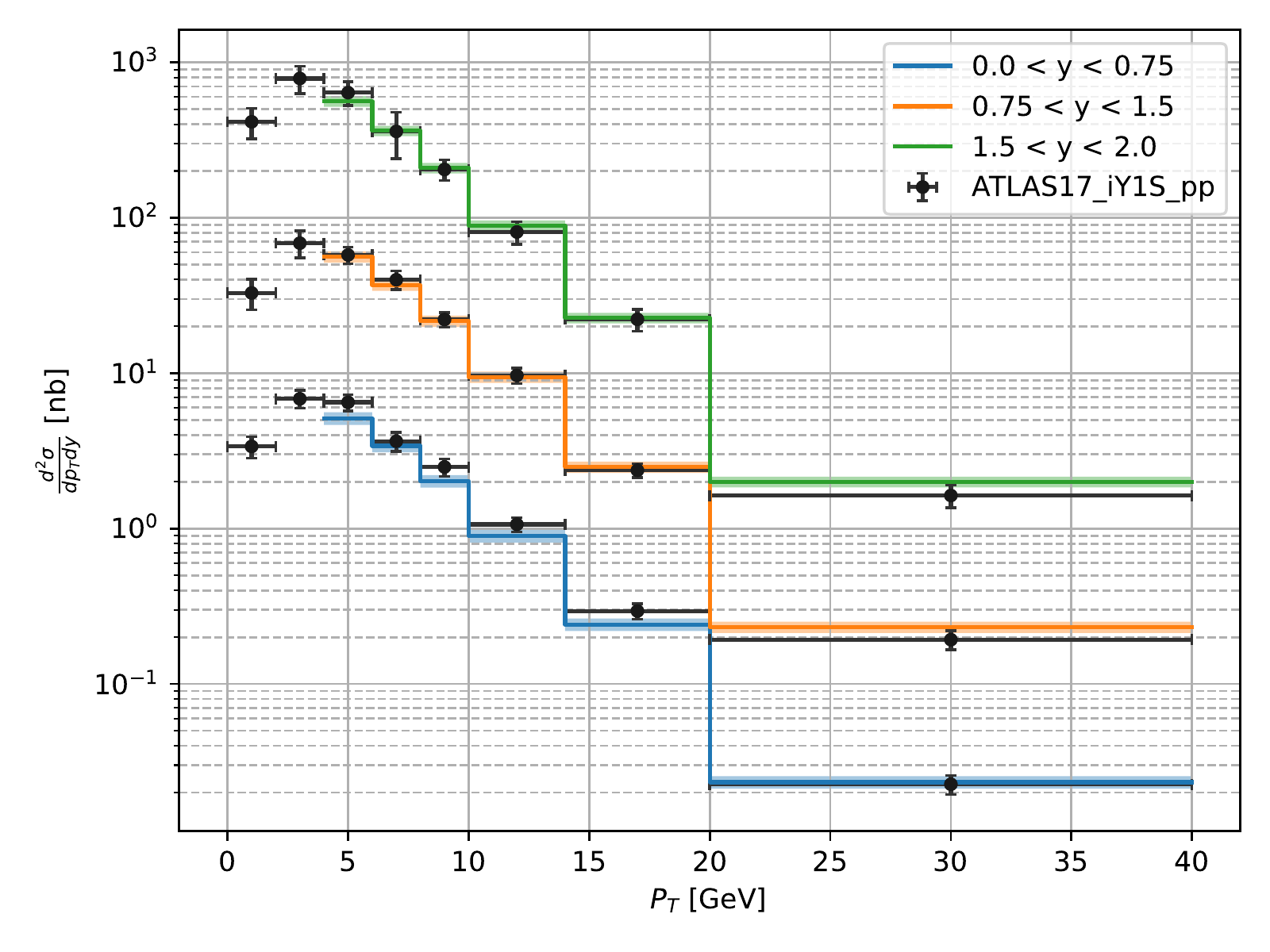}
	\includegraphics[width=0.48\textwidth]{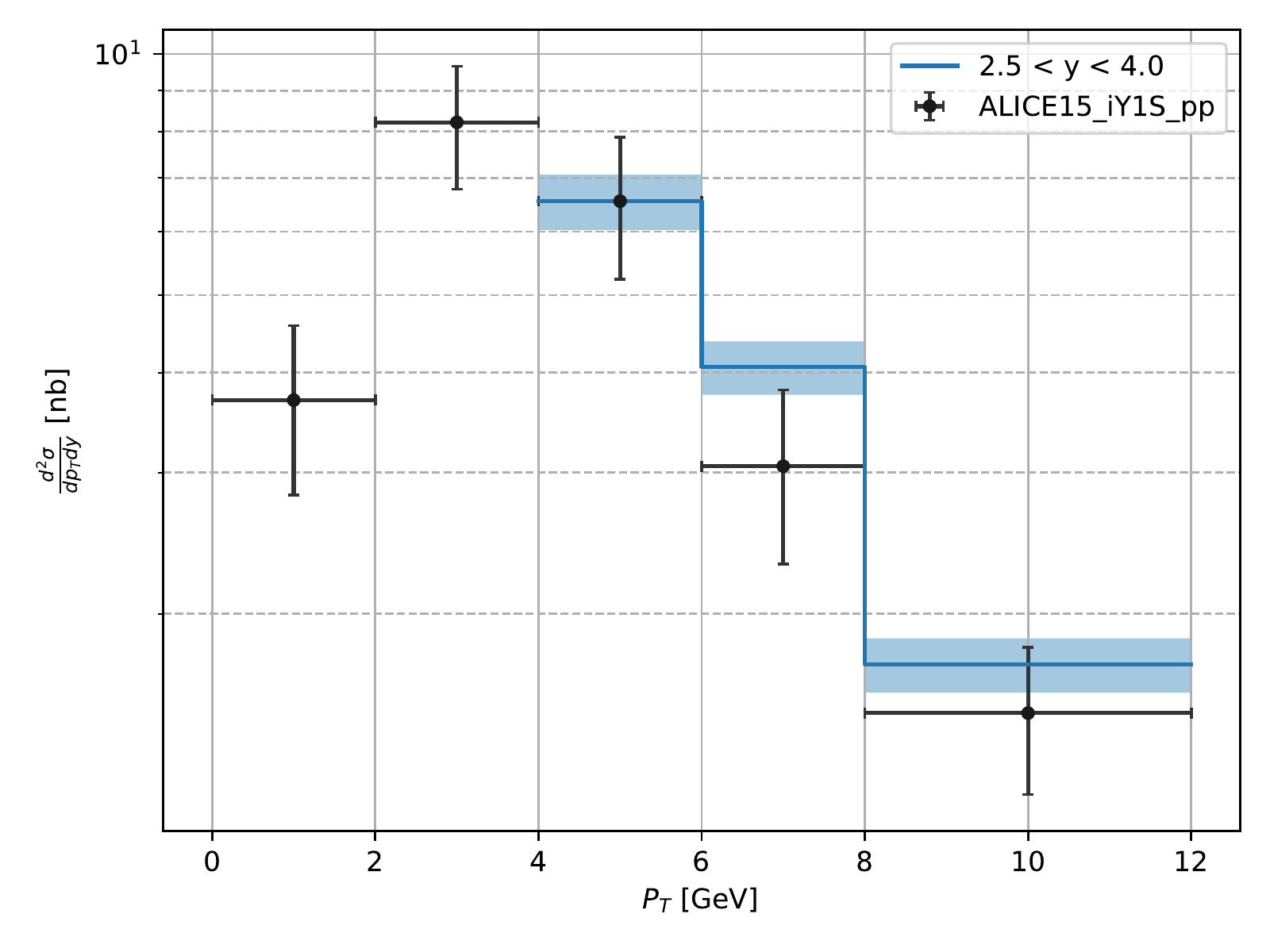}
	\caption{Predictions for $\Upsilon(1S)$ production in proton-proton collisions with uncertainties from the Crystal Ball fit. Different rapidity bins are separated by multiplying the cross sections by powers of ten for visual clarity.}
	\label{fig:ppBaselineY1S2}
\end{figure*}

\begin{figure*}[htbp!]
	\centering
	\includegraphics[width=0.48\textwidth]{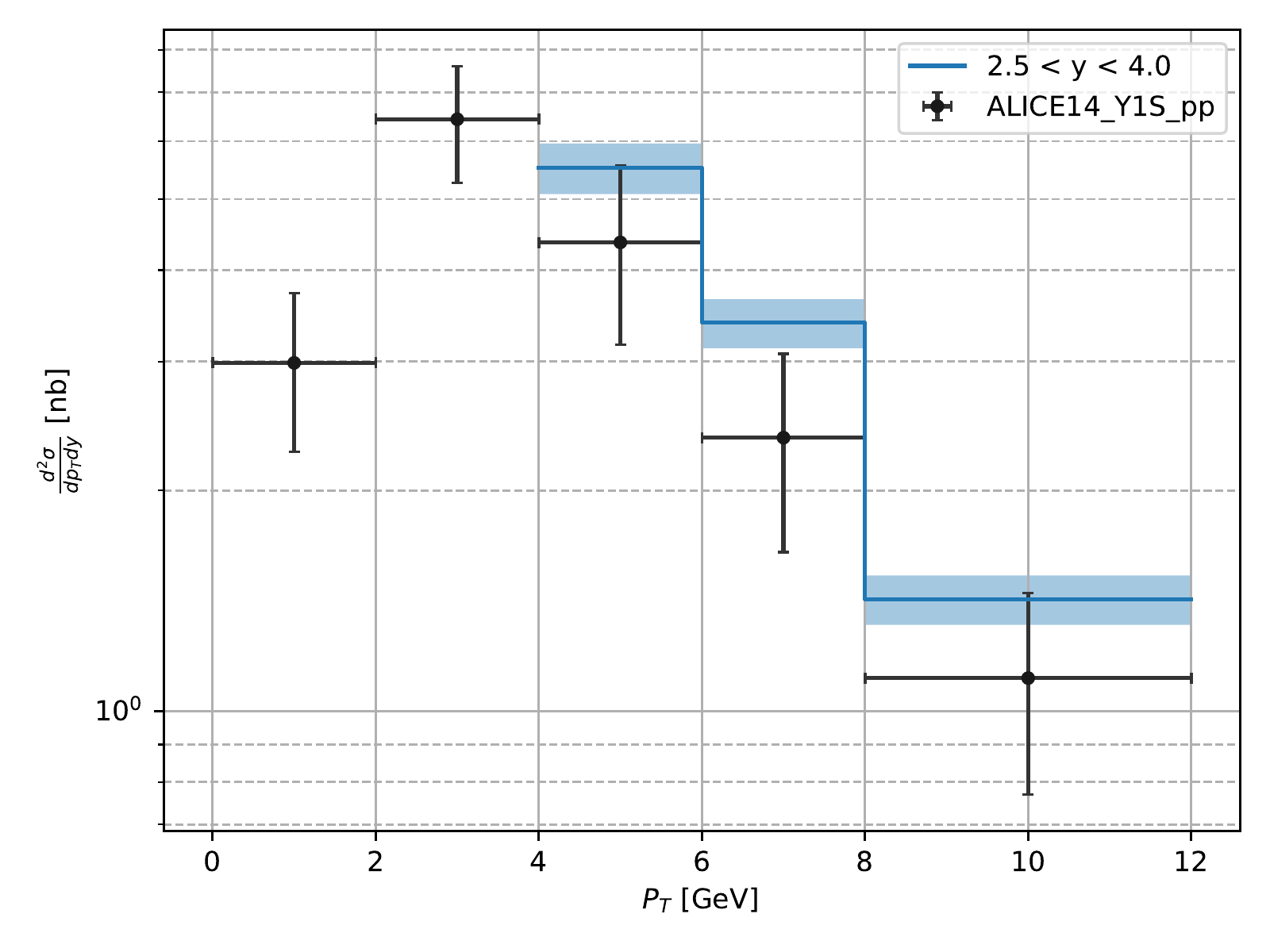}
	\includegraphics[width=0.48\textwidth]{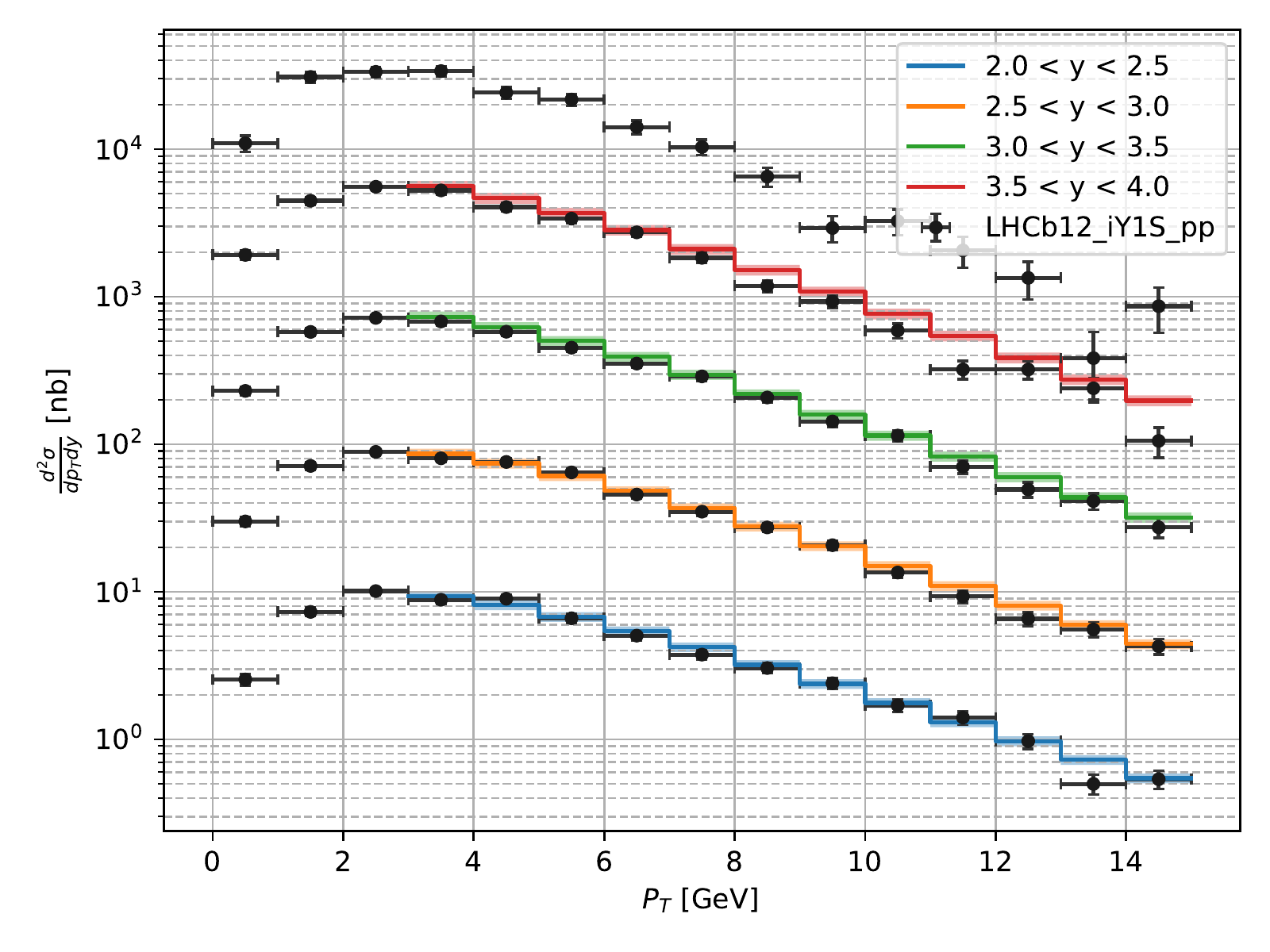}
	\includegraphics[width=0.48\textwidth]{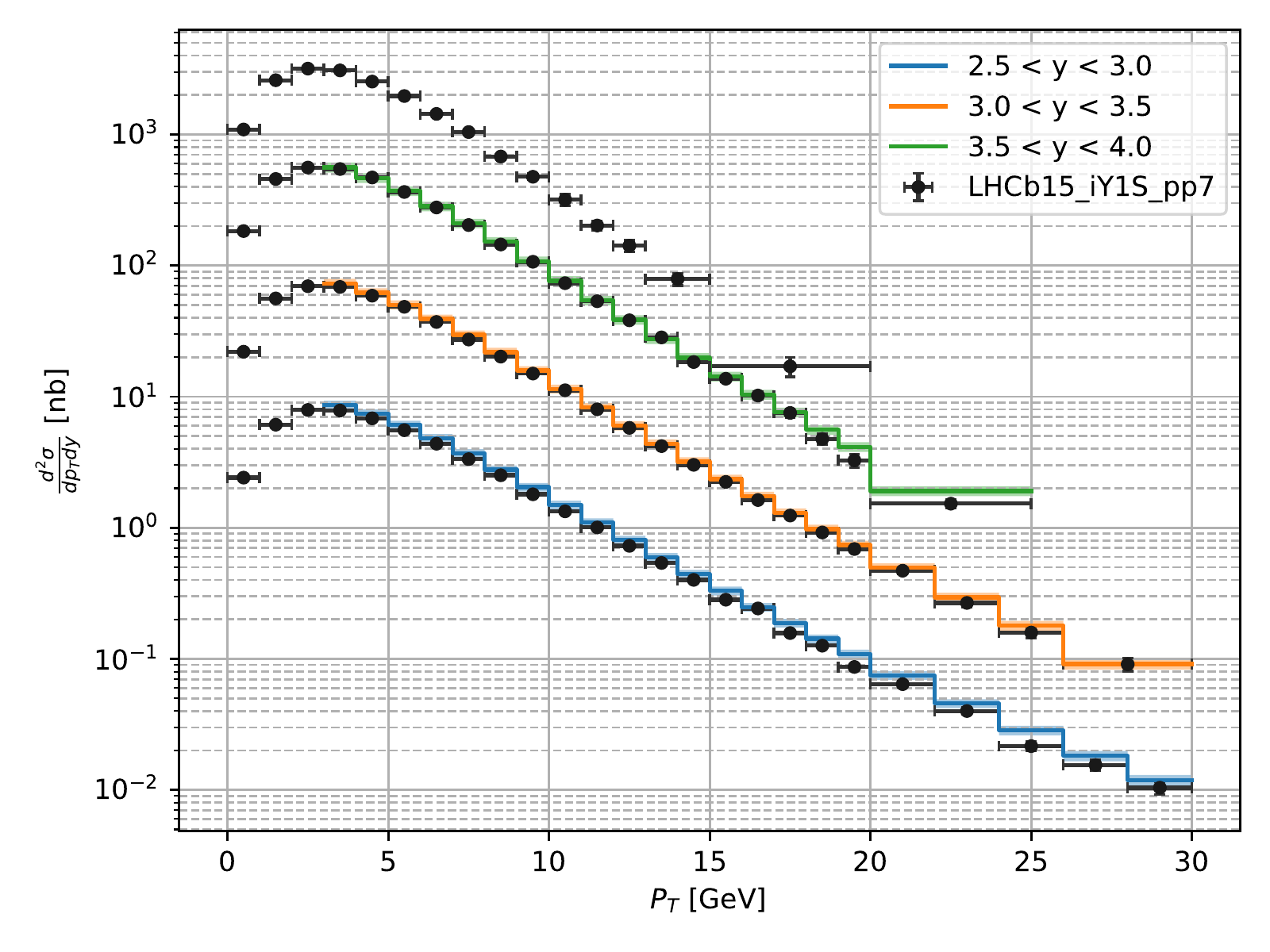}
	\includegraphics[width=0.48\textwidth]{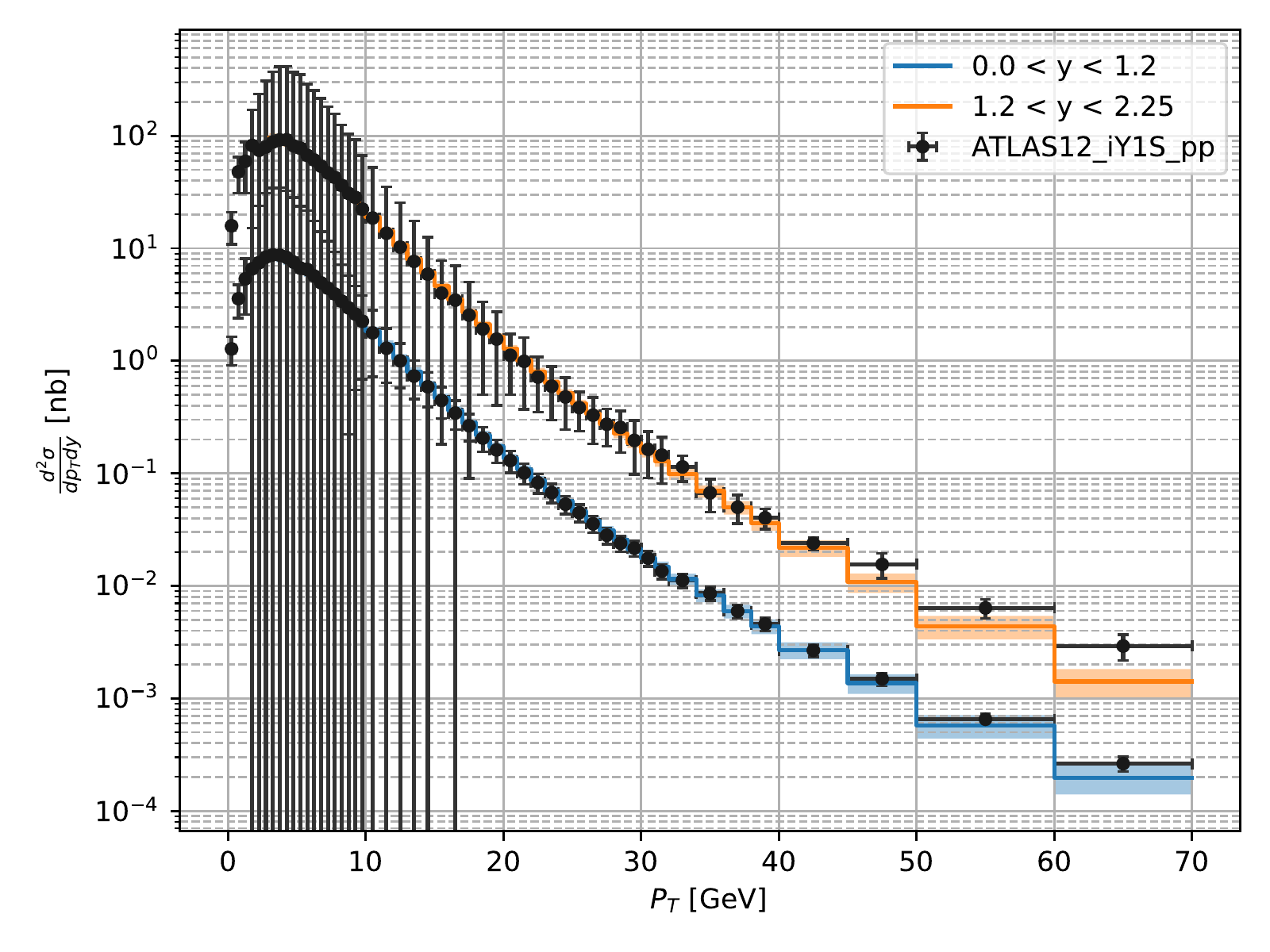}
	\includegraphics[width=0.48\textwidth]{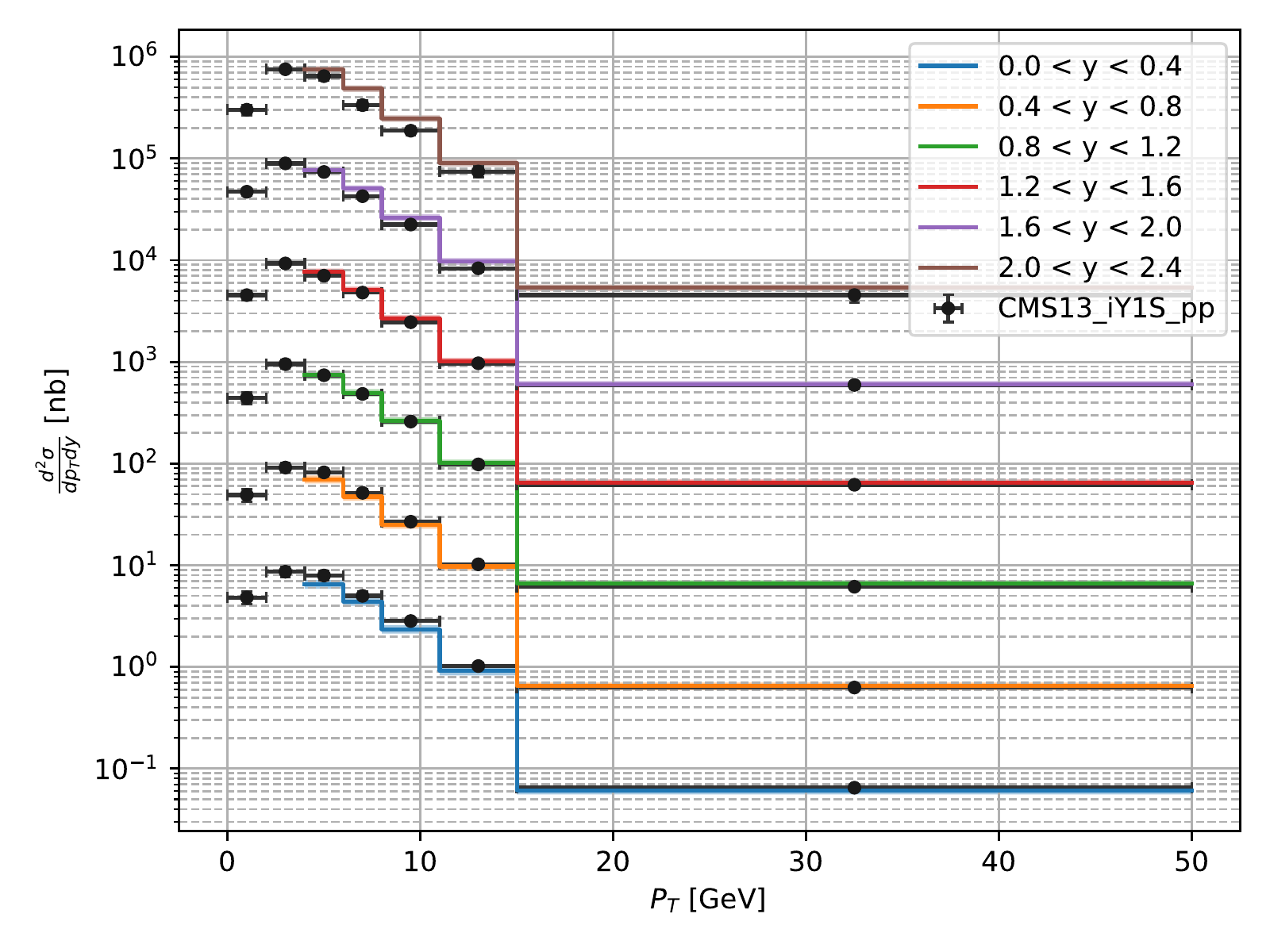}
	\includegraphics[width=0.48\textwidth]{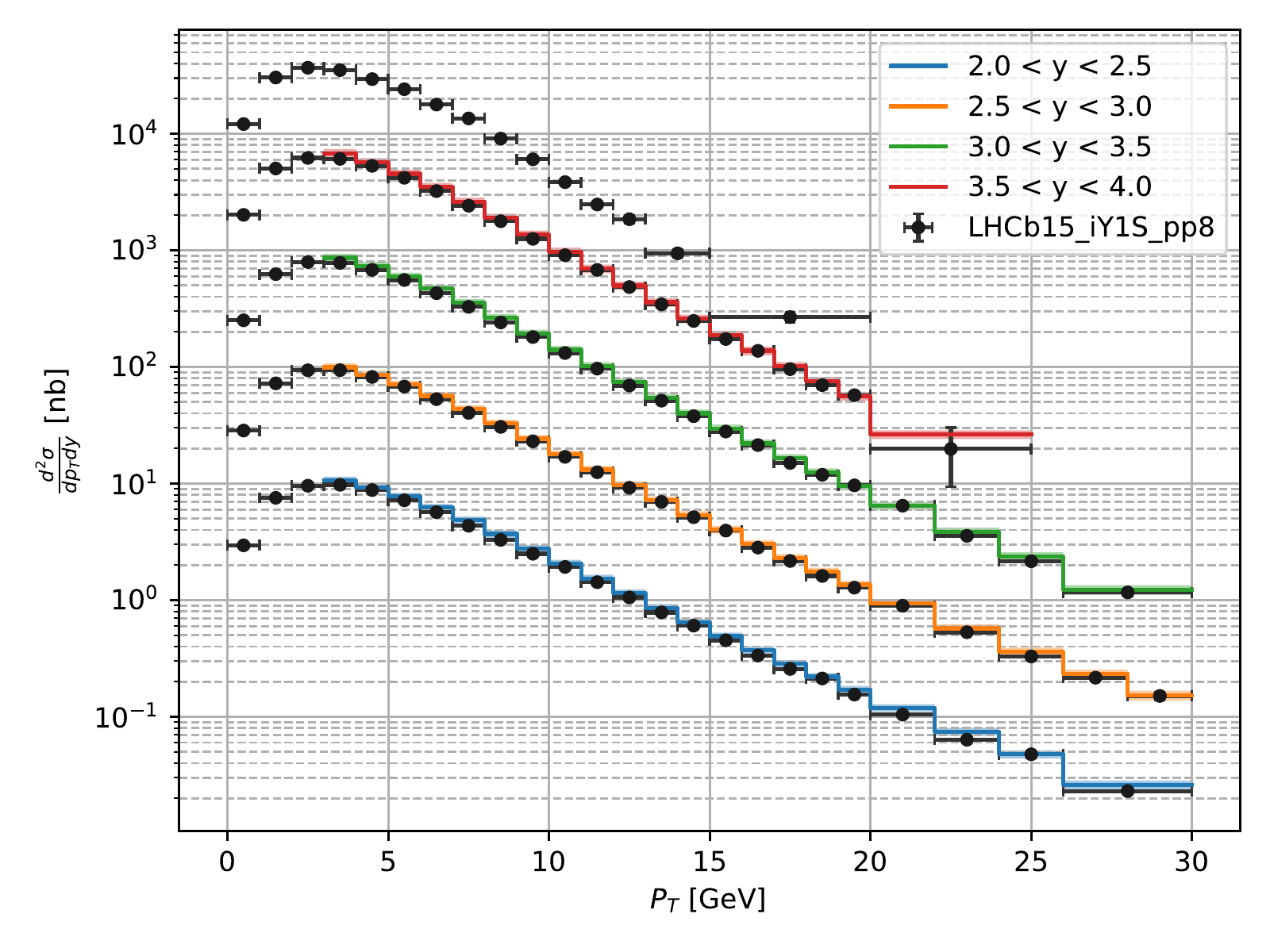}
	\caption{Predictions for $\Upsilon(1S)$ production in proton-proton collisions with uncertainties from the Crystal Ball fit. Different rapidity bins are separated by multiplying the cross sections by powers of ten for visual clarity.}
	\label{fig:ppBaselineY1S}
\end{figure*}

\begin{figure*}[htbp!]
	\centering
	\includegraphics[width=0.48\textwidth]{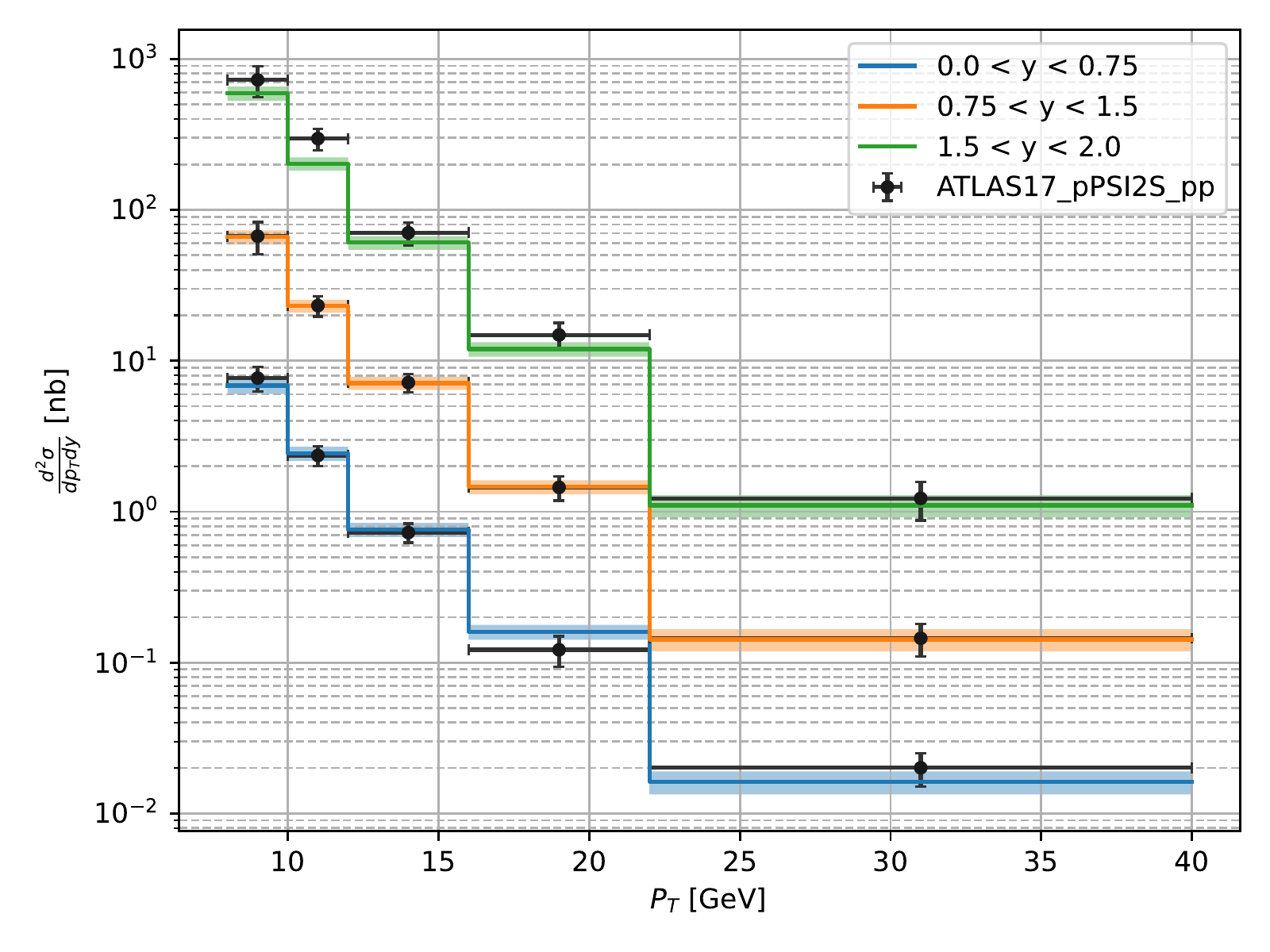}
	\includegraphics[width=0.48\textwidth]{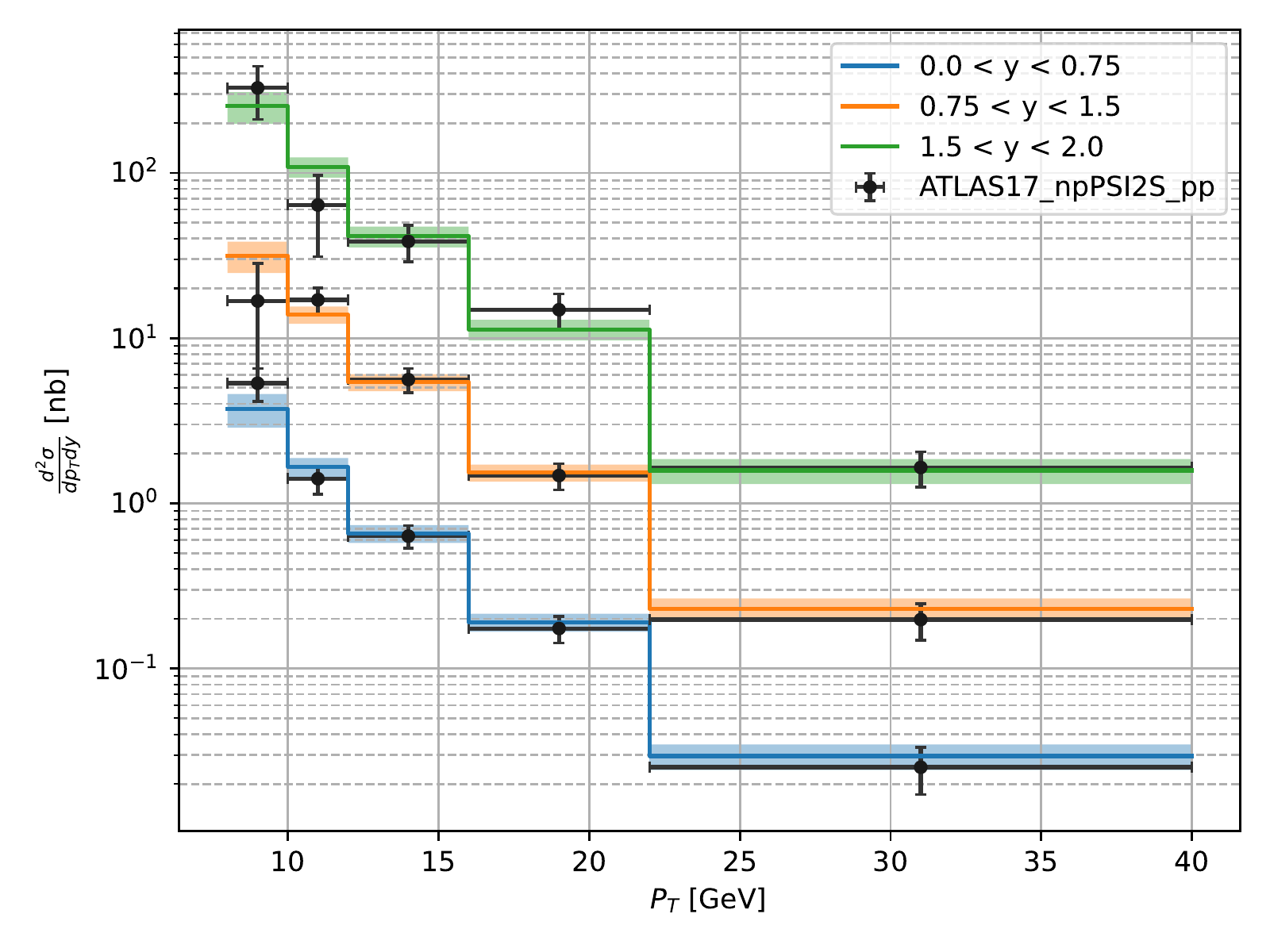}
	\includegraphics[width=0.48\textwidth]{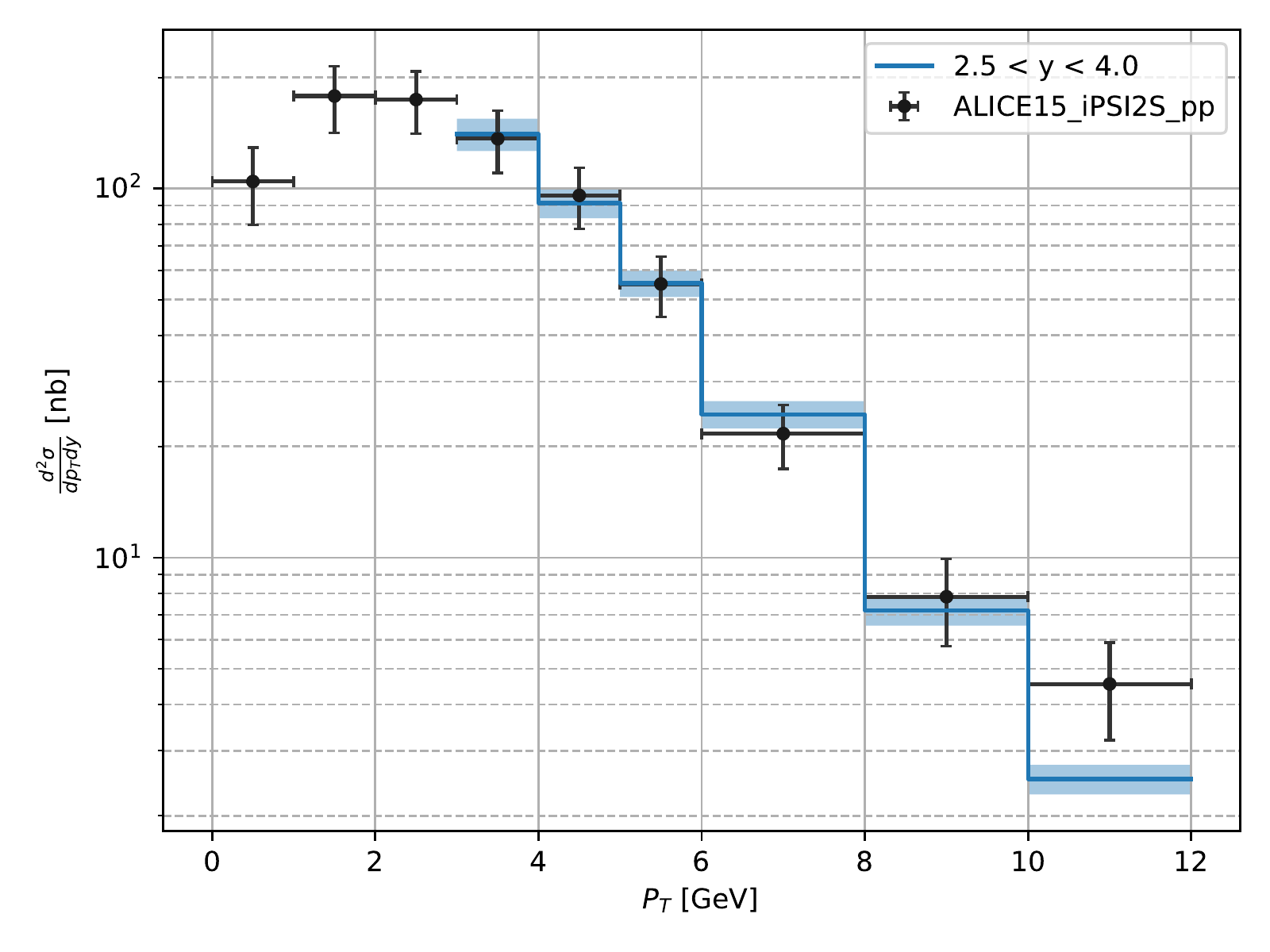}
	\includegraphics[width=0.48\textwidth]{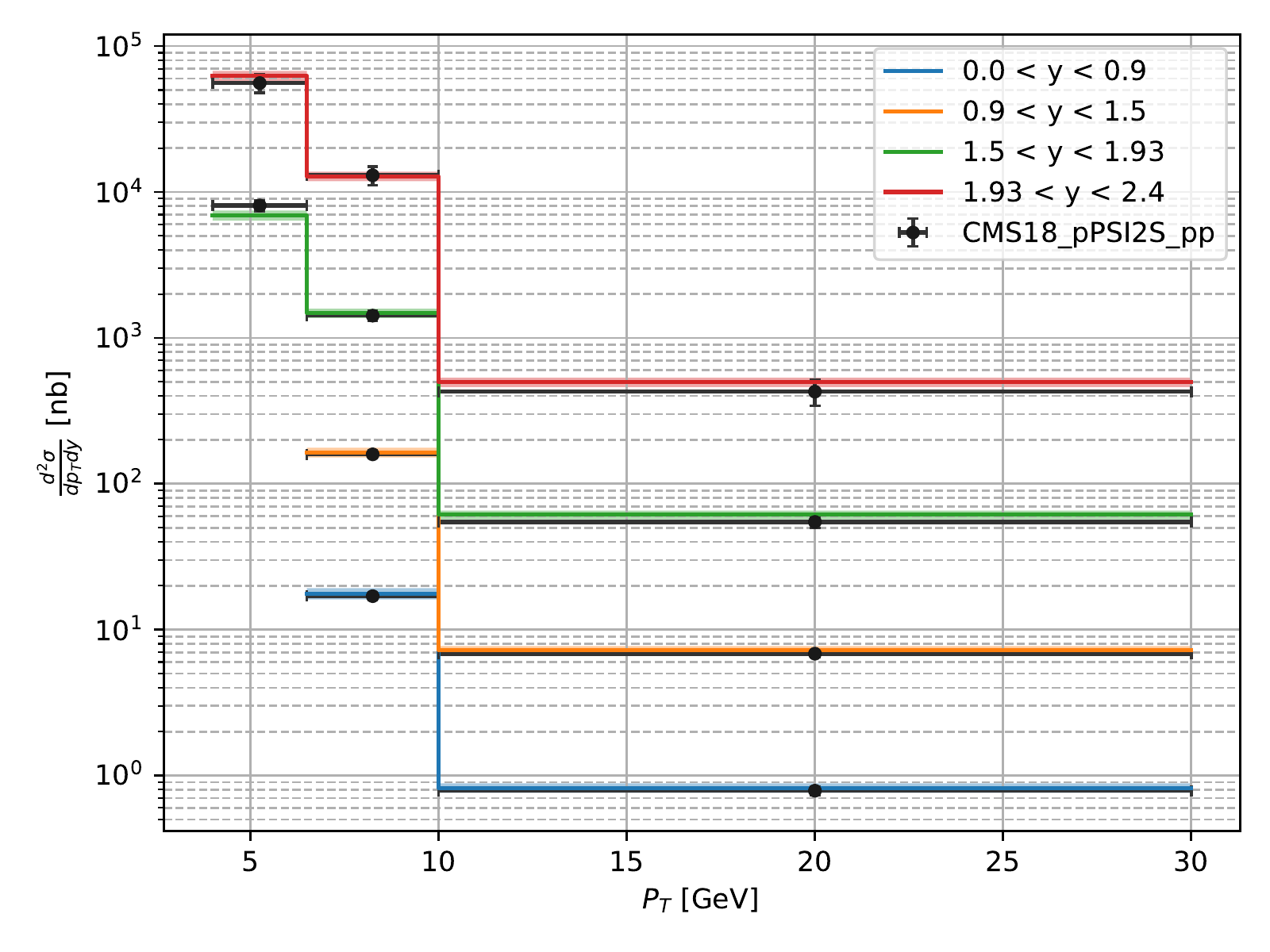}
	\includegraphics[width=0.48\textwidth]{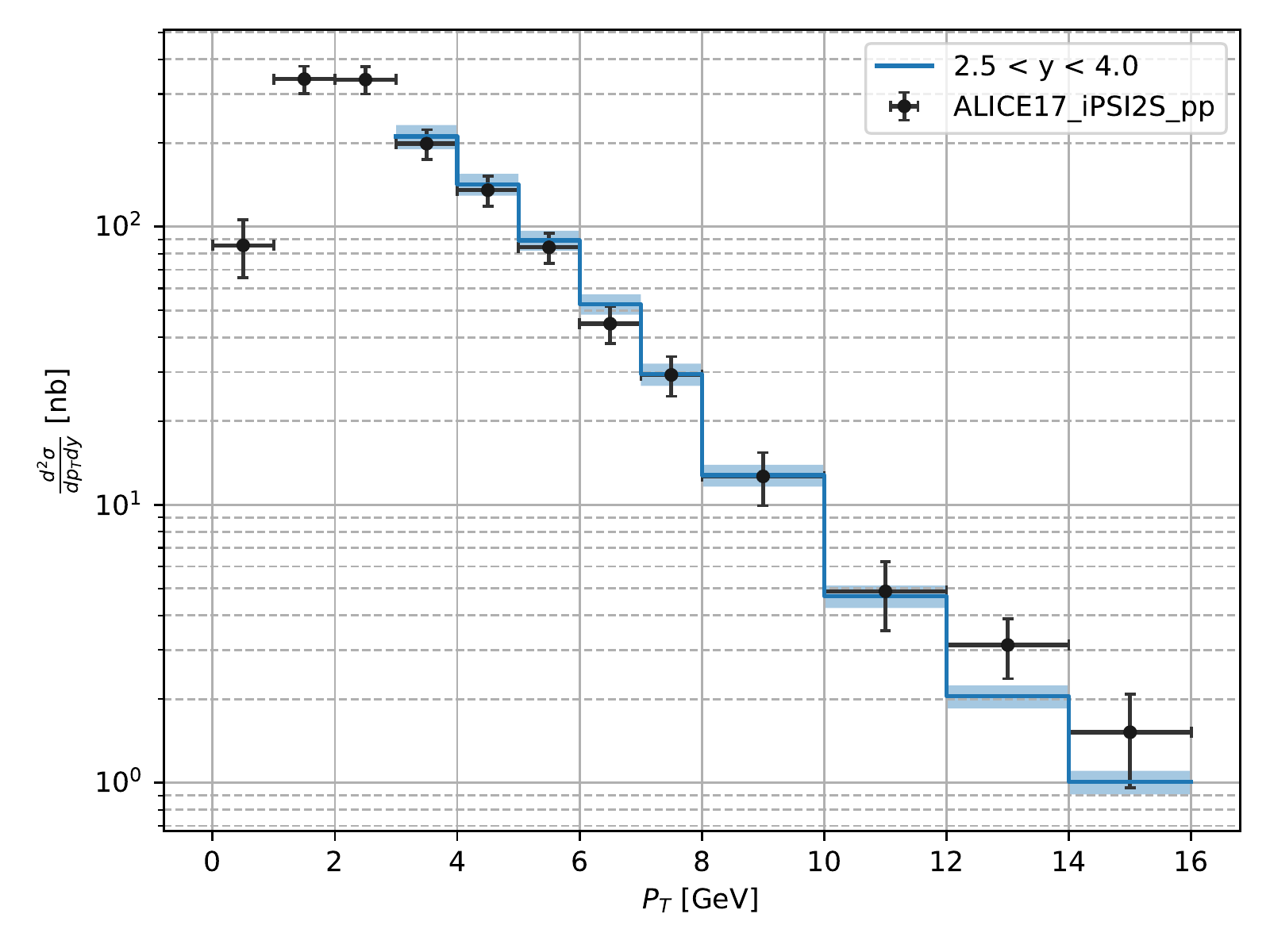}
	\caption{Predictions for $\psi(2S)$ production in proton-proton collisions with uncertainties from the Crystal Ball fit. Different rapidity bins are separated by multiplying the cross sections by powers of ten for visual clarity.}
	\label{fig:ppBaselinePSI2S}
\end{figure*}

\begin{figure*}[htbp!]
	\centering
	\includegraphics[width=0.48\textwidth]{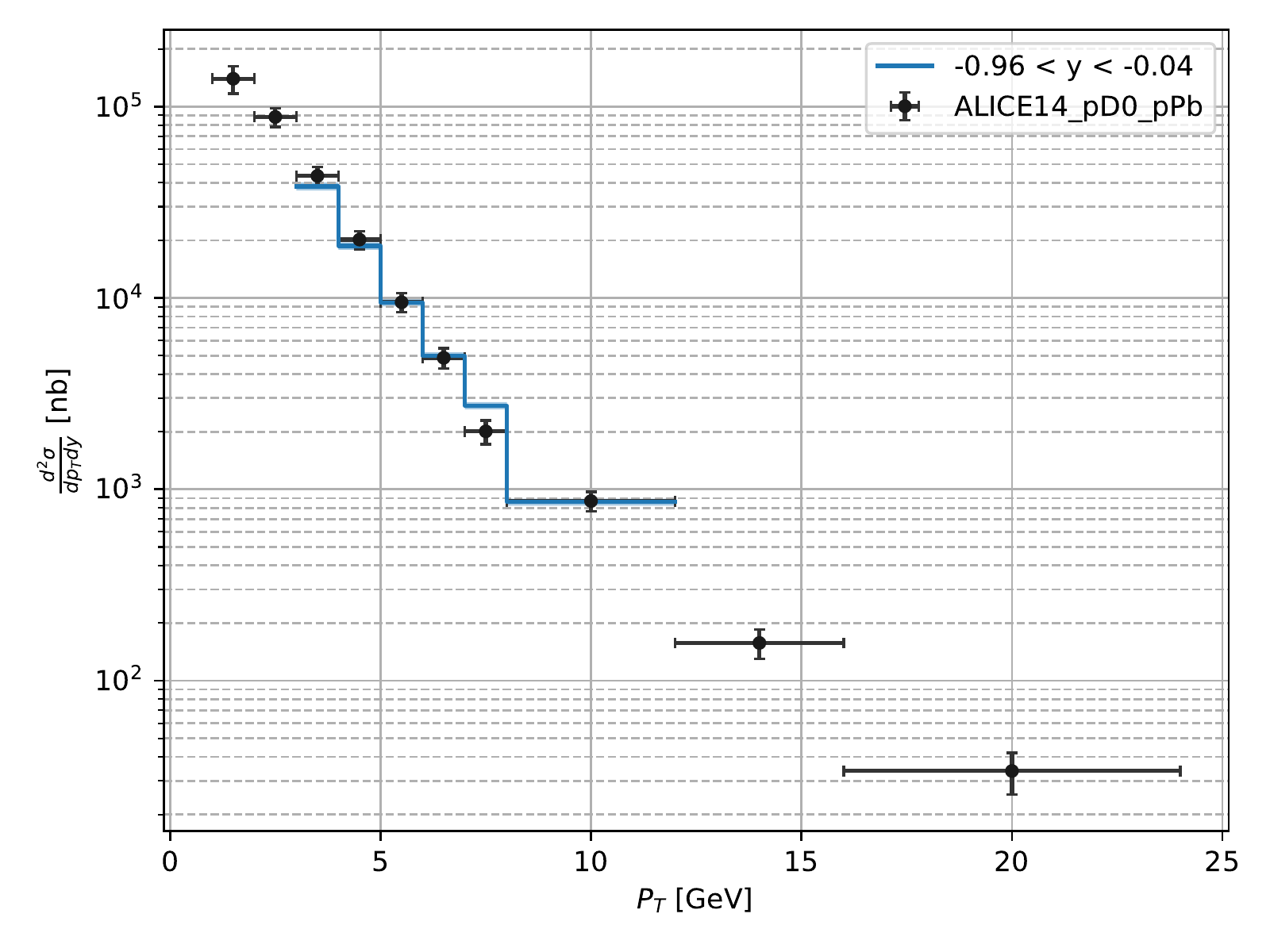}
	\includegraphics[width=0.48\textwidth]{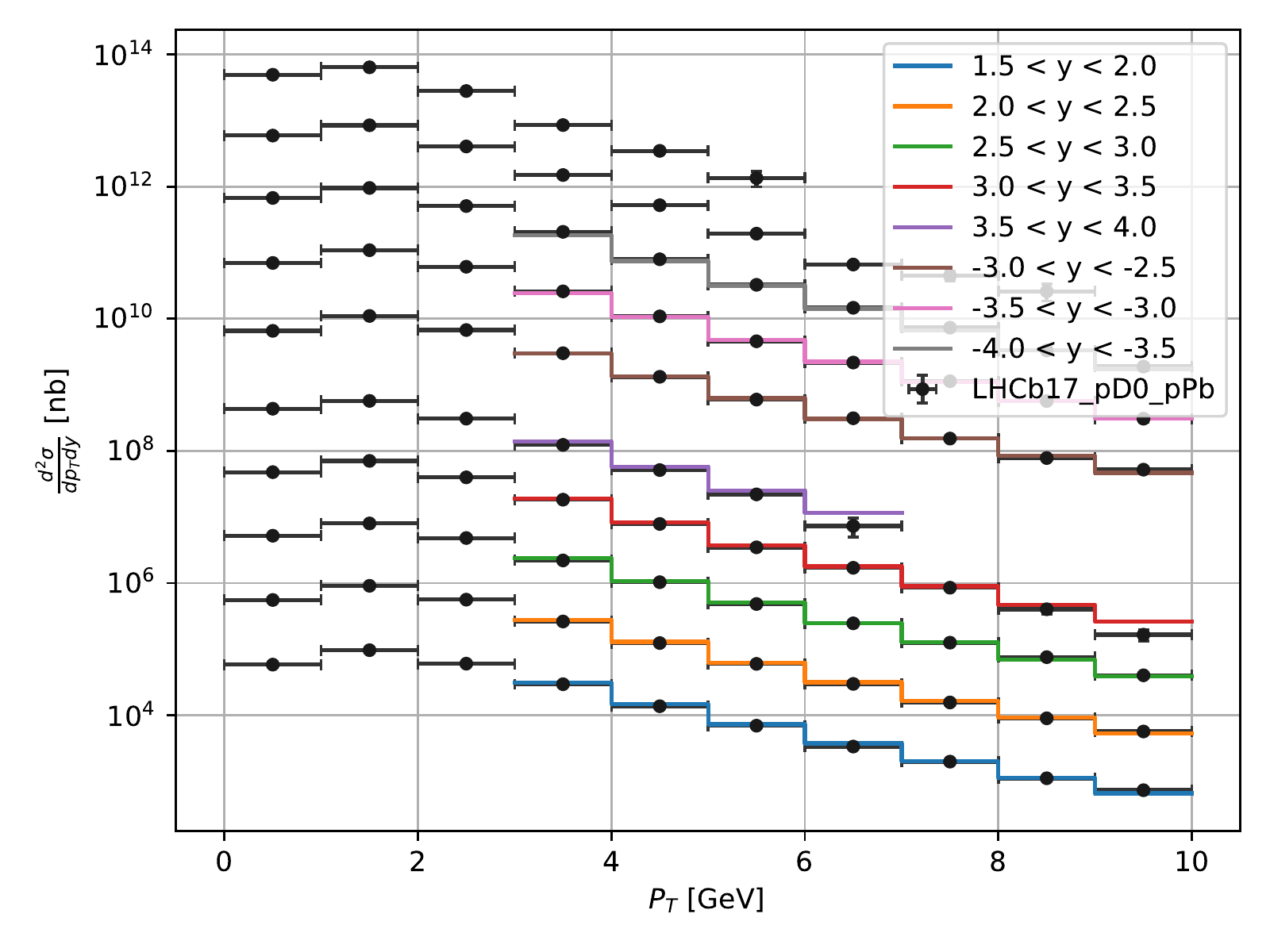}
	\includegraphics[width=0.48\textwidth]{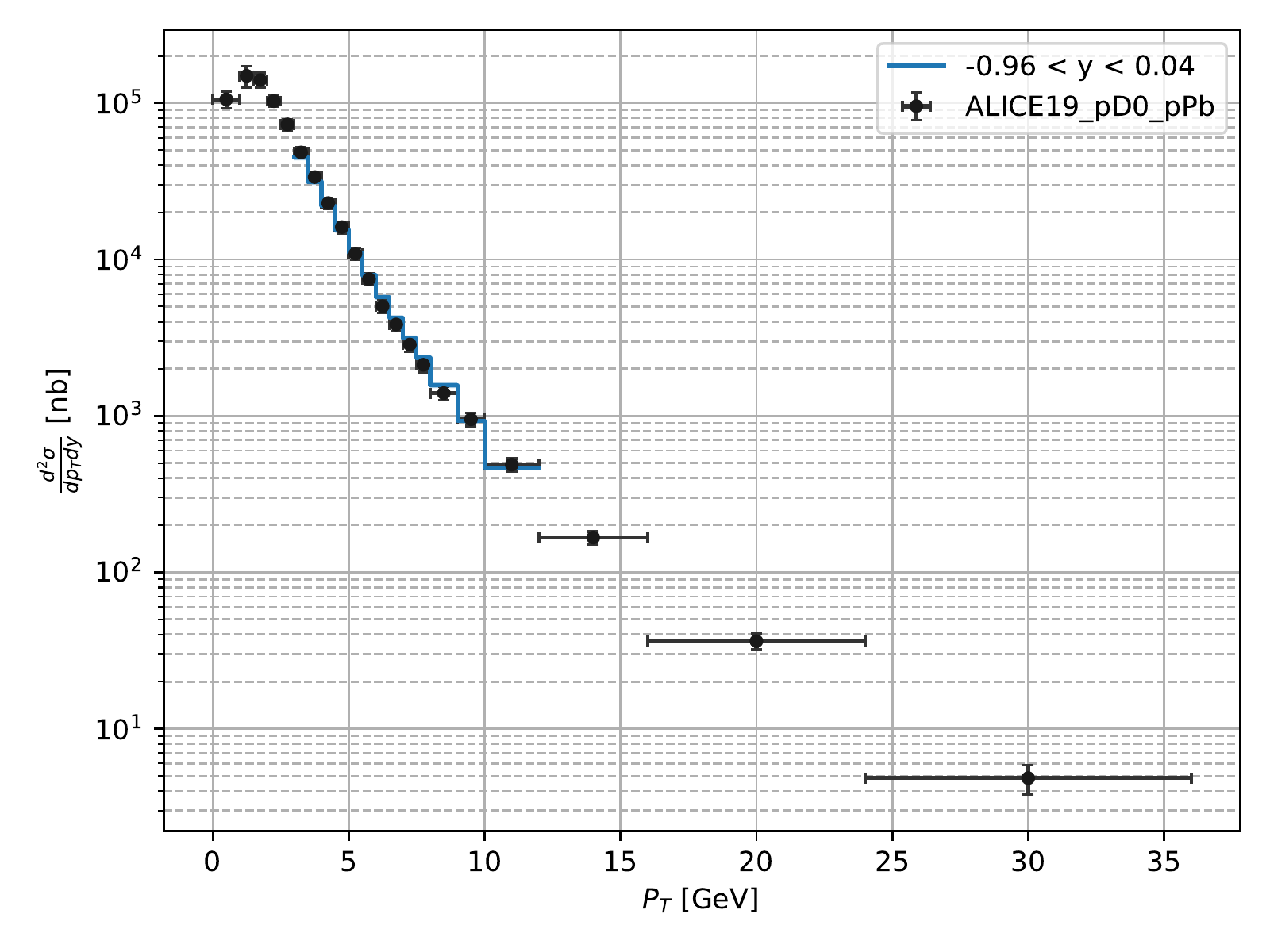}
	\includegraphics[width=0.48\textwidth]{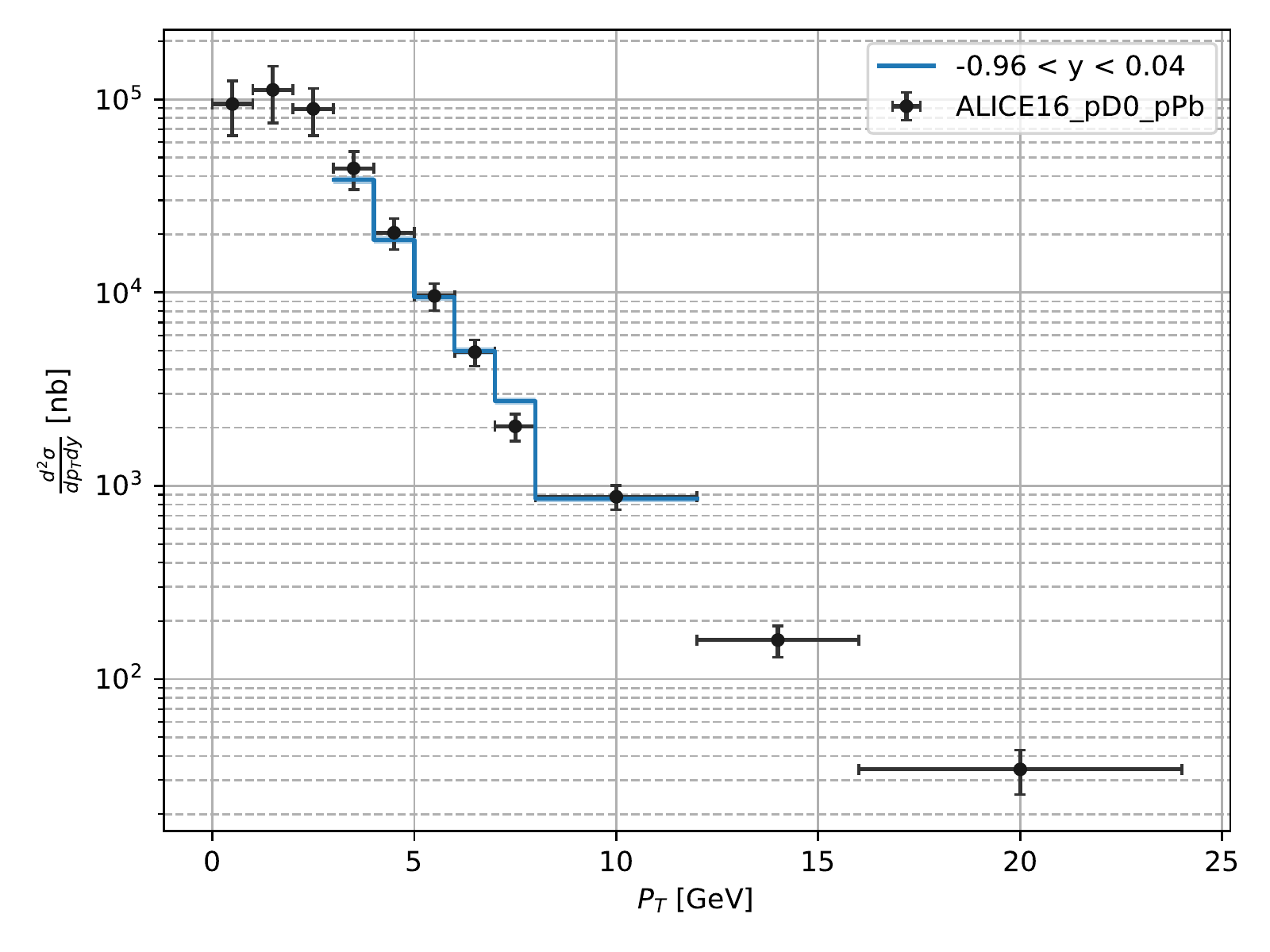}
	\caption{Predictions for $D^0$ production in proton-lead collisions with PDF uncertainties of the nCTEQ15HQ fit. Different rapidity bins are separated by multiplying the cross sections by powers of ten for visual clarity.}
	\label{fig:pPbFitD0}
\end{figure*}

\begin{figure*}[htbp!]
	\centering
	\includegraphics[width=0.48\textwidth]{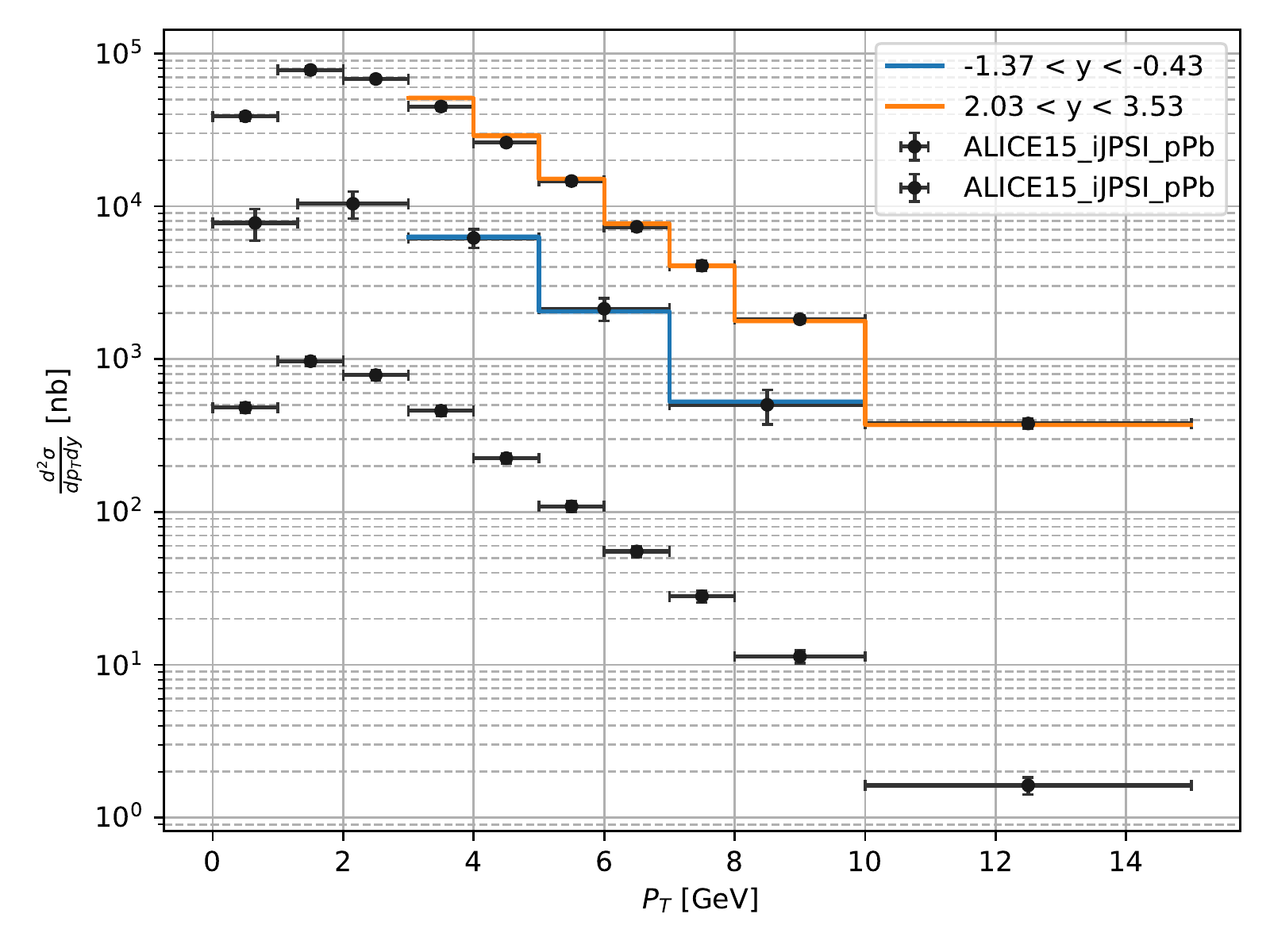}
	\includegraphics[width=0.48\textwidth]{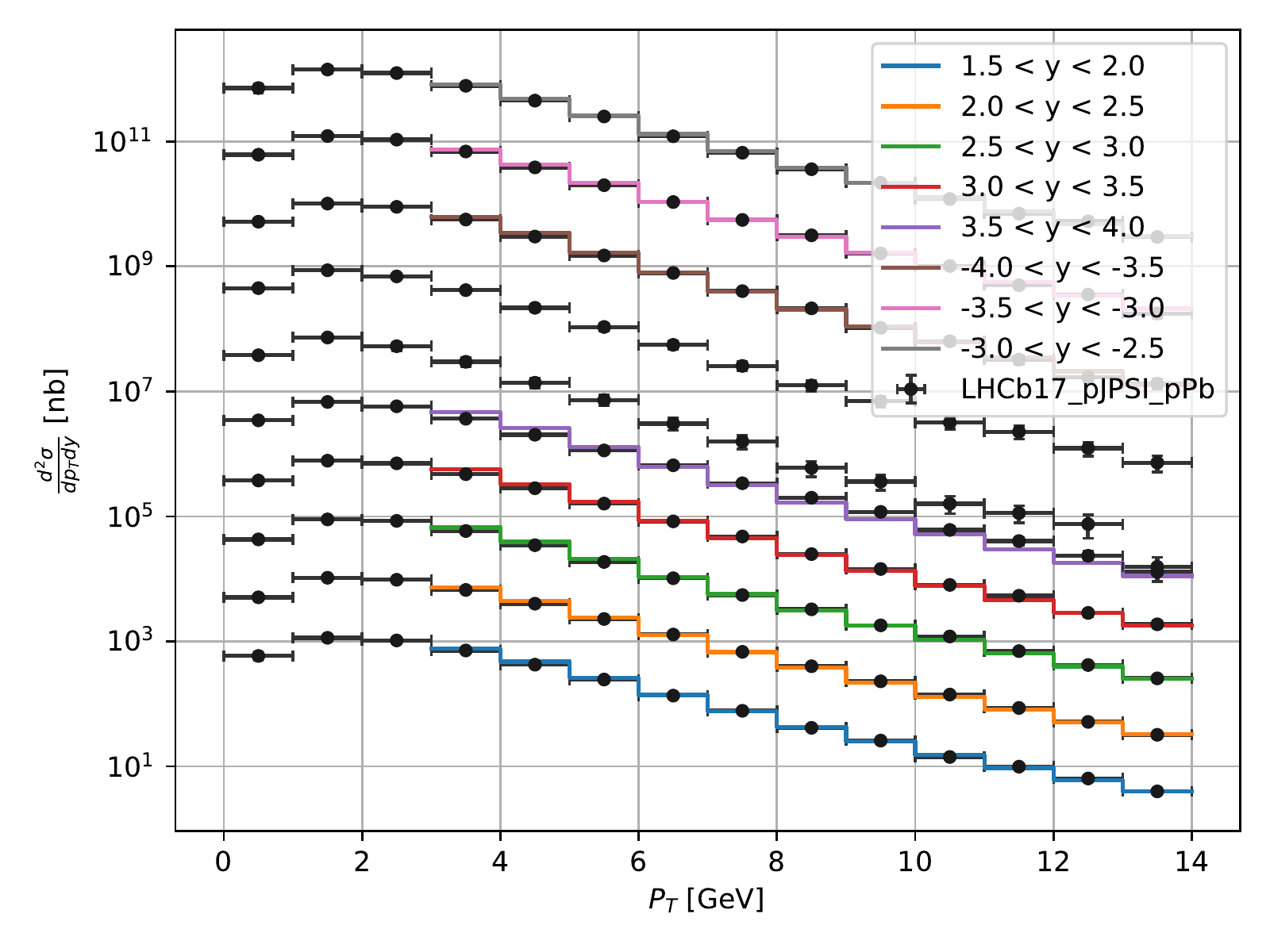}
	\includegraphics[width=0.48\textwidth]{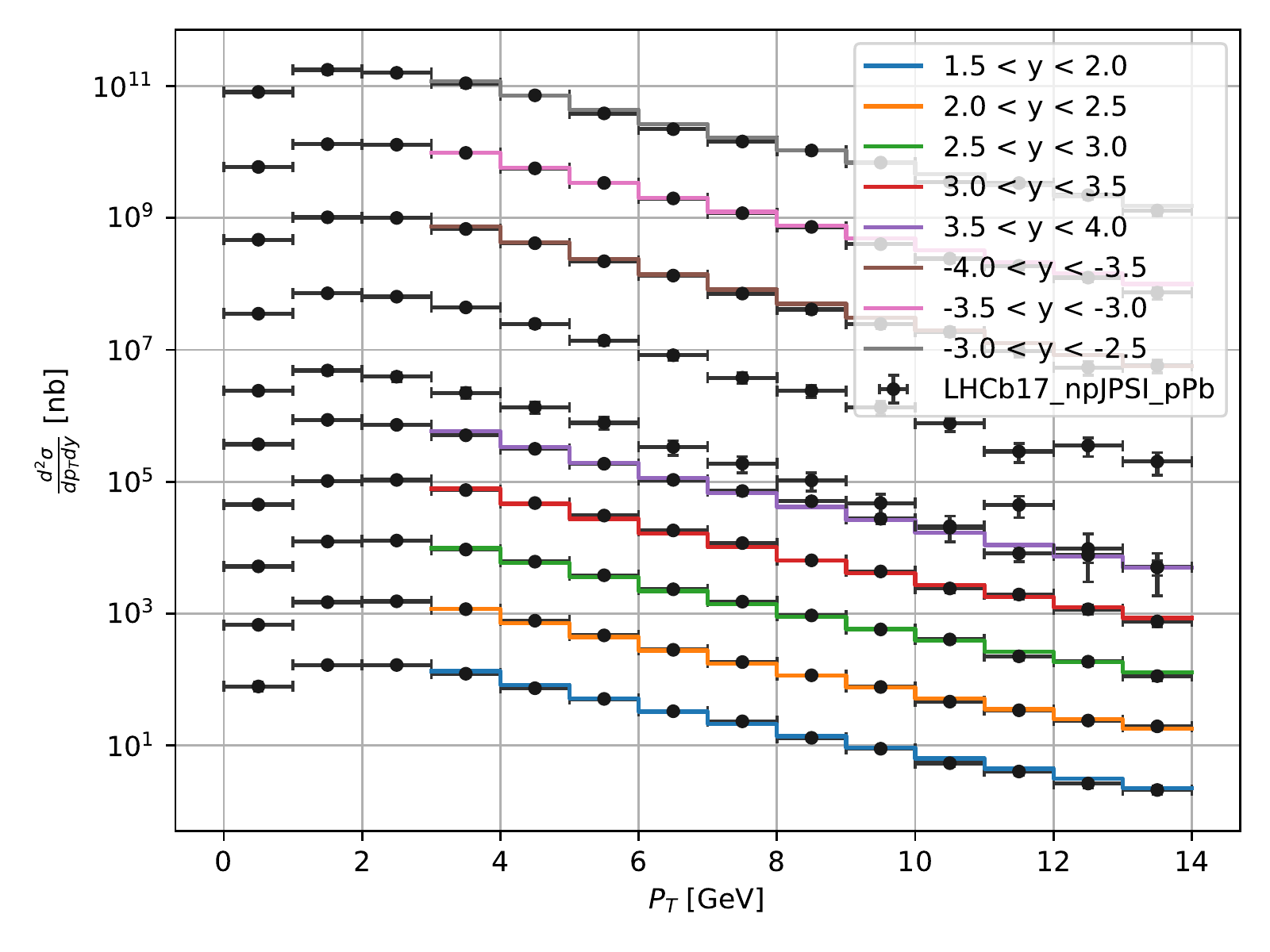}
	\includegraphics[width=0.48\textwidth]{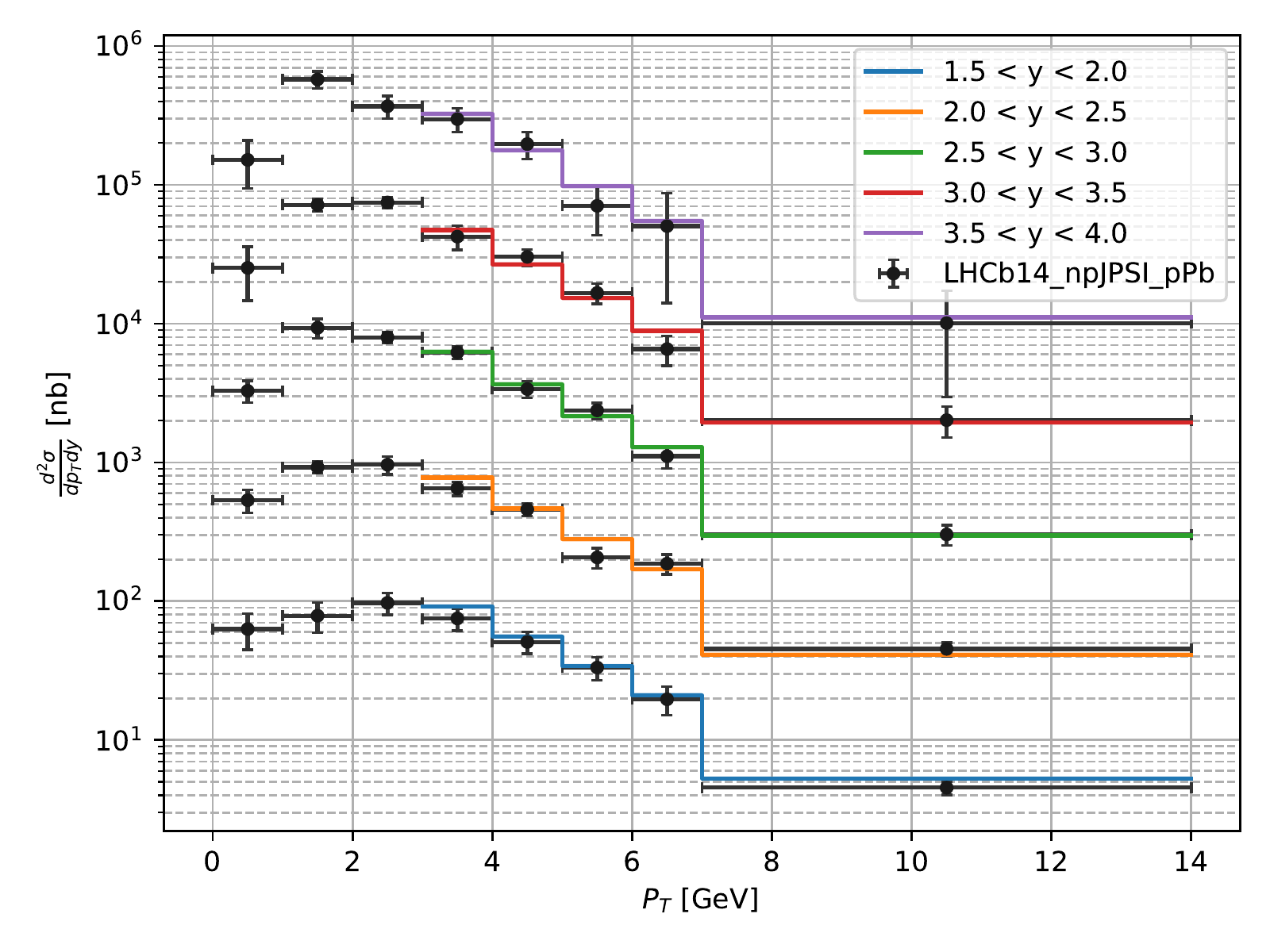}
	\includegraphics[width=0.48\textwidth]{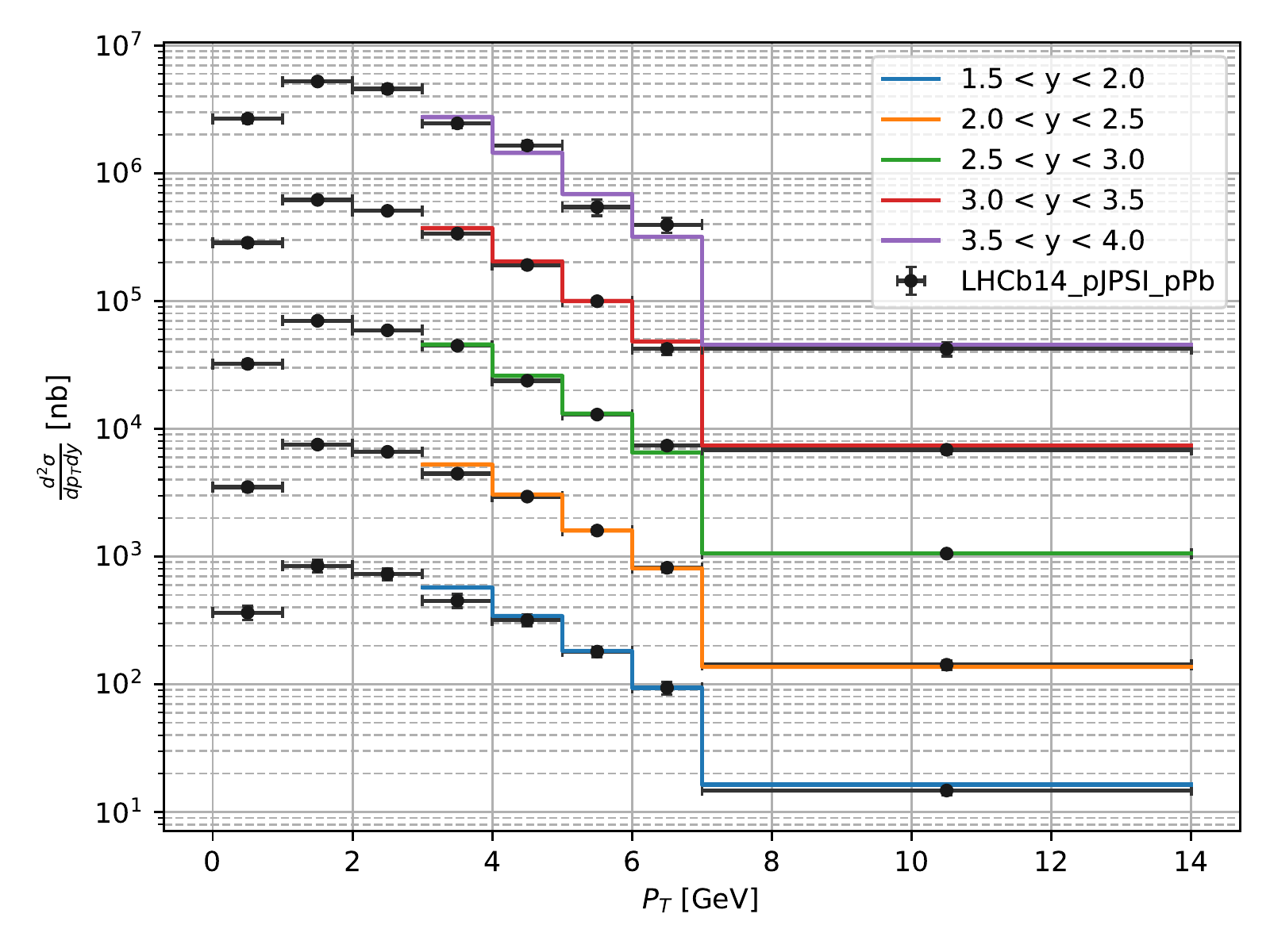}
	\includegraphics[width=0.48\textwidth]{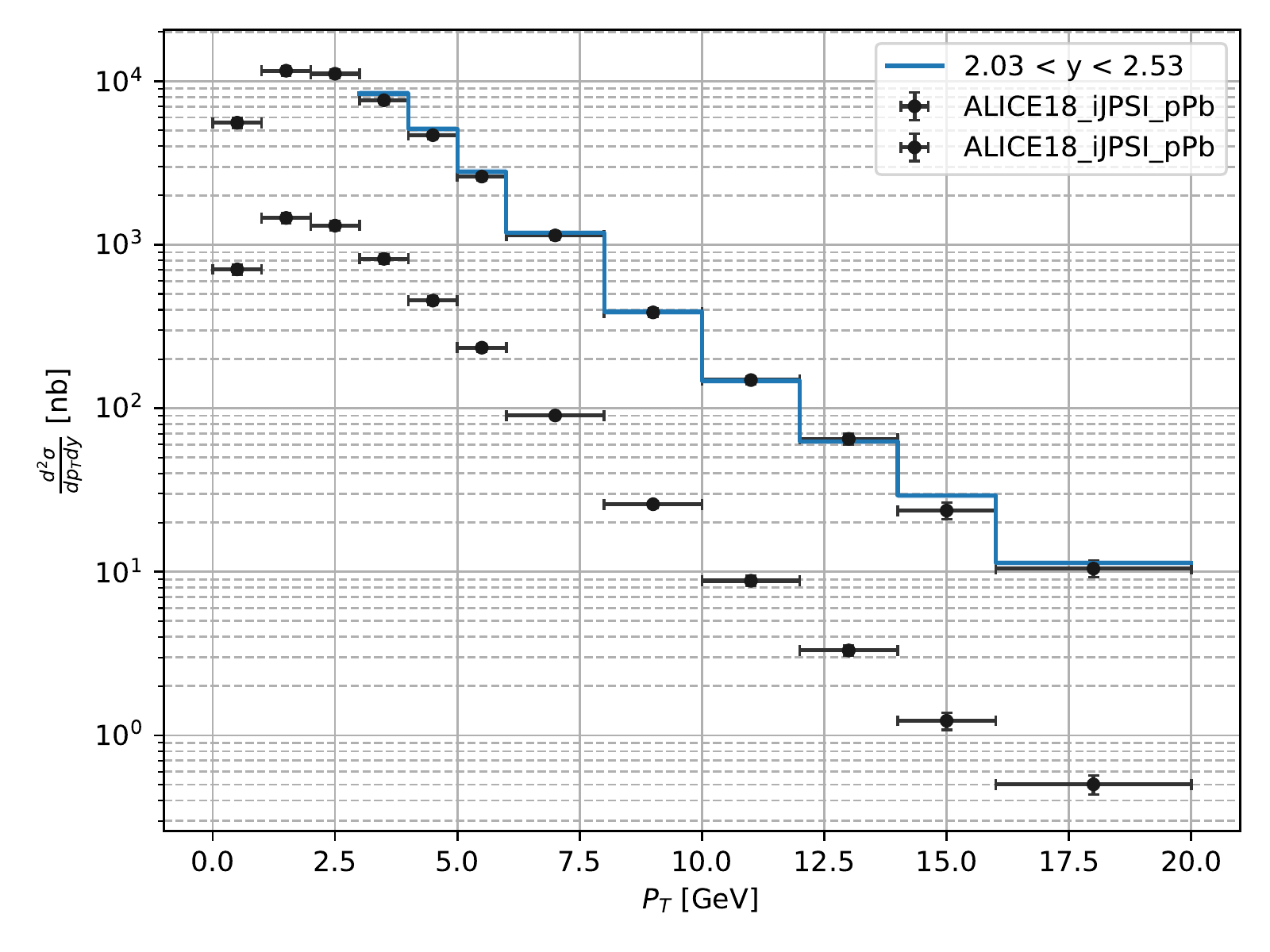}
	\caption{Predictions for $J/\psi$ production in proton-lead collisions with PDF uncertainties of the nCTEQ15HQ fit. Different rapidity bins are separated by multiplying the cross sections by powers of ten for visual clarity.}
	\label{fig:pPbFitJPSI}
\end{figure*}

\begin{figure*}[htbp!]
	\centering
	\includegraphics[width=0.48\textwidth]{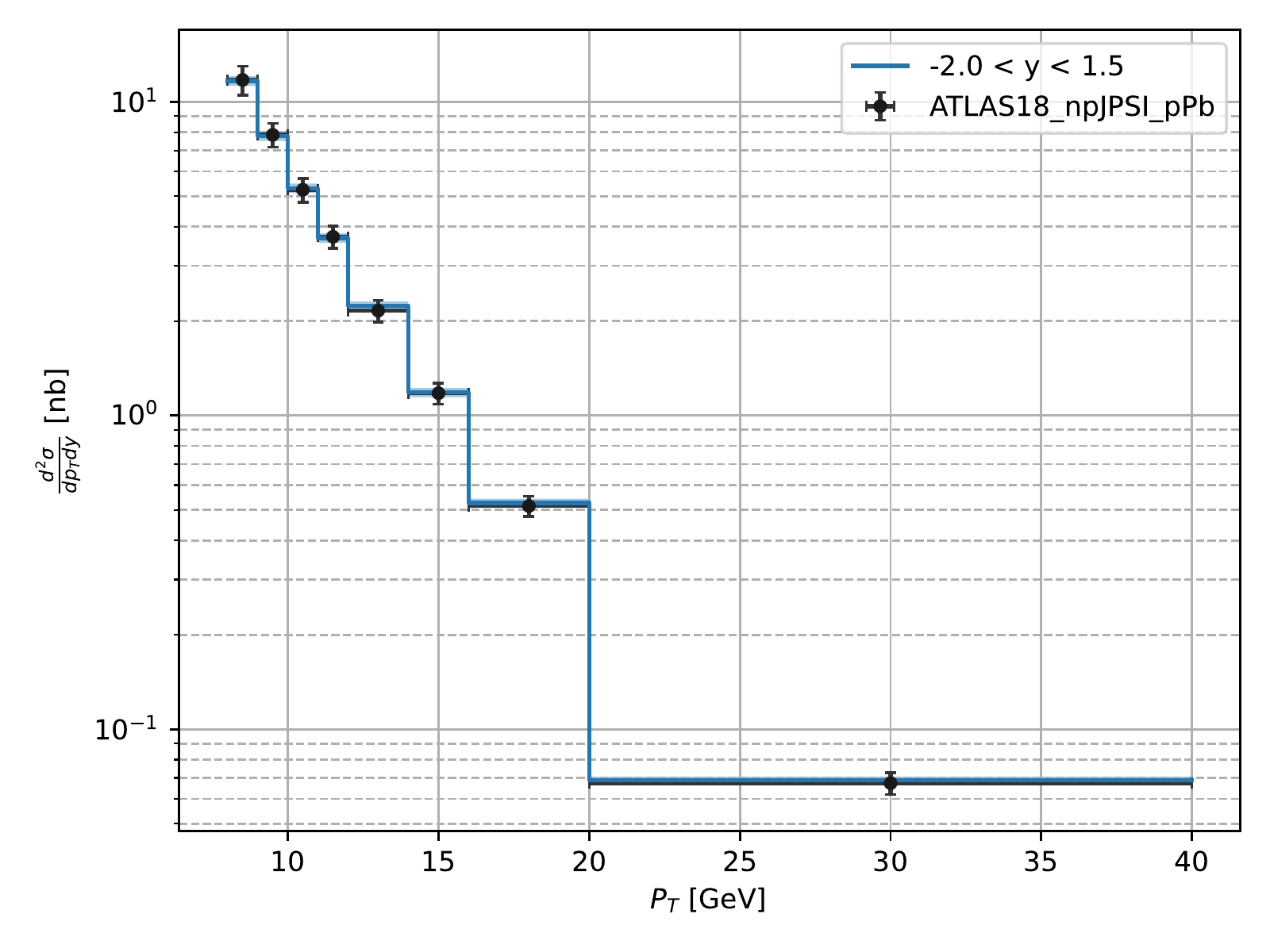}
	\includegraphics[width=0.48\textwidth]{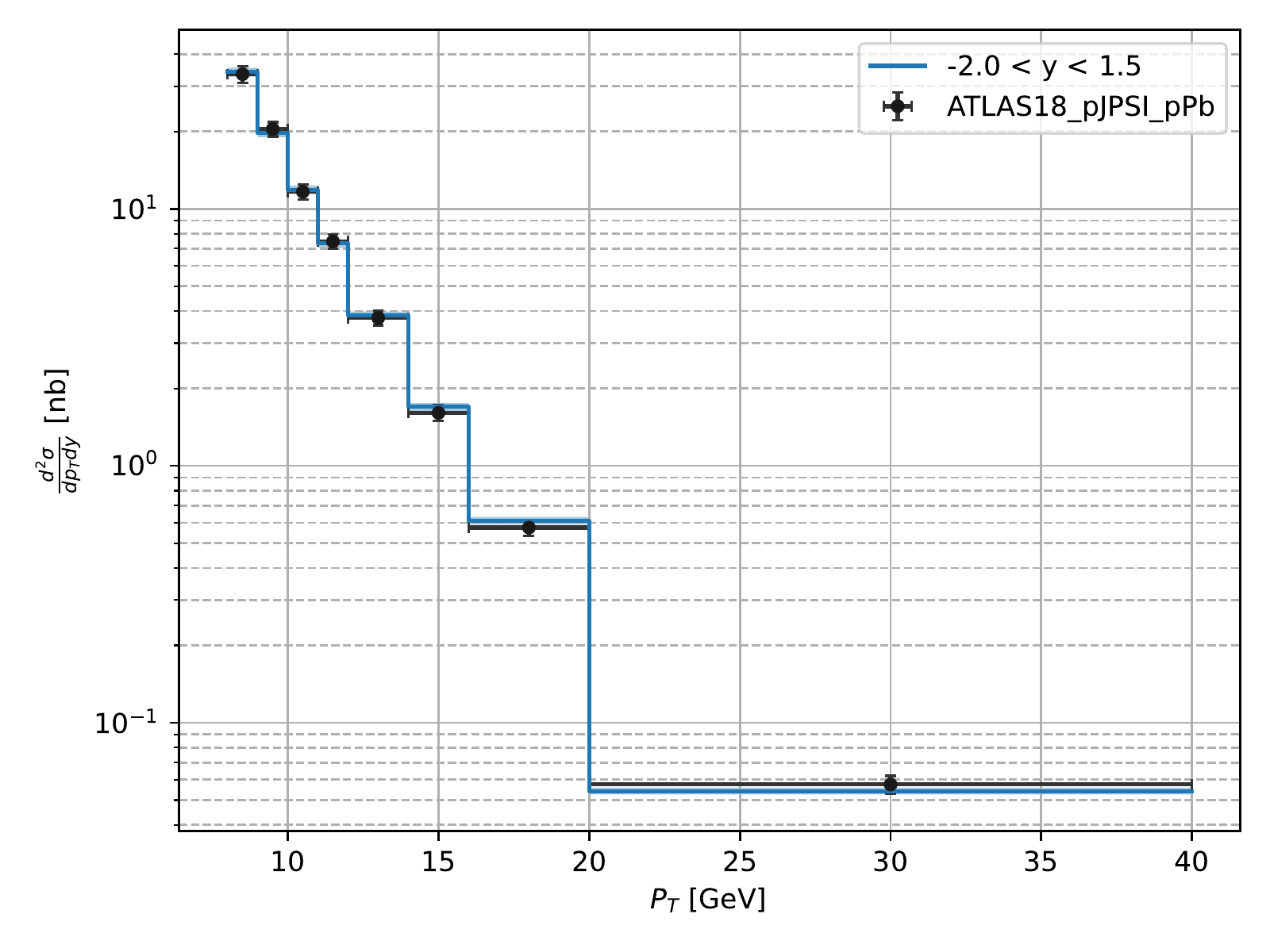}
	\includegraphics[width=0.48\textwidth]{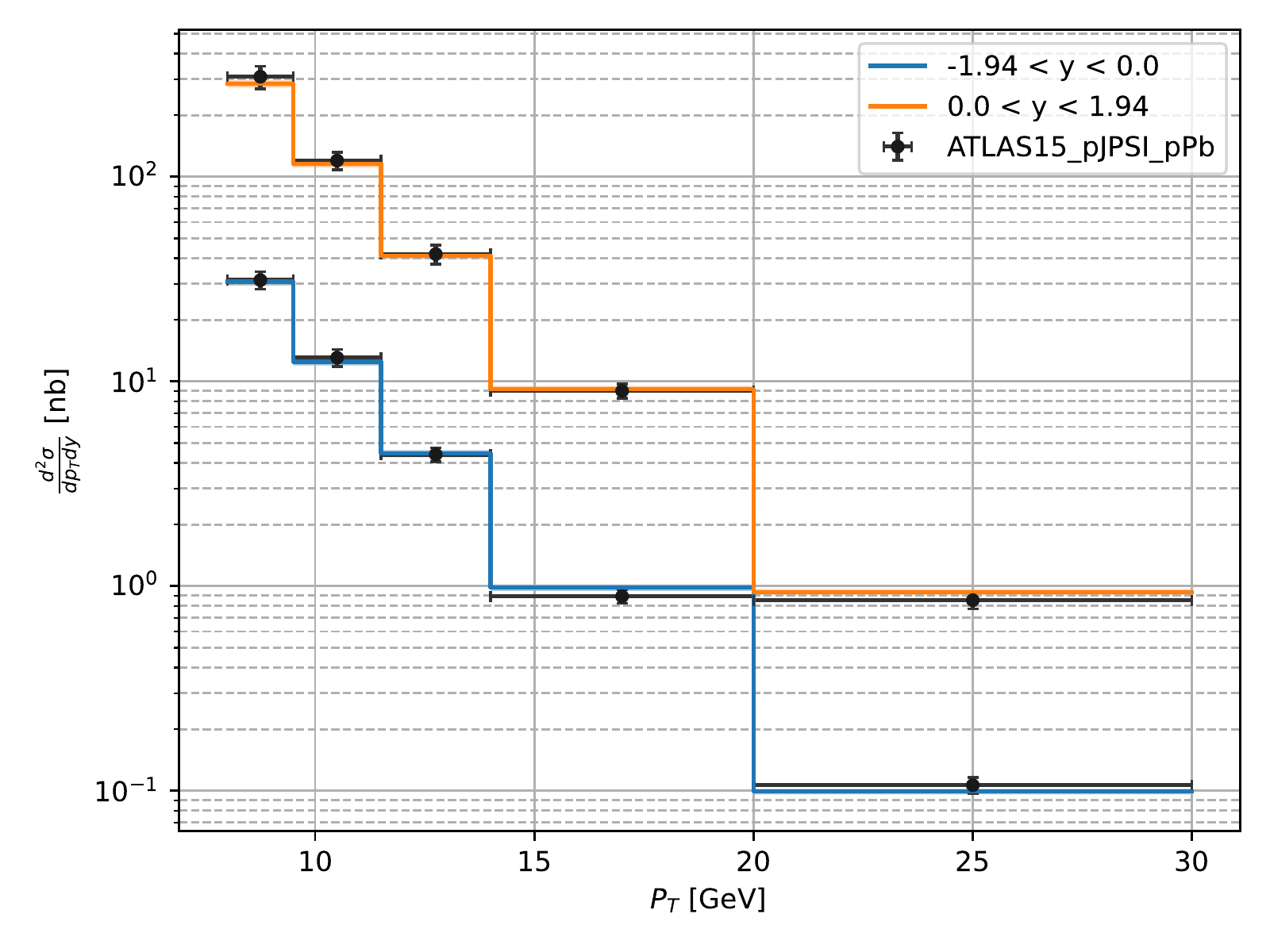}
	\includegraphics[width=0.48\textwidth]{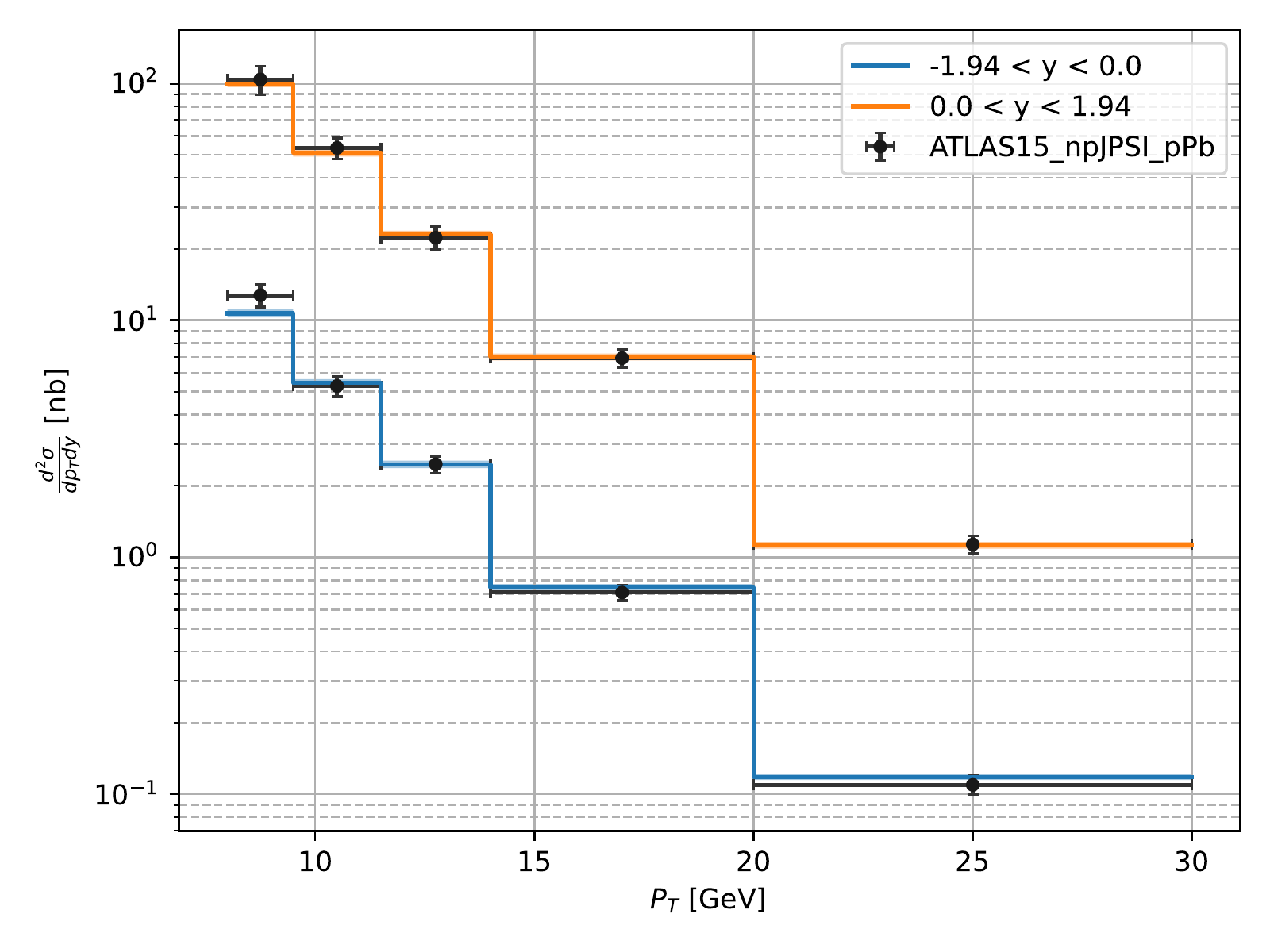}
	\includegraphics[width=0.48\textwidth]{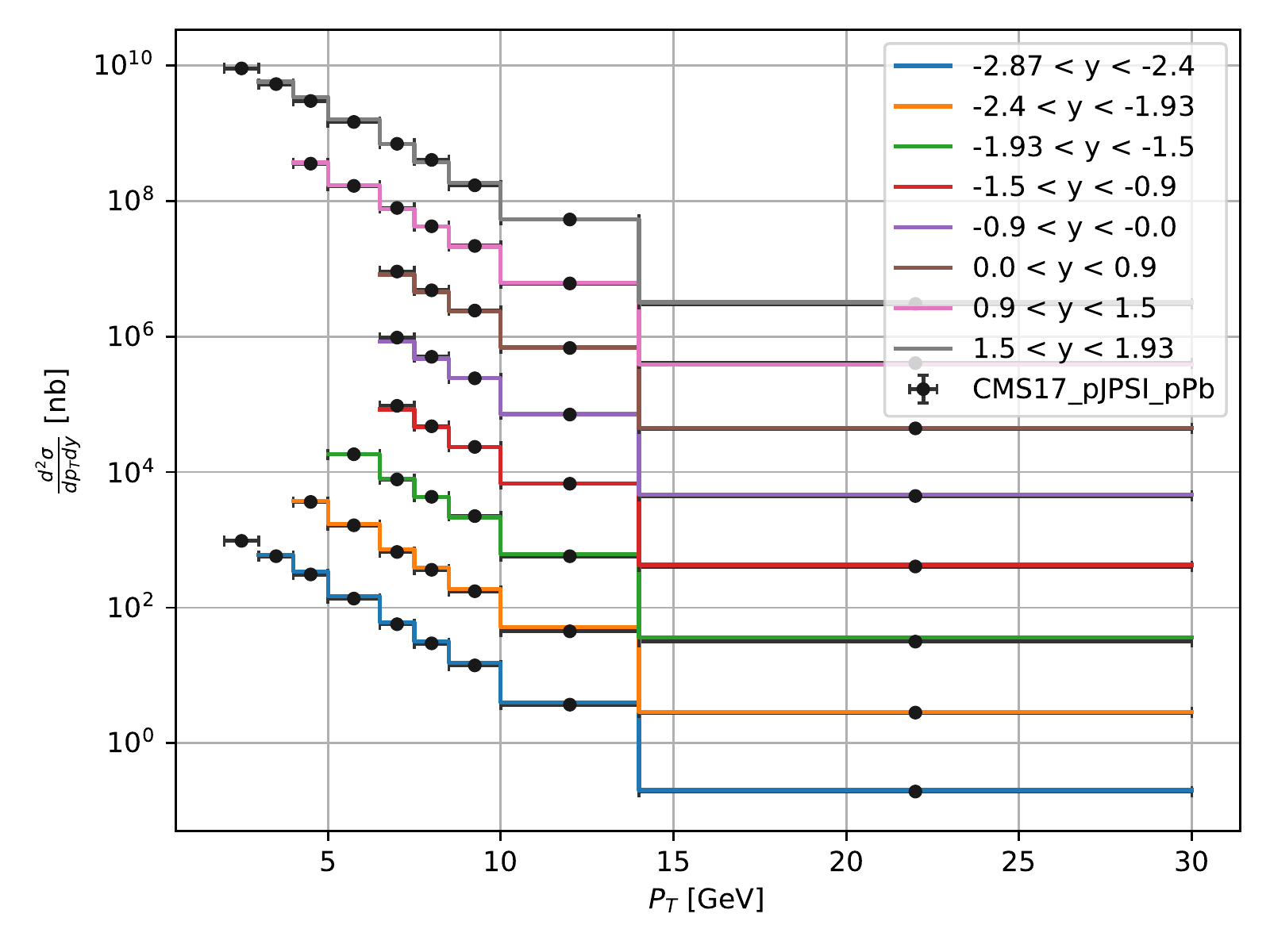}
	\includegraphics[width=0.48\textwidth]{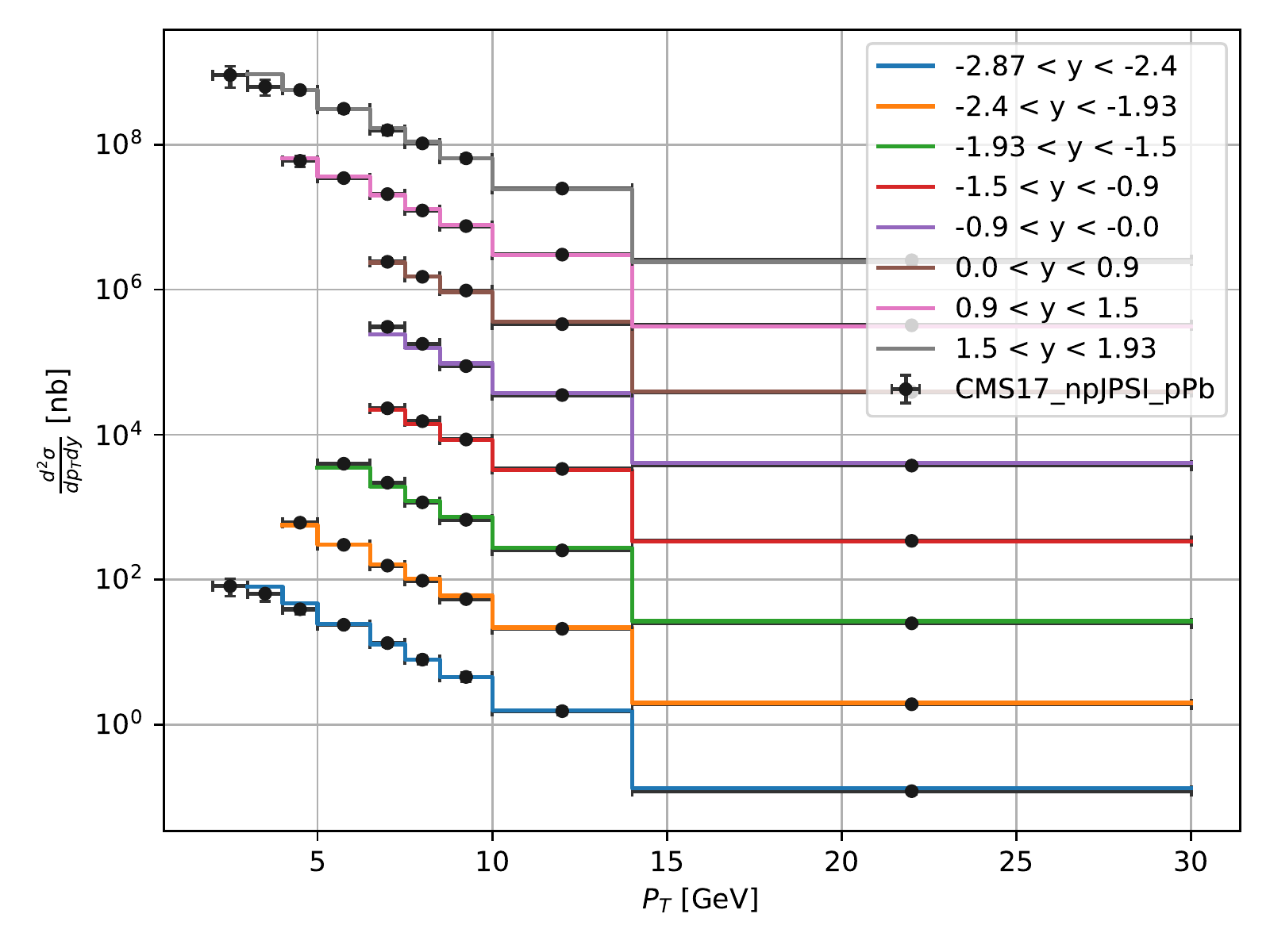}
	\caption{Predictions for $J/\psi$ production in proton-lead collisions with PDF uncertainties of the nCTEQ15HQ fit. Different rapidity bins are separated by multiplying the cross sections by powers of ten for visual clarity.}
	\label{fig:pPbFitJPSI2}
\end{figure*}

\begin{figure*}[htbp!]
	\centering
	\includegraphics[width=0.48\textwidth]{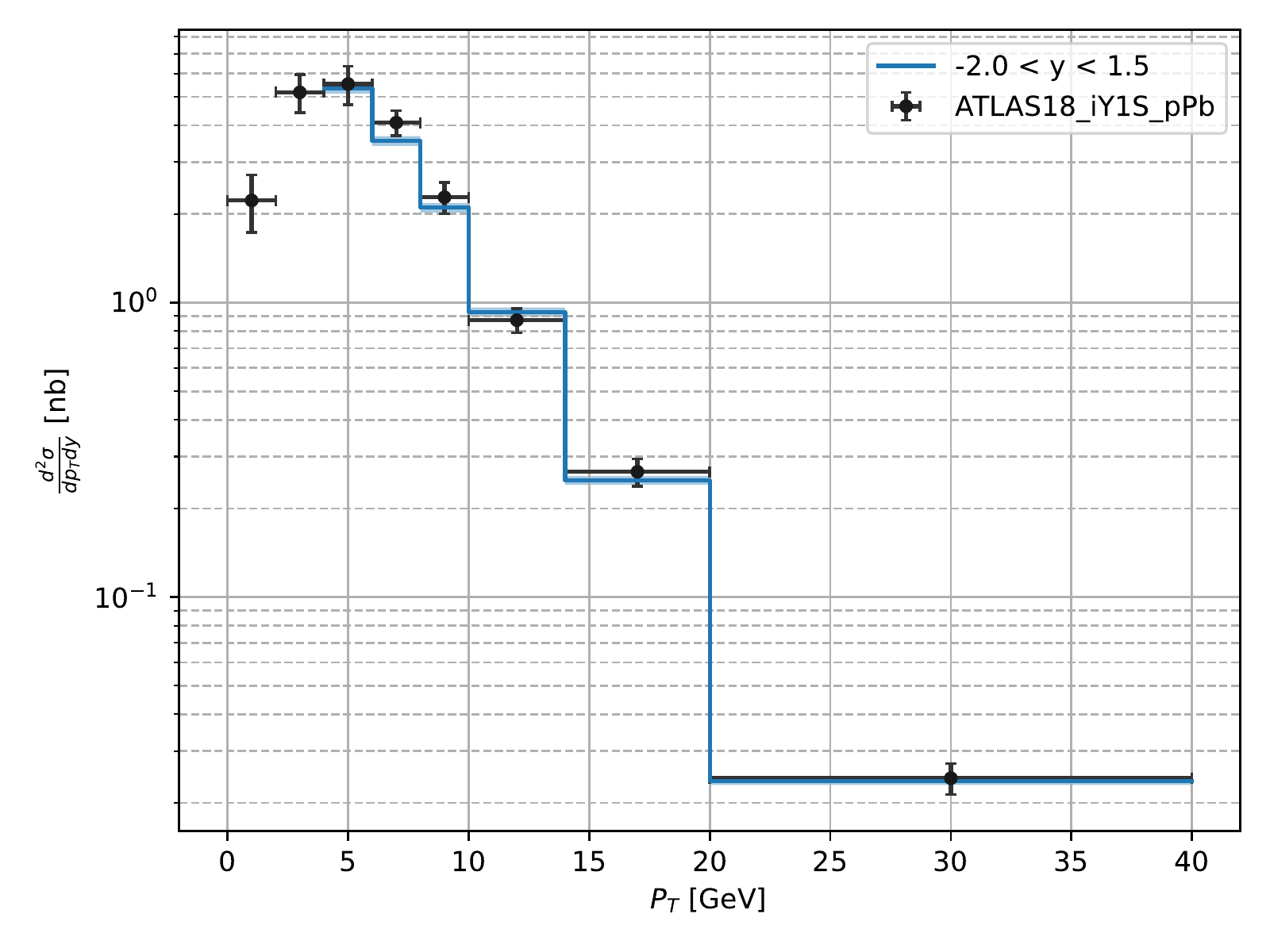}
	\includegraphics[width=0.48\textwidth]{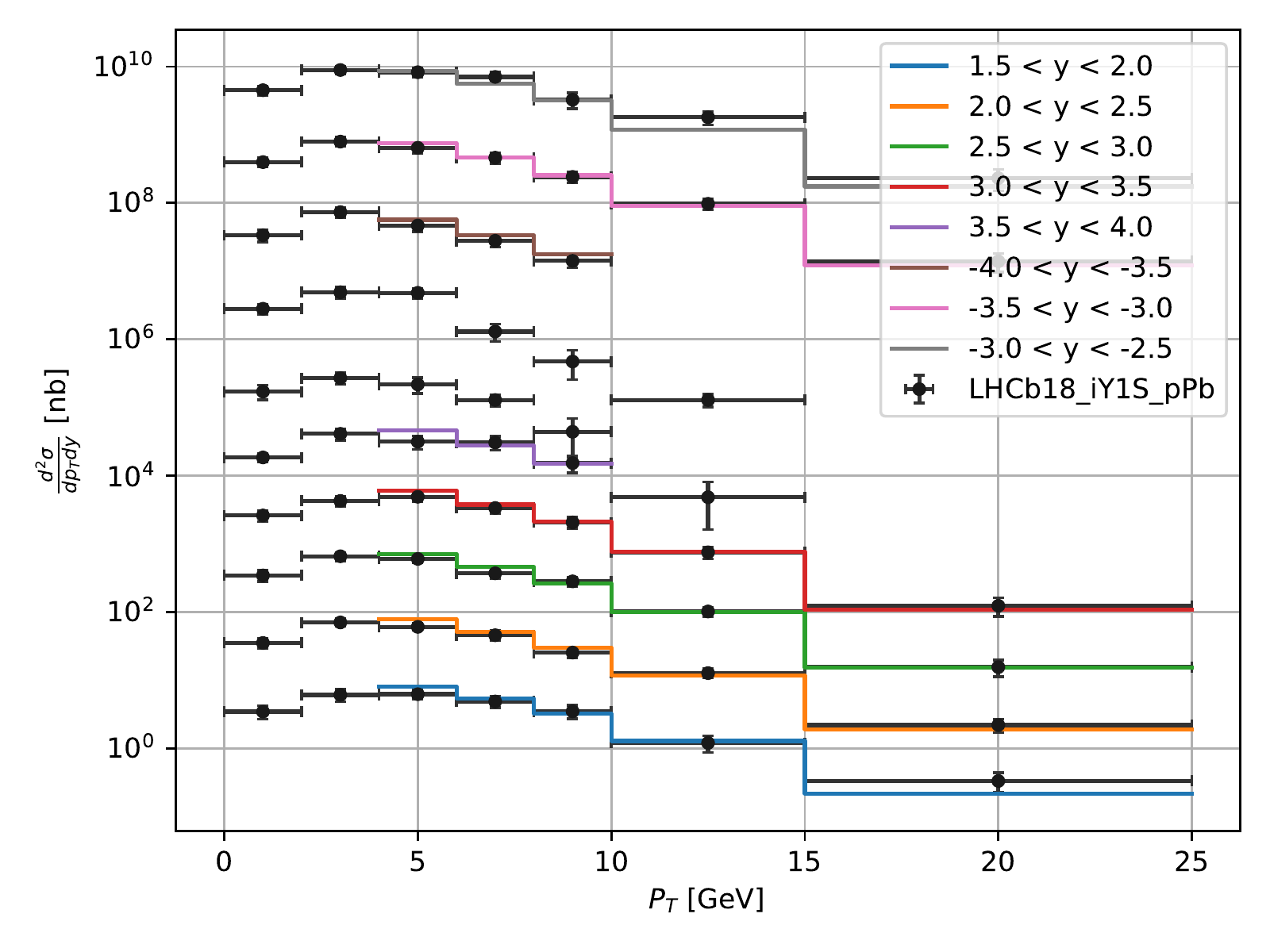}
	\includegraphics[width=0.48\textwidth]{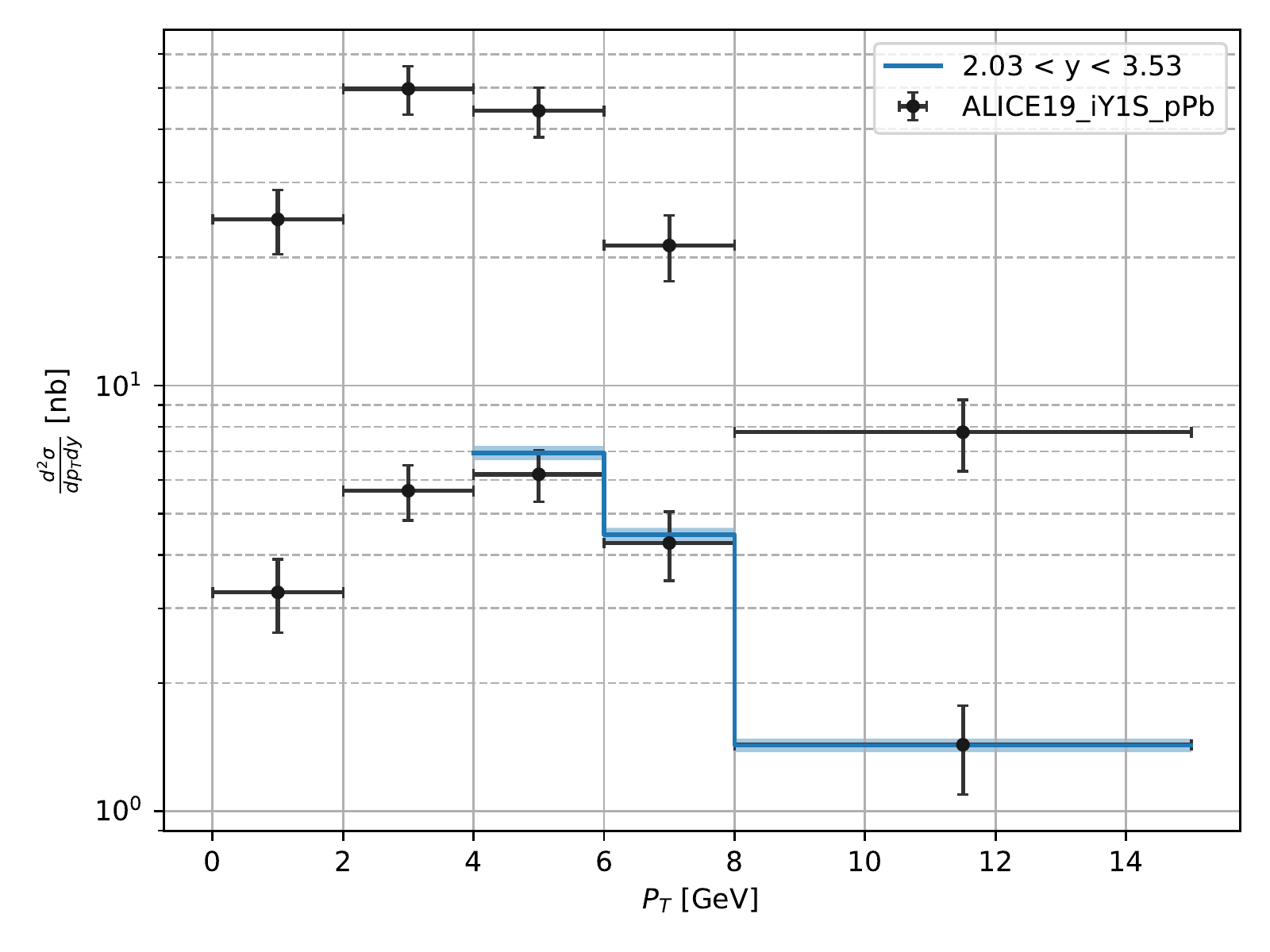}
	\caption{Predictions for $\Upsilon(1S)$ production in proton-lead collisions with PDF uncertainties of the nCTEQ15HQ fit. Different rapidity bins are separated by multiplying the cross sections by powers of ten for visual clarity.}
	\label{fig:pPbFitY1S}
\end{figure*}

\begin{figure*}[htbp!]
	\centering
	\includegraphics[width=0.48\textwidth]{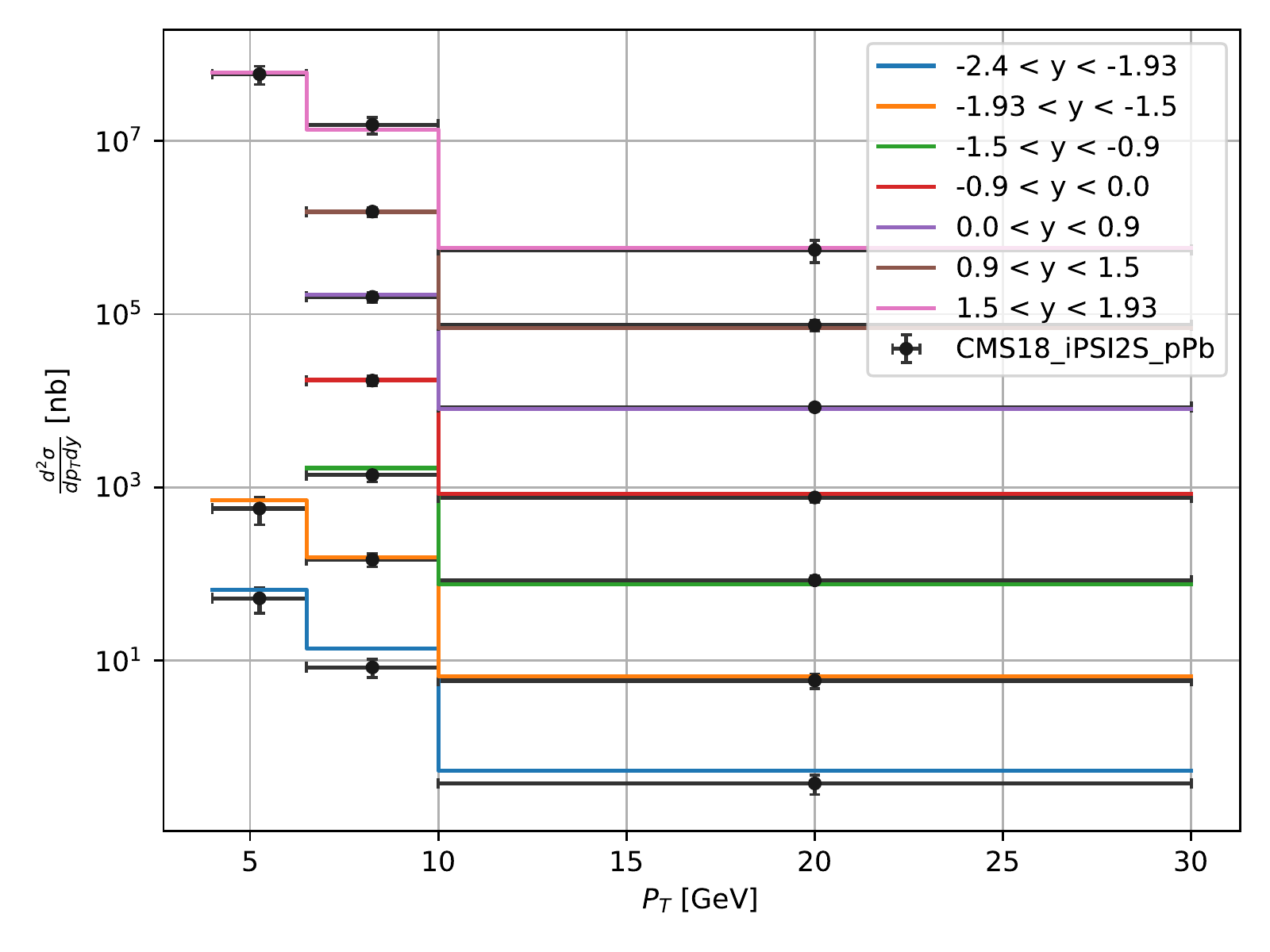}
	\includegraphics[width=0.48\textwidth]{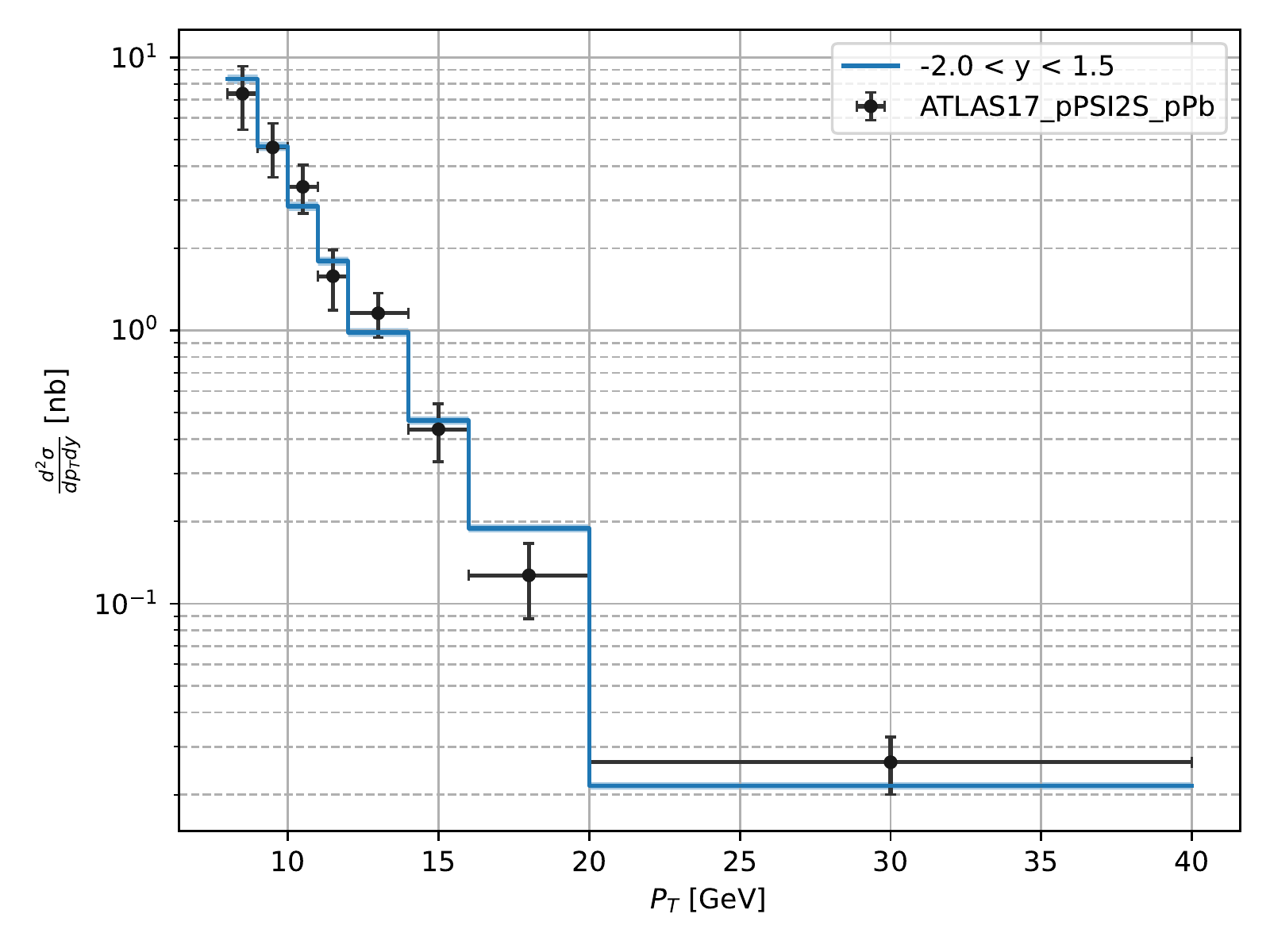}
	\includegraphics[width=0.48\textwidth]{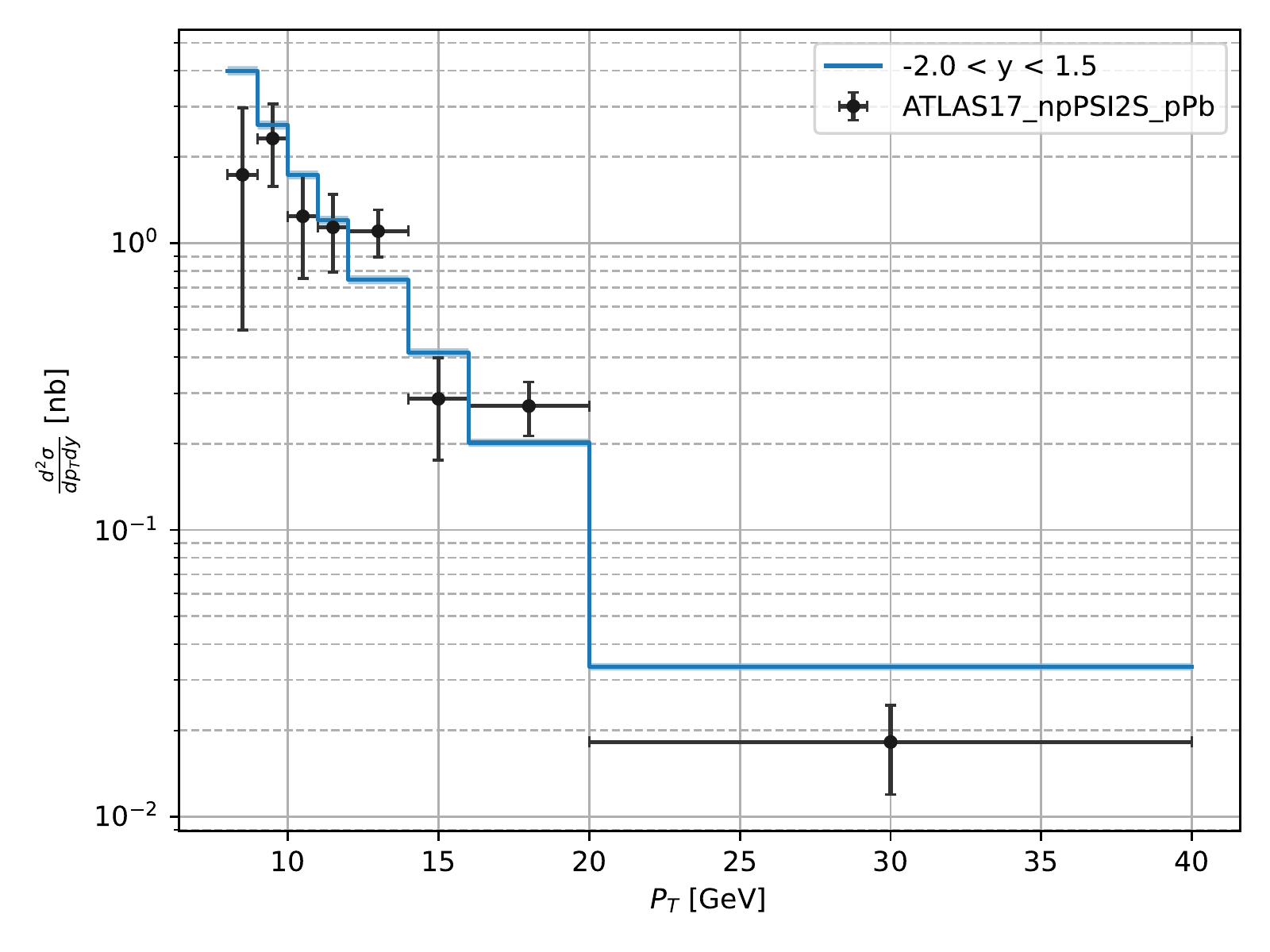}
	\includegraphics[width=0.48\textwidth]{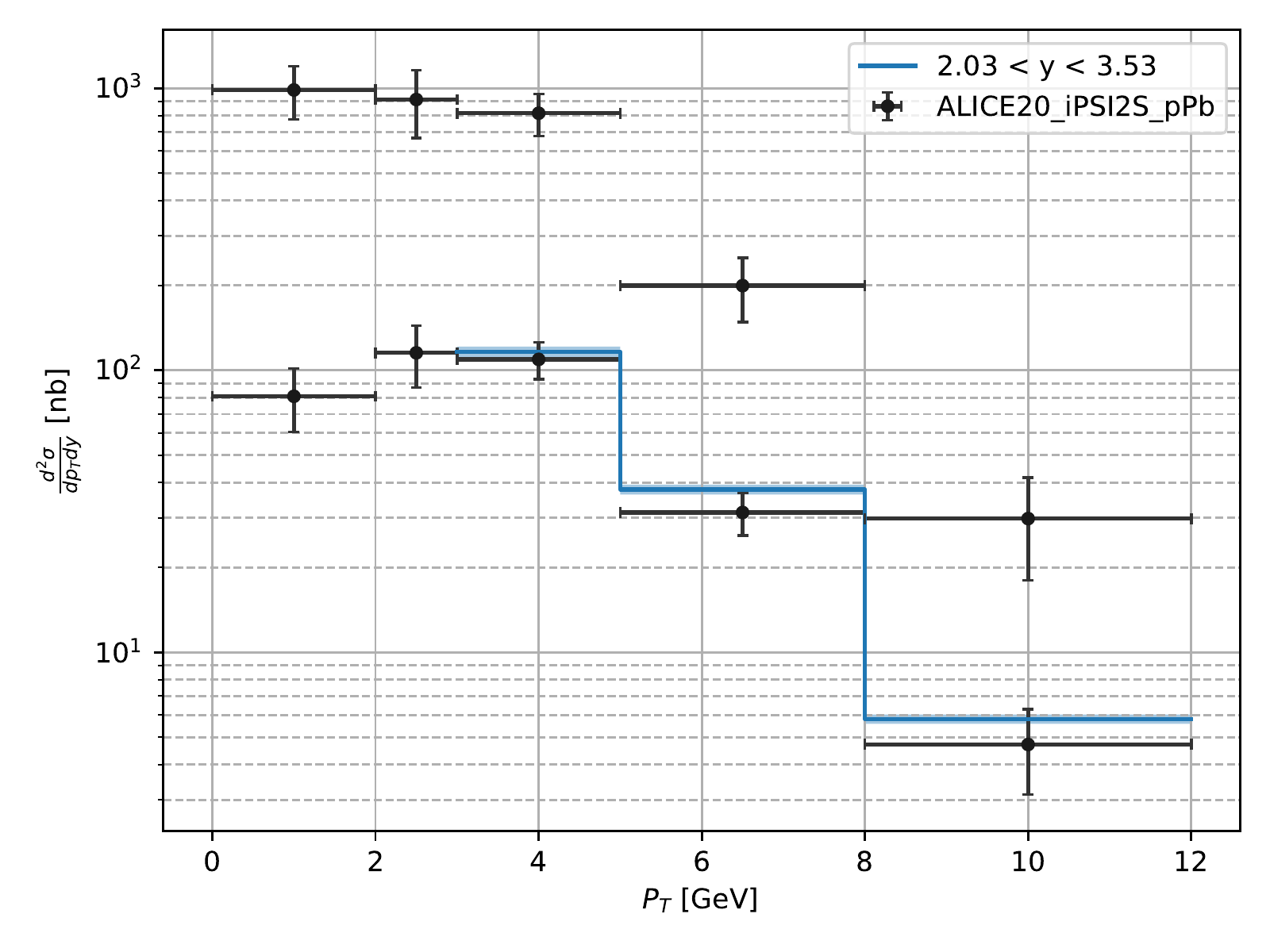}
	\includegraphics[width=0.48\textwidth]{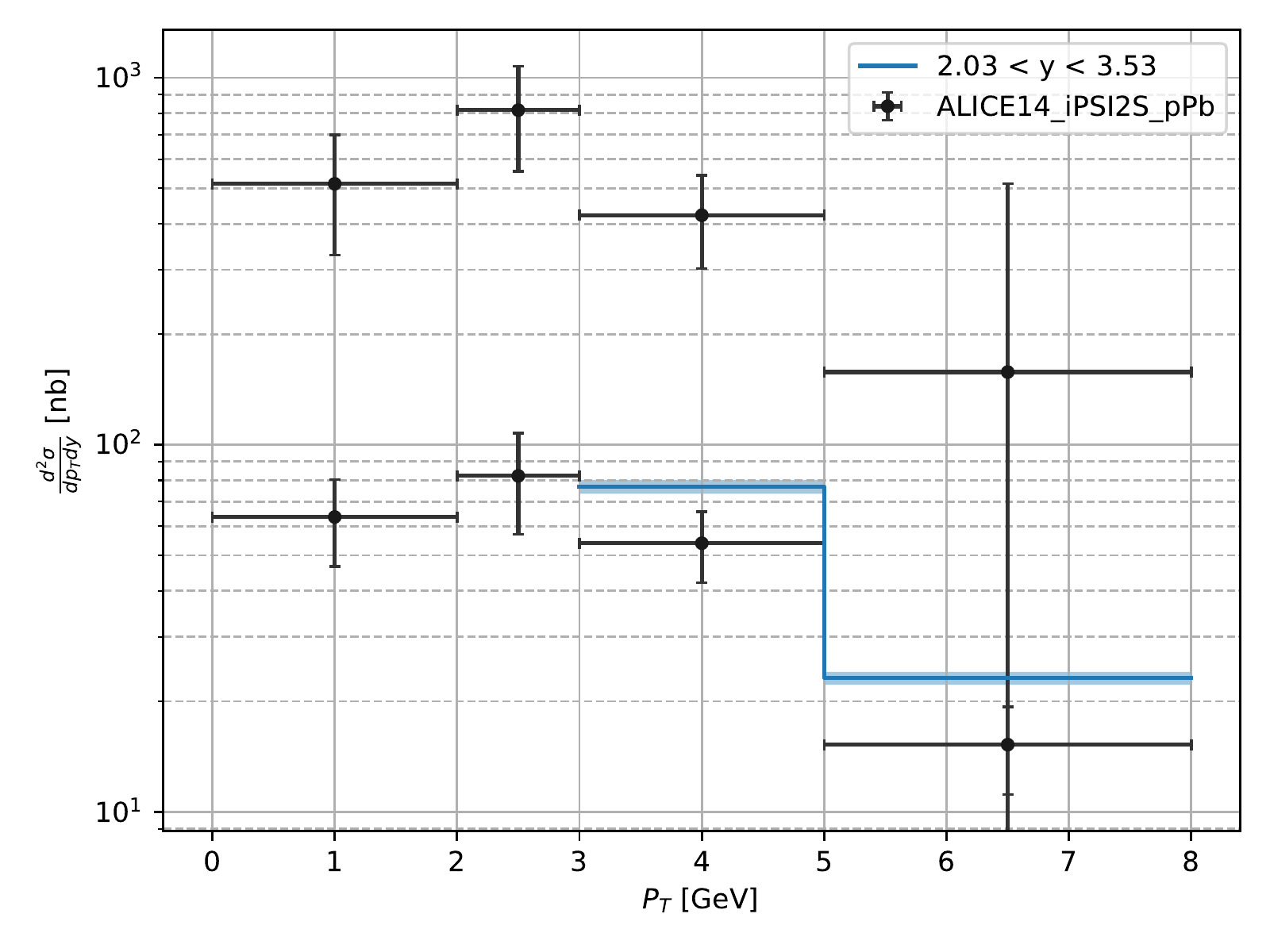}
	\caption{Predictions for $\psi(2S)$ production in proton-lead collisions with PDF uncertainties of the nCTEQ15HQ fit. Different rapidity bins are separated by multiplying the cross sections by powers of ten for visual clarity.}
	\label{fig:pPbFitPSI2S}
\end{figure*}

This appendix contains comparisons between data and predictions made with the Crystal Ball fit. Figs.~\ref{fig:ppBaselineD0}-\ref{fig:ppBaselinePSI2S} show the predictions for the fitted proton-proton data with the uncertainties of the Crystal Ball fit, while Figs.~\ref{fig:pPbFitD0}-\ref{fig:pPbFitPSI2S} show those for the proton-lead data with the uncertainties of the nCTEQ15HQ nuclear PDFs. Note that the proton-lead data is scaled by a factor $\frac{1}{208}$ to give the cross section per nucleon, which is used in the fit.%


\FloatBarrier

\bibliographystyle{utphys}
\bibliography{refs.bib,extra.bib}
\printtables 
\printfigures

\end{document}